\documentclass[seceq]{ptptex}

\usepackage{amsmath}
\usepackage{amsfonts}
\usepackage{latexsym}
\usepackage{amssymb} 
\usepackage{verbatim}
\usepackage{amsthm}
\usepackage{graphicx}
\usepackage{wrapft}

\title{
Perturbations of Matter Fields in the Second-order
Gauge-invariant Cosmological Perturbation Theory
}

\author{
  Kouji \textsc{Nakamura}%
}

\inst{
  Department of Astronomical Science,
  the Graduate University for Advanced Studies,
  Mitaka, Tokyo 181-8588, Japan.
}

\abst{
  Some formulae for the perturbations of the matter fields are
  summarized within the framework of the second-order
  gauge-invariant cosmological perturbation theory in a four 
  dimensional homogeneous isotropic universe, which is developed
  in the papers [K.~Nakamura, Prog.~Theor.~Phys.\ {\bf 117}
  (2007), 17.].
  We derive the formulae for the perturbations of the energy
  momentum tensors and equations of motion for a perfect fluid,
  an imperfect fluid, and a signle scalar field, and show that
  all equations are derived in terms of gauge-invariant
  variables without any gauge fixing.
}

\begin{document}

\maketitle

\section{Introduction}
\label{sec:intro}


The general relativistic second-order cosmological perturbation
theory is one of topical subjects in the recent cosmology. 
By the recent observation of the Cosmic Microwave Background
(CMB) by Wilkinson Microwave Anisotropy Probe\cite{WMAP}, the
first order approximation of our universe from a homogeneous
isotropic one was revealed.
The observational results suggest that the fluctuations of our
universe are adiabatic and Gaussian at least in the first order
approximation.
As a next step, the clarifications of the accuracy of these
observational results are actively discussed both in the
observational\cite{Non-Gaussianity-observation} and the
theoretical
side\cite{Non-Gaussianity-inflation,Non-Gaussianity-in-CMB}
through the non-Gaussianity, the non-adiabaticity, and so on.
With the increase of precision of the CMB data, the study of
relativistic cosmological perturbations beyond linear order is a
topical subject especially to study the generation of the
primordial non-Gaussianity in inflationary
scenarios\cite{Non-Gaussianity-inflation} and the non-Gaussian 
component in CMB anisotropy\cite{Non-Gaussianity-in-CMB}.
The second-order cosmological perturbation theory is one of
such perturbation theories beyond linear order.


According to this physical motivation, we proposed a clear
gauge-invariant formulation of the second-order general
relativistic cosmological perturbation
theory\cite{kouchan-cosmo-second}. 
In this paper, we refer these works as KN2007.
This gauge-invariant formulation of the second-order
cosmological perturbations is a natural extension of the
first-order gauge-invariant cosmological perturbation
theory\cite{Bardeen-1980,Kodama-Sasaki-1984,Mukhanov-Feldman-Brandenberger-1992}. 
The formulation in KN2007 is one of the applications of the
gauge-invariant formulation of the second-order perturbation
theory on the generic background spacetime developed in the
papers by the present
author\cite{kouchan-gauge-inv,kouchan-second}.
These papers are referred in this paper as
KN2003\cite{kouchan-gauge-inv} and KN2005\cite{kouchan-second}.
This general formulation is a by-product of the investigations
of the oscillatory behaviors of self-gravitating Nambu-Goto
membranes\cite{kouchan-papers}.


In KN2007, we defined the complete set of the gauge-invariant
variables of the second-order cosmological perturbations in the
Friedmann-Robertson-Walker universe based on the formulation
developed in the papers KN2003 and KN2005.
We considered the two cases of the Friedmann-Robertson-Walker
universe: one is the universe filled with the single perfect
fluid and another is the universe filled with the single scalar
field.
We also derived the second-order Einstein equations of
cosmological perturbations in terms of these gauge-invariant
variables without any gauge fixing in these two cases.
We have also found that the procedure to find gauge invariant
variables proposed in KN2003 plays a crucial role in the
derivations.
Rather, we can use the formulae proposed in KN2003 to check
whether the resulting formulae are correct or not.


This paper is the second part of KN2007.
In this paper, we summarize the formulae for the components of
the first- and the second-order perturbations of the energy
momentum tensors and the equations of motion which are derived
from the divergence of the energy momentum tensors.
As the matter contents, we consider the three matter fields: a
single perfect fluid; a single imperfect fluid which includes 
additional terms of the energy flux and anisotropic stress to
the perfect fluid; and a single scalar field.
In the early universe, photon and neutrinos should be described
by the Boltzmann distribution 
functions\cite{Kodama-Sasaki-1984,anisotropic-stress-in-cosmology}.
Photon's interaction with baryon and the free streaming of
neutrinos lead anisotropic stress and these effects will be
important in the recent cosmology.
Although the energy flux and the anisotropic stress in the
imperfect fluid are determined through these micro-physical
processes, we just phenomenologically treat these terms in this 
paper.


Although the perturbative expressions of the energy momentum
tensor and equations of motion were also derived in some
literatures\cite{Noh-Hwang-2004}, in this paper, we show
alternative derivations of these perturbations.
In our derivations, we respect the gauge invariance of the
perturbative variables.
We again show that the formulae of gauge invariant-variables
proposed in KN2003 [Eqs.~(\ref{eq:matter-gauge-inv-decomp-1.0})
and (\ref{eq:matter-gauge-inv-decomp-2.0}) in this paper] also
play crucial roles in the derivations of perturbative
expressions of the equations for matter fields.
The first- and the second-order perturbations of the equations
of motion are decomposed into gauge-invariant and gauge-variant
parts as Eqs.~(\ref{eq:matter-gauge-inv-decomp-1.0}) and
(\ref{eq:matter-gauge-inv-decomp-2.0}), respectively. 
In these derivations, we do not fix any gauge degree of freedom.
In spite of this no gauge-fixing, we show all perturbations of
the equations of motion are given in gauge-invariant forms
through the lower order perturbations of the equations of motion
for matter fields.
In this sense, we may say that the general framework of the
second-order gauge invariant perturbations proposed in KN2003
and KN2005 does work not only in the perturbations of the
Einstein equations but also in the equations of motion for the
matter fields.
The main purpose of this paper is to show this.


Further, in this paper, we do not ignore the first-order vector-
and tensor-modes which are ignored in KN2007.
Moreover, in the derivation of the perturbations of the energy
momentum tensors and the equations of motion, we do not use any
information of the Einstein equations.
Therefore, the formulae derived in this paper are valid even if
we consider any other theory of gravity than the Einstein theory.


The organization of this paper is as follows. In
\S\ref{sec:Second-order-cosmological-perturbatios},
we briefly review the definitions of the gauge-invariant
variables for the first- and second-order perturbations which
were defined by KN2007\cite{kouchan-cosmo-second}.
In
\S\ref{sec:Generic-form-of-perturbations-of-energy-momentum-tensors-and-equations-of-motion},
we derive the first- and the second-order perturbations of the
energy momentum tensors and equations of motion for a perfect
fluid, an imperfect fluid, and a scalar field.
In the derivation in this section, we do not specify the
background spacetime.
Therefore, the formulae summarized in this section are valid in
perturbation theories on any background spacetime.
In \S\ref{sec:Explicit}, we derive the explicit expression of the
components of the energy momentum tensors and the equations of
motion for matter fields.
The final section, \S\ref{sec:summary}, is devoted to the
summary and the discussions.
Further, in Appendix
\ref{sec:Perturbations-of-accerelation-expansion-shear-and-rotation}, 
we explicitly give the components of the perturbations of the
acceleration, expansion, shear, and rotation associated with the
fluid four-velocity, which are necessary to derive the results
in
\S\S\ref{sec:Generic-form-of-perturbations-of-energy-momentum-tensors-and-equations-of-motion}
and \ref{sec:Explicit}.


We employ the notation of our previous papers
KN2003\cite{kouchan-gauge-inv}, KN2005\cite{kouchan-second}, and
KN2007\cite{kouchan-cosmo-second} and use the abstract
index notation\cite{Wald-book}.
We also employ the natural unit in which the light velocity is
denoted by $c=1$.


\section{Gauge-invariant variables in the second-order perturbations}
\label{sec:Second-order-cosmological-perturbatios}


In this section, we briefly review the definitions of the
gauge-invariant variables.  
First, in
\S\ref{sec:Second-order-cosmological-perturbatios-gauge}, we
review the gauge degree of freedom and the first- and the 
second-order gauge transformation rules.
Then, we briefly explain the gauge-invariant variables for the 
metric (in
\S\ref{sec:Second-order-cosmological-perturbatios-metric}) and
matter perturbations (in
\S\ref{sec:Second-order-cosmological-perturbatios-matter-fundamental}).


\subsection{Gauge degree of freedom}
\label{sec:Second-order-cosmological-perturbatios-gauge}


In any perturbation theory, we always treat two spacetime
manifolds.
One is the physical spacetime ${\cal M}={\cal M}_{\lambda}$ and
the other is the background spacetime ${\cal M}_{0}$.
The physical spacetime ${\cal M}_{\lambda}$ is our nature
itself which we want to describe through the perturbations.
On the other hand, the background spacetime ${\cal M}_{0}$ is
just a reference spacetime for the calculations of
perturbations.
Although this background spacetime has nothing to do with our
nature, to calculate perturbations, it is necessary to
introduce this reference spacetime ${\cal M}_{0}$ by hand.
Since these two spacetime manifolds are distinct from each
other, we have to introduce a point-identification map 
${\cal X}_{\lambda}: {\cal M}_{0} \rightarrow {\cal M}_{\lambda}$.
This point-identification map ${\cal X}_{\lambda}$ called a
gauge choice in perturbation theories.
Through the pull-back ${\cal X}^{*}_{\lambda}$ of the gauge choice
${\cal X}_{\lambda}$, any physical variable $\bar{Q}_{\lambda}$
on the physical manifold ${\cal M}_{\lambda}$ is pulled back to 
${\cal X}^{*}_{\lambda}\bar{Q}_{\lambda}$ on the background
spacetime ${\cal M}_{0}$.
The pull-back ${\cal X}^{*}_{\lambda}\bar{Q}_{\lambda}$ is a
representation on the background spacetime ${\cal M}_{0}$ of the
physical variable $\bar{Q}_{\lambda}$ on the physical spacetime
${\cal M}_{\lambda}$.
Although we do not know about the physical spacetime 
${\cal M}_{\lambda}$ at the starting point of the perturbation
theory, we can treat the physical variable $\bar{Q}_{\lambda}$
on the physical manifold ${\cal M}_{\lambda}$ as the variable 
${\cal X}^{*}_{\lambda}\bar{Q}_{\lambda}$ on the background
spacetime ${\cal M}_{0}$ through this pull-back
${\cal X}^{*}_{\lambda}$.


In the case of the perturbations in the theory with general
covariance, the above strategy of the perturbation theory
includes an important trouble.
This is the fact that the gauge choice ${\cal X}_{\lambda}$ is
not unique by virtue of general covariance.
Rather, there is degree of freedom in the gauge choice
${\cal X}_{\lambda}$, i.e., we may apply the different
point-identification map ${\cal Y}_{\lambda}$ from
${\cal X}_{\lambda}$ as a gauge choice.
In this case, the representation 
${\cal Y}^{*}_{\lambda}\bar{Q}_{\lambda}$ on the
background spacetime ${\cal M}_{0}$ of the physical variable
$\bar{Q}_{\lambda}$ on the physical spacetime
${\cal M}_{\lambda}$ is different from the representation
${\cal X}^{*}_{\lambda}\bar{Q}_{\lambda}$. 
This difference is unphysical because it has nothing to do with
the nature of the physical spacetime ${\cal M}_{\lambda}$.
We can also consider the transformation rule from a gauge choice 
${\cal X}_{\lambda}$ to another one ${\cal Y}_{\lambda}$, which
is called gauge transformation. 
The gauge transformation 
${\cal X}_{\lambda} \rightarrow {\cal Y}_{\lambda}$ is induced
by the diffeomorphism 
$\Phi_{\lambda} := ({\cal X}_{\lambda})^{-1}\circ {\cal Y}_{\lambda}$.
Actually, the diffeomorphism $\Phi_{\lambda}$ changes the
point-identification maps from ${\cal X}_{\lambda}$ to
${\cal Y}_{\lambda}$.
Further, the pull-back $\Phi_{\lambda}^{*}$ of the
diffeomorphism $\Phi_{\lambda}$ changes the representation 
${\cal X}^{*}_{\lambda}\bar{Q}_{\lambda}$ of the physical variable
$\bar{Q}_{\lambda}$ to another representation 
${\cal Y}^{*}_{\lambda}\bar{Q}_{\lambda}$:
\begin{eqnarray}
  {\cal Y}^{*}_{\lambda}\bar{Q}_{\lambda}
  = {\cal Y}^{*}_{\lambda} \left(
    {\cal X}_{\lambda} \circ {\cal X}_{\lambda}^{-1}
  \right)^{*} \bar{Q}_{\lambda}
  = {\cal Y}^{*}_{\lambda} ({\cal X}_{\lambda}^{-1})^{*}
  {\cal X}_{\lambda}^{*} \bar{Q}_{\lambda}
  = \Phi^{*}_{\lambda} {\cal X}_{\lambda}^{*}\bar{Q}_{\lambda}.
\end{eqnarray}


Since $\Phi_{\lambda}$ is the diffeomorphism on the background
manifold ${\cal M}_{0}$, the Taylor expansion of the pull-back
$\Phi^{*}_{\lambda}$ is given by
\begin{eqnarray}
  \Phi^{*}_{\lambda} {\cal X}_{\lambda}^{*}\bar{Q}_{\lambda}
  = {\cal X}_{\lambda}^{*}\bar{Q}_{\lambda}
  + \lambda {\pounds}_{\xi_{1}}{\cal X}_{\lambda}^{*}\bar{Q}_{\lambda}
  + \frac{1}{2} \lambda^{2} \left(
    {\pounds}_{\xi_{2}} + {\pounds}_{\xi_{1}}^{2}
  \right)
  {\cal X}_{\lambda}^{*}\bar{Q}_{\lambda}
  + O(\lambda^{3}),
  \label{eq:Taylor-Phi-pull-back}
\end{eqnarray}
where $\xi_{1}^{a}$ and $\xi_{2}^{a}$ are generators of the
diffeomorphism $\Phi^{*}_{\lambda}$\cite{S.Sonego-M.Bruni-CMP1998}.
On the other hand, we consider the perturbative expansion
\begin{equation}
  {\cal X}^{*}_{\lambda}\bar{Q}_{\lambda}
  =
  Q_{0}
  + \lambda {}^{(1)}_{{\cal X}}\!Q
  + \frac{1}{2} \lambda^{2} {}^{(2)}_{{\cal X}}\!Q
  + O(\lambda^{3})
  \label{eq:perturbative-expansion-def}
\end{equation}
of the representation ${\cal X}_{\lambda}^{*}\bar{Q}_{\lambda}$
under the gauge choice ${\cal X}_{\lambda}$, where $Q_{0}$ is
the background value of the variable $\bar{Q}_{\lambda}$.
The first- and the second-order perturbations 
${}^{(1)}_{{\cal X}}\!Q$ and ${}^{(2)}\!_{{\cal X}}Q$ are
defined by this equation (\ref{eq:perturbative-expansion-def}).
From
Eqs.~(\ref{eq:Taylor-Phi-pull-back}) and
(\ref{eq:perturbative-expansion-def}), we can easily derive the
gauge transformation rule of each order: 
\begin{eqnarray}
  \label{eq:gauge-trans-first-order-generic}
  {}^{(1)}_{\;\cal Y}\!Q - {}^{(1)}_{\;\cal X}\!Q &=& 
  {\pounds}_{\xi_{(1)}} Q_{0} 
  ,
  \\
  \label{eq:gauge-trans-second-order-generic}
  {}^{(2)}_{\;\cal Y}\!Q - {}^{(2)}_{\;\cal X}\!Q &=& 
  2 {\pounds}_{\xi_{(1)}} {}^{(1)}_{\;\cal X}\!Q 
  +\left\{{\pounds}_{\xi_{(2)}}+{\pounds}_{\xi_{(1)}}^{2}\right\} Q_{0}.
\end{eqnarray}
Further, we introduce the concept of ``{\it the order by order
  gauge invariance}''. 
We call the $p$th-order perturbation ${}^{(p)}_{\;\cal X}\!Q$ is
gauge invariant iff
\begin{equation}
  {}^{(p)}_{\;\cal Y}\!Q = {}^{(p)}_{\;\cal X}\!Q
\end{equation}
for any gauge choice ${\cal X}_{\lambda}$ and
${\cal Y}_{\lambda}$.
We have been considering the concept of ``{\it the gauge
  invariant up to order $n$}'' in the series of the papers
KN2003\cite{kouchan-gauge-inv}, KN2005\cite{kouchan-second}, and
KN2007\cite{kouchan-cosmo-second}, following the idea by Bruni
and Sonego\cite{M.Bruni-S.Soonego-CQG1997}.
However, we should regard the gauge invariance in this series
of the papers is this ``order by order gauge invariance'' rather
than ``the gauge invariance up to order $n$''.
This notion of the order by order gauge invariance is weaker
than the notion of the gauge invariance up to $n$, since we do
not say anything about the gauge invariance of the other orders
in the above order by order gauge invariance.


Employing the idea of this order by order gauge invariance, we
proposed a procedure to construct gauge-invariant variables of
higher-order perturbations in KN2003\cite{kouchan-gauge-inv}.
Inspecting the gauge transformation rules
(\ref{eq:gauge-trans-first-order-generic}) and
(\ref{eq:gauge-trans-second-order-generic}), we can define the
gauge-invariant variables for a metric perturbation and for
arbitrary matter fields.


\subsection{Gauge-invariant variables for metric perturbations}
\label{sec:Second-order-cosmological-perturbatios-metric}


Following the expansion form
(\ref{eq:perturbative-expansion-def}), we also expand the
pulled-back ${\cal X}^{*}_{\lambda}\bar{g}_{ab}$ of the metric
$\bar{g}_{ab}$ on the physical spacetime ${\cal M}_{\lambda}$:
\begin{eqnarray}
  {\cal X}^{*}_{\lambda}\bar{g}_{ab}
  &=&
  g_{ab} + \lambda {}_{{\cal X}}\!h_{ab} 
  + \frac{\lambda^{2}}{2} {}_{{\cal X}}\!l_{ab}
  + O(\lambda^{3}),
  \label{eq:metric-expansion}
\end{eqnarray}
where $g_{ab}$ is the metric on the background spacetime 
${\cal M}_{0}$. 
Although the expansion (\ref{eq:metric-expansion}) of the metric
depends entirely on the gauge choice  ${\cal X}_{\lambda}$,
henceforth, we do not explicitly express the index of the gauge
choice ${\cal X}_{\lambda}$ in expressions if there is no
possibility of confusion.
As shown in KN2007\cite{kouchan-cosmo-second}, at least in the
cosmological perturbation case, the first-order metric
perturbation $h_{ab}$ is decomposed as
\begin{eqnarray}
  h_{ab} =: {\cal H}_{ab} + {\pounds}_{X}g_{ab},
  \label{eq:linear-metric-decomp}
\end{eqnarray}
where ${\cal H}_{ab}$ and $X^{a}$ are the gauge-invariant and
gauge-variant parts of the linear-order metric
perturbations\cite{kouchan-gauge-inv}, i.e., under the gauge
transformation (\ref{eq:gauge-trans-first-order-generic}), these
are transformed as 
\begin{equation}
  {}_{{\cal Y}}\!{\cal H}_{ab} - {}_{{\cal X}}\!{\cal H}_{ab} =  0, 
  \quad
  {}_{\quad{\cal Y}}\!X^{a} - {}_{{\cal X}}\!X^{a} = \xi^{a}_{(1)}. 
  \label{eq:linear-metric-decomp-gauge-trans}
\end{equation}
Further, the second-order metric perturbation $l_{ab}$ is also
decomposed as 
\begin{eqnarray}
  \label{eq:second-metric-decomp}
  l_{ab}
  =:
  {\cal L}_{ab} + 2 {\pounds}_{X} h_{ab}
  + \left(
      {\pounds}_{Y}
    - {\pounds}_{X}^{2} 
  \right)
  g_{ab},
\end{eqnarray}
where ${\cal L}_{ab}$ and $Y^{a}$ are the gauge-invariant and
gauge-variant parts of the second-order metric perturbations,
i.e., these are transformed as 
\begin{eqnarray}
  {}_{{\cal Y}}\!{\cal L}_{ab} - {}_{{\cal X}}\!{\cal L}_{ab} = 0,
  \quad
  {}_{{\cal Y}}\!Y^{a} - {}_{{\cal X}}\!Y^{a}
  = \xi_{(2)}^{a} + [\xi_{(1)},X]^{a}
  \label{eq:second-metric-decomp-gauge-trans}
\end{eqnarray}
under the gauge transformation
(\ref{eq:gauge-trans-second-order-generic}).


In KN2007\cite{kouchan-cosmo-second}, the details of the
derivation of this gauge-invariant part of the second-order
metric perturbation are explained in the context of cosmological
perturbations.
In the case of the cosmological perturbations,
we consider the homogeneous isotropic background spacetime whose
metric is given by  
\begin{eqnarray}
  \label{eq:background-metric}
  g_{ab} = a^{2}\left\{
    - (d\eta)_{a}(d\eta)_{b}
    + \gamma_{ij} (dx^{i})_{a} (dx^{j})_{b}
  \right\},
\end{eqnarray}
where $\gamma_{ab} := \gamma_{ij} (dx^{i})_{a} (dx^{j})_{b}$ is
the metric on the maximally symmetric three space.
As shown in KN2007, the decomposition
(\ref{eq:linear-metric-decomp}) is accomplished if we assume the
existence of the Green functions
$\Delta^{-1}:=(D^{i}D_{i})^{-1}$, 
$(\Delta + 2 K)^{-1}$, and $(\Delta + 3 K)^{-1}$, where $D_{i}$
is the covariant derivative associated with the metric
$\gamma_{ij}$ on the maximally symmetric three space and $K$ is
the curvature constant of this maximally symmetric three space.
We also showed in KN2007 that we may choose the components
of the gauge-invariant part ${\cal H}_{ab}$ of the first-order
metric perturbation as
\begin{eqnarray}
  \label{eq:components-calHab}
  {\cal H}_{ab}
  &=& a^{2} \left\{
    - 2 \stackrel{(1)}{\Phi} (d\eta)_{a}(d\eta)_{b}
    + 2 \stackrel{(1)}{\nu}_{i} (d\eta)_{(a}(dx^{i})_{b)}
  \right.
  \nonumber\\
  && \quad\quad\quad\quad
  \left.
    + 
    \left( - 2 \stackrel{(1)}{\Psi} \gamma_{ij} 
      + \stackrel{(1)}{\chi}_{ij} \right)
    (dx^{i})_{a}(dx^{j})_{b}
  \right\},
\end{eqnarray}
where $\stackrel{(1)}{\nu}_{i}$ and $\stackrel{(1)}{\chi_{ij}}$
satisfy the properties
\begin{eqnarray}
  && D^{i}\stackrel{(1)}{\nu}_{i} :=
  \gamma^{ij}D_{i}\stackrel{(1)}{\nu}_{j} = 0, \quad
  \stackrel{(1)}{\chi_{i}^{\;\;i}}
  := \gamma^{ij} \stackrel{(1)}{\chi_{ij}} = 0, \quad
   D^{i}\stackrel{(1)}{\chi}_{ij} = 0.
\end{eqnarray}
Further, we may also choose the components of the
gauge-invariant part ${\cal L}_{ab}$ of the second-order metric
perturbation as 
\begin{eqnarray}
  \label{eq:components-calLab}
  {\cal L}_{ab}
  &=& a^{2} \left\{
    - 2 \stackrel{(2)}{\Phi} (d\eta)_{a}(d\eta)_{b}
    + 2 \stackrel{(2)}{\nu}_{i} (d\eta)_{(a}(dx^{i})_{b)}
  \right.
  \nonumber\\
  && \quad\quad\quad\quad
  \left.
    + 
    \left( - 2 \stackrel{(2)}{\Psi} \gamma_{ij} 
      + \stackrel{(2)}{\chi}_{ij} \right)
    (dx^{i})_{a}(dx^{j})_{b}
  \right\},
\end{eqnarray}
where $\stackrel{(2)}{\nu}_{i}$ and $\stackrel{(2)}{\chi_{ij}}$ satisfy
the properties
\begin{eqnarray}
  D^{i}\stackrel{(2)}{\nu}_{i} = 0, \quad
  \stackrel{(2)}{\chi_{i}^{\;\;i}} = 0, \quad
   D^{i}\stackrel{(2)}{\chi_{ij}} = 0.
\end{eqnarray}


Here, we also note the fact that the definitions
(\ref{eq:linear-metric-decomp}) and
(\ref{eq:second-metric-decomp}) of the gauge-invariant variables
are not unique.  
This comes from the fact that we can always construct new
gauge-invariant quantities by the combination of the
gauge-invariant variables.
For example,  using the gauge-invariant variables
$\stackrel{(1)}{\Phi}$ and $\stackrel{(1)}{\nu_{i}}$ of the
first-order metric perturbation, we can define a vector field
$Z_{a}$ by
\begin{equation}
  Z_{a} := a \stackrel{(1)}{\Phi} (d\eta)_{a} + a
  \stackrel{(1)}{\nu_{i}} (dx^{i})_{a}.
  \label{eq:artificial-gauge-invariant-vector}
\end{equation}
Although there is no specific physical meaning in this vector
field $Z_{a}$, at least, we can say that the vector field
$Z_{a}$ defined by (\ref{eq:artificial-gauge-invariant-vector})
is gauge-invariant.
Through this gauge-invariant vector field $Z_{a}$, we can
rewrite the decomposition formula
(\ref{eq:linear-metric-decomp}) for the linear-order metric
perturbation as
\begin{eqnarray}
  h_{ab}
  &=&
  {\cal H}_{ab} - {\pounds}_{Z}g_{ab}
  + {\pounds}_{Z}g_{ab} + {\pounds}_{X}g_{ab},
  \nonumber\\
  &=:&
  {\cal K}_{ab} + {\pounds}_{X+Z}g_{ab},
\end{eqnarray}
where we have defined new gauge-invariant variable 
${\cal K}_{ab}$ by
\begin{eqnarray}
  {\cal K}_{ab} := {\cal H}_{ab} - {\pounds}_{Z}g_{ab}.
\end{eqnarray}
Clearly, ${\cal K}_{ab}$ is gauge-invariant and the vector field
$X^{a}+Z^{a}$ satisfies the gauge transformation rule
(\ref{eq:linear-metric-decomp-gauge-trans}) for the
gauge-variant part of the first-order metric perturbations. 
Although the definition of the gauge-invariant variables
is not unique, we can specify the components of the
gauge-variant part $X_{a}$ without ambiguities if we specify the
components of the gauge-invariant part ${\cal H}_{ab}$ as shown
in KN2007.
In this paper, we specify the components of the tensor 
${\cal H}_{ab}$ as Eq.~(\ref{eq:components-calHab}), which is
the gauge-invariant part of the linear-order metric perturbation
associated with the longitudinal gauge.


\subsection{Gauge-invariant variables for matter fields}
\label{sec:Second-order-cosmological-perturbatios-matter-fundamental}


As shown in KN2003, using the above first- and
second-order gauge-variant parts, $X^{a}$ and $Y^{a}$, of the
metric perturbations, we can define the gauge-invariant
variables for an arbitrary field $Q$ other than the metric.
These definitions imply that the first- and the second-order
perturbations ${}^{(1)}\!Q$ and ${}^{(2)}\!Q$ are always
decomposed into gauge-invariant part and gauge-variant part as 
\begin{eqnarray}
  \label{eq:matter-gauge-inv-decomp-1.0}
  {}^{(1)}\!Q &=:& {}^{(1)}\!{\cal Q} + {\pounds}_{X}Q_{0}
  , \\ 
  \label{eq:matter-gauge-inv-decomp-2.0}
  {}^{(2)}\!Q  &=:& {}^{(2)}\!{\cal Q} + 2 {\pounds}_{X} {}^{(1)}Q 
  + \left\{ {\pounds}_{Y} - {\pounds}_{X}^{2} \right\} Q_{0}
  ,
\end{eqnarray}
respectively, where ${}^{(1)}\!{\cal Q}$ and 
${}^{(2)}\!{\cal Q}$ are gauge-invariant parts of the first- and
the second-order perturbations of ${}^{(1)}\!Q$ and
${}^{(2)}\!Q$, respectively.


Through the formulae (\ref{eq:matter-gauge-inv-decomp-1.0}) and
(\ref{eq:matter-gauge-inv-decomp-2.0}), we can define the
gauge-invariant variables for the matter field.
In this paper, we consider the cases of a perfect fluid; an
imperfect fluid; and a scalar field.
All these matter fields consist of fundamental quantities.
For example, we regard the energy density, the pressure, and the
four-velocity as fundamental variables for a perfect fluid.
Here, we show the definitions of the gauge-invariant variables
for these fundamental quantities.
These definitions are just following to the formulae 
(\ref{eq:matter-gauge-inv-decomp-1.0}) and
(\ref{eq:matter-gauge-inv-decomp-2.0}).
However, based on these definitions of gauge-invariant variables,
we will show that all perturbative quantities are decomposed into
gauge-invariant and gauge-variant parts as
Eqs.~(\ref{eq:matter-gauge-inv-decomp-1.0}) and
(\ref{eq:matter-gauge-inv-decomp-2.0}).


\subsubsection{Perfect fluid}
\label{sec:Second-order-cosmological-perturbatios-matter-perfect-fundamental}


Here, we consider the definitions of the gauge-invariant
variables for the fundamental variables of a perfect fluid.
The total energy momentum tenor of the fluid is given by
\begin{equation}
  {}^{(p)}\!\bar{T}_{a}^{\;\;b} 
  = (\bar{\epsilon} + \bar{p}) \bar{u}_{a} \bar{u}^{b} 
  + \bar{p} \delta_{a}^{\;\;b},
  \label{eq:MFB-5.2-again}
\end{equation}
where the fundamental variables for a perfect fluid are the
energy density $\bar{\epsilon}$, the pressure $\bar{p}$, and 
the four-velocity $\bar{u}^{a}$.
We expand these fundamental variables as 
\begin{eqnarray}
  \bar{\epsilon}
  &:=&
  \epsilon 
  + \lambda \stackrel{(1)}{\epsilon}
  + \frac{1}{2} \lambda^{2} \stackrel{(2)}{\epsilon} 
  + O(\lambda^{3})
  \label{eq:energy-density-expansion}
  , \\
  \bar{p}
  &:=&
  p
  + \lambda\stackrel{(1)}{p}
  + \frac{1}{2} \lambda^{2} \stackrel{(2)}{p}
  + O(\lambda^{3})
  \label{eq:pressure-expansion}
  , \\
  \bar{u}_{a} 
  &:=& 
  u_{a}
  + \lambda \stackrel{(1)}{(u_{a})}
  + \frac{1}{2} \lambda^{2} \stackrel{(2)}{(u_{a})}
  + O(\lambda^{3}),
  \label{eq:four-velocity-expansion}
\end{eqnarray}
where $\bar{\epsilon}$, $\bar{p}$, and $\bar{u}_{a}$
characterize the pull-back of the fluid on the physical
spacetime to the background spacetime through an appropriate
gauge choice ${\cal X}_{\lambda}$, while $\epsilon$, $p$,
and $u_{a}$ are their background values on the background
spacetime.
Following to Eqs.~(\ref{eq:matter-gauge-inv-decomp-1.0})
and (\ref{eq:matter-gauge-inv-decomp-2.0}), we define the
gauge-invariant variable for the perturbation of the fluid
components $\bar{\epsilon}$, $\bar{p}$, and $\bar{u_{a}}$: 
\begin{eqnarray}
  \label{eq:kouchan-16.13}
  &&
  \stackrel{(1)}{{\cal E}} 
  :=
  \stackrel{(1)}{\epsilon}
  - {\pounds}_{X}\epsilon,
  \quad
  \stackrel{(1)}{{\cal P}}
  :=
  \stackrel{(1)}{p}
  - {\pounds}_{X}p,
  \quad
  \stackrel{(1)}{{\cal U}_{a}}
  :=
  \stackrel{(1)}{(u_{a})}
  - {\pounds}_{X}u_{a}, \\
  &&
  \stackrel{(2)}{{\cal E}} 
  := \stackrel{(2)}{\epsilon} 
  - 2 {\pounds}_{X} \stackrel{(1)}{\epsilon}
  - \left\{
    {\pounds}_{Y}
    -{\pounds}_{X}^{2}
  \right\} \epsilon
  , \quad
  \label{eq:kouchan-16.17}
  \stackrel{(2)}{{\cal P}}
  := \stackrel{(2)}{p}
  - 2 {\pounds}_{X} \stackrel{(1)}{p}
  - \left\{
    {\pounds}_{Y}
    -{\pounds}_{X}^{2}
  \right\} p
  , \\
  &&
  \stackrel{(2)}{{\cal U}_{a}}
  :=
  \stackrel{(2)}{(u_{a})}
  - 2 {\pounds}_{X} \stackrel{(1)}{u_{a}}
  - \left\{
    {\pounds}_{Y}
    -{\pounds}_{X}^{2}
  \right\} u_{a}.
  \label{eq:kouchan-16.18}
\end{eqnarray}


\subsubsection{Imperfect fluid}
\label{sec:Second-order-cosmological-perturbatios-matter-imperfect-fundamental}


Here, we consider the generic case of an imperfect fluid.
The energy-momentum tensor is decomposed into fluid quantities
based on the orthogonality to the four-vector field
$\bar{u}^{a}$ as
\begin{eqnarray}
  \bar{T}_{a}^{\;\;b} 
  &=& 
  \bar{\epsilon} \bar{u}_{a} \bar{u}^{b} 
  + \bar{p} \left(\delta_{a}^{\;\;b} + \bar{u}_{a} \bar{u}^{b}\right)
  + \bar{q}_{a} \bar{u}^{b} 
  + \bar{u}_{a} \bar{q}^{b}
  + \bar{\pi}_{a}^{\;\;b}
  \label{eq:imperfect-fluid-ene-mom-full-1}
  \\
  &=& 
  {}^{(p)}\!\bar{T}_{a}^{\;\;b} 
  + \bar{g}^{bc} \bar{q}_{a} \bar{u}_{c} 
  + \bar{g}^{bc} \bar{u}_{a} \bar{q}_{c}
  + \bar{g}^{bc} \bar{\pi}_{ac},
  \label{eq:imperfect-fluid-ene-mom-full-2}
\end{eqnarray}
where
\begin{eqnarray}
  \label{eq:kouchan-16.58}
  &&
  \bar{u}^{a} \bar{q}_{a} = 0, \\
  \label{eq:kouchan-16.59}
  &&
  \bar{\pi}_{[ab]} = 0, \quad
  \bar{u}^{a} \bar{\pi}_{ab} = 0, \quad
  \bar{\pi}_{a}^{\;\;a} = \bar{g}^{ab} \bar{\pi}_{ab} = 0.
\end{eqnarray}
The energy density $\bar{\epsilon}$, the isotropic pressure
$\bar{p}$, and the four-velocity $\bar{u}_{a}$ of the imperfect
fluid in Eq.~(\ref{eq:imperfect-fluid-ene-mom-full-1}) are
expanded as
Eqs.~(\ref{eq:energy-density-expansion})--(\ref{eq:four-velocity-expansion})
and the gauge-invariant variables for their perturbations are
defined by
Eqs.~(\ref{eq:kouchan-16.13})--(\ref{eq:kouchan-16.18}) as in
\S\ref{sec:Second-order-cosmological-perturbatios-matter-perfect-fundamental}.
In addition to these fluid components, in the imperfect fluid
case, we add the energy flux $\bar{q}_{a}$ and the anisotropic
stress $\bar{\pi}_{ab}$ associated with the vector field
$\bar{u}_{a}$ as fluid components.
Since the first two terms in
Eq.~(\ref{eq:imperfect-fluid-ene-mom-full-1}) coincide with the
energy momentum tensor for a perfect fluid, we call these terms
as ``{\it the perfect part}'' and denote them by
${}^{(p)}\!\bar{T}_{a}^{\;\;b}$ as in
Eq.~(\ref{eq:imperfect-fluid-ene-mom-full-2}).
On the other hand, we call the remaining terms in 
Eq.~(\ref{eq:imperfect-fluid-ene-mom-full-2})
\begin{eqnarray}
  {}^{(i)}\!\bar{T}_{a}^{\;\;b} 
  :=
    \bar{q}_{a} \bar{u}^{b} 
  + \bar{u}_{a} \bar{q}^{b}
  + \bar{\pi}_{a}^{\;\;b}
  \label{eq:imperfect-part-ene-mom-full}
  ,
\end{eqnarray}
as ``{\it the imperfect part}'' of the energy momentum tensor
for an imperfect fluid.


Now, we consider the perturbative expansion of the energy flux
$\bar{q}_{a}$ and the anisotropic stress $\bar{\pi}_{ab}$.
Although these quantities should be given through the
micro-physical process, in this paper, we regard these variables
as fundamental quantities for an imperfect fluid and expand
these variables as 
\begin{eqnarray}
  \label{eq:kouchan-16.72}
  \bar{q}_{a} 
  &=:& 
  q_{a}
  + \lambda \stackrel{(1)}{(q_{a})}
  + \frac{1}{2} \lambda^{2} \stackrel{(2)}{(q_{a})}
  + O(\lambda^{3})
  , \\
  \label{eq:kouchan-16.74}
  \bar{\pi}_{ab} 
  &=:& 
  \pi_{ab}
  + \lambda \stackrel{(1)}{(\pi_{ab})}
  + \frac{1}{2} \lambda^{2} \stackrel{(2)}{(\pi_{ab})}
  + O(\lambda^{3})
  .
\end{eqnarray}
Further, we introduce gauge-invariant variables for the
perturbations of the energy flux $q_{a}$ and the anisotropic
stress $\pi_{ab}$.
Following to the decompositions
(\ref{eq:matter-gauge-inv-decomp-1.0}) and
(\ref{eq:matter-gauge-inv-decomp-2.0}) for an arbitrary matter
field, the first- and the second-order perturbations of the
energy flux and the anisotropic stress are decomposed into
gauge-invariant and gauge-variant parts as
\begin{eqnarray}
  \label{eq:kouchan-16.92}
  \stackrel{(1)}{(q_{a})}
  &=:&
  \stackrel{(1)}{{\cal Q}_{a}} + {\pounds}_{X}q_{a}
  , \quad
  \stackrel{(2)}{(q_{a})}
  =:
  \stackrel{(2)}{{\cal Q}_{a}}
  + 2 {\pounds}_{X} \stackrel{(1)}{(q_{a})}
  + \left\{
       {\pounds}_{Y}
    -  {\pounds}_{X}^{2}
  \right\} q_{a}
  , \\
  \label{eq:kouchan-16.93}
  \stackrel{(1)}{(\pi_{ab})}
  &=:&
  \stackrel{(1)}{\Pi_{ab}} + {\pounds}_{X}\pi_{ab}
  , \quad
  \stackrel{(2)}{(\pi_{ab})}
  =:
  \stackrel{(2)}{\Pi_{ab}}
  + 2 {\pounds}_{X} \stackrel{(1)}{(\pi_{ab})}
  + \left\{
      {\pounds}_{Y}
    - {\pounds}_{X}^{2}
  \right\} \pi_{ab}
  .
\end{eqnarray}
If we represent the multi-fluid system as an imperfect fluid
system or we consider the micro-physical process, these energy
flux and the anisotropic stress are related to the other fluid
components\cite{Kodama-Sasaki-1984,anisotropic-stress-in-cosmology}.    
In this case, the gauge transformations (\ref{eq:kouchan-16.92})
and (\ref{eq:kouchan-16.93}) should be derived from this
relation\cite{Pitrou:2007jy}.
However, in this paper, we regard these energy flux and the
anisotropic stress as fundamental variables for an imperfect
fluid, phenomenologically.


\subsubsection{Scalar field}
\label{sec:Second-order-cosmological-perturbatios-matter-scalar-fundamental}


Finally, we briefly summarize the perturbations of a scalar
field $\bar{\varphi}$ whose energy momentum tensor is given by
\begin{eqnarray}
  \bar{T}_{a}^{\;\;b} = 
  \bar{g}^{bc} \bar{\nabla}_{a}\bar{\varphi} \bar{\nabla}_{c}\bar{\varphi} 
  - \frac{1}{2} \delta_{a}^{\;\;b}
  \left(
    \bar{g}^{cd} \bar{\nabla}_{c}\bar{\varphi}\bar{\nabla}_{d}\bar{\varphi}
    + 2 V(\bar{\varphi})
  \right),
  \label{eq:MFB-6.2-again}
\end{eqnarray}
where $V(\bar{\varphi})$ is the potential of the scalar field
$\bar{\varphi}$.
The fundamental variable of this system is the scalar field
$\bar{\varphi}$ itself.
In perturbation theory, we also expand this scalar field
$\bar{\varphi}$ as 
\begin{eqnarray}
  \bar{\varphi}
  =
  \varphi
  + \lambda \hat{\varphi}_{1}
  + \frac{1}{2} \lambda^{2} \hat{\varphi}_{2}
  + O(\lambda^{3}),
  \label{eq:scalar-field-expansion-second-order}
\end{eqnarray}
where $\varphi$ is the background value of the scalar field
$\bar{\varphi}$.
Further, as in the cases of the fluids, each order perturbations
of the scalar field $\varphi$ is decomposed into the
gauge-invariant part and gauge-variant part as 
\begin{eqnarray}
  \label{eq:varphi-1-def}
  \hat{\varphi}_{1} &=:& \varphi_{1} + {\pounds}_{X}\varphi, \\
  \label{eq:varphi-2-def}
  \hat{\varphi}_{2} &=:& \varphi_{2} 
  + 2 {\pounds}_{X}\hat{\varphi}_{1} 
  + \left( {\pounds}_{Y} - {\pounds}_{X}^{2} \right) \varphi, 
\end{eqnarray}
where $\varphi_{1}$ and $\varphi_{2}$ are the first-order and
the second-order gauge-invariant perturbation of the scalar
field.


\section{Generic form of perturbations of energy momentum tensors and equations of motion}
\label{sec:Generic-form-of-perturbations-of-energy-momentum-tensors-and-equations-of-motion}


Here, we consider the generic expression of the perturbations of
the energy momentum tensors and equations of motion for a perfect
fluid (\S\ref{sec:Perfect-fluid-generic}), an imperfect fluid 
(\S\ref{sec:imperfect-fluid-generic}),
and a single scalar field (\S\ref{sec:scalar-field-generic}).
We derive these perturbative expressions in terms of
gauge-invariant variables defined in the last section.
We also show that all perturbative variables are given in the
same form as Eqs.~(\ref{eq:matter-gauge-inv-decomp-1.0}) and
(\ref{eq:matter-gauge-inv-decomp-2.0}).


We note that we do not explicitly use any background values of
the metric and matter fields in the derivations within this
section.
Further, we also note that we do not use any information of the
Einstein equation nor the equations of state of the matter
fields. 
Therefore, the ingredients of this section will be valid for any 
background spacetime and many perturbation theories of gravity
with general covariance if the decomposition formula
(\ref{eq:linear-metric-decomp}) is correct.


\subsection{Perfect fluid}
\label{sec:Perfect-fluid-generic}



The perturbative expressions of the energy momentum tensor
for a perfect fluid are already discussed in
KN2007\cite{kouchan-cosmo-second}. 
Therefore, in this subsection, we just summarize the definitions
and results in KN2007 for the perturbations of the energy
momentum tensor in \S\ref{sec:Perfect-fluid-ene-mon-generic}.
In addition to the results in KN2007, we also show the
perturbative expression of the equations of motion, i.e., the
equation of continuity
(\S\ref{sec:Perfect-fluid-continuity-equation-generic}) and the
Euler equation
(\S\ref{sec:Perfect-fluid-Euler-equation-generic}), which are
derived from the perturbations of the divergence of the energy
momentum tensor.


\subsubsection{Perturbations of the energy momentum tensor}
\label{sec:Perfect-fluid-ene-mon-generic}


The perturbative expansion of the energy momentum tensor
(\ref{eq:MFB-5.2-again}) is given by  
\begin{eqnarray}
  {}^{(p)}\!\bar{T}_{a}^{\;\;b}
  &=:& 
  {}^{(p)}\!T_{a}^{\;\;b}
  + \lambda \stackrel{(1)}{{}^{(p)}\!T_{a}^{\;\;b}}
  + \frac{1}{2} \lambda^{2} \stackrel{(2)}{{}^{(p)}\!T_{a}^{\;\;b}}
  + O(\lambda^{3}).
  \label{eq:energy-momentum-tensor-expansion-perfect}
\end{eqnarray}
The background energy momentum tensor for a perfect fluid is
given by  
\begin{equation}
  {}^{(p)}\!T_{a}^{\;\;b}
  =
  (\epsilon + p) u_{a} u^{b} + p \delta_{a}^{\;\;b}.
  \label{eq:background-energy-momentum-tensor-perfect}
\end{equation}
The first- and the second-order perturbations
$\stackrel{(1)}{{}^{(p)}\!T_{a}^{\;\;b}}$ and
$\stackrel{(2)}{{}^{(p)}\!T_{a}^{\;\;b}}$ of the energy 
momentum tensor are also decomposed into the form as
Eqs.~(\ref{eq:matter-gauge-inv-decomp-1.0}) and
(\ref{eq:matter-gauge-inv-decomp-2.0}), respectively, i.e., 
\begin{eqnarray}
  \label{eq:first-energy-momentum-tensor-perfect-decomposed}
  \stackrel{(1)}{{}^{(p)}\!T_{a}^{\;\;b}}
  &=:&
  \stackrel{(1)}{{}^{(p)}\!{\cal T}_{a}^{\;\;b}}
  + {\pounds}_{X}{}^{(p)}\!T_{a}^{\;\;b}
  ,\\
  \label{eq:second-energy-momentum-tensor-perfect-decomposed}
  \stackrel{(2)}{{}^{(p)}\!T_{a}^{\;\;b}}
  &=:&
  \stackrel{(2)}{{}^{(p)}\!{\cal T}_{a}^{\;\;b}}
  + 2 {\pounds}_{X} \stackrel{(1)}{{}^{(p)}\!T_{a}^{\;\;b}}
  + \left\{
    {\pounds}_{Y}
    - {\pounds}_{X}^{2}
  \right\} {}^{(p)}\!T_{a}^{\;\;b},
\end{eqnarray}
where gauge-invariant parts
$\stackrel{(1)}{{}^{(p)}\!{\cal T}_{a}^{\;\;b}}$ and 
$\stackrel{(2)}{{}^{(p)}\!{\cal T}_{a}^{\;\;b}}$ of the first-
and the second-order perturbations are given by 
\begin{eqnarray}
  \stackrel{(1)}{{}^{(p)}\!{\cal T}_{a}^{\;\;b}}
  &:=&
  \left(
    \stackrel{(1)}{{\cal E}}
    + \stackrel{(1)}{{\cal P}}
  \right) u_{a} u^{b}
  + \stackrel{(1)}{{\cal P}} \delta_{a}^{\;\;b}
  + \left( \epsilon + p \right) \left(
    u_{a} \stackrel{(1)}{{\cal U}^{b}} 
    - {\cal H}^{bc} u_{c} u_{a} 
    + \stackrel{(1)}{{\cal U}_{a}} u^{b}
  \right)
  \label{eq:first-energy-momentum-tensor-perfect-gauge-inv}
  , \\
  \stackrel{(2)}{{}^{(p)}\!{\cal T}_{a}^{\;\;b}}
  &:=&
  \left( 
    \stackrel{(2)}{{\cal E}} 
    + \stackrel{(2)}{{\cal P}}
  \right) u_{a} u^{b}
  + 2 \left( 
    \stackrel{(1)}{{\cal E}} + \stackrel{(1)}{{\cal P}} 
  \right) 
  u_{a}
  \left(
    \stackrel{(1)}{{\cal U}^{b}} - {\cal H}^{bc} u_{c}
  \right)
  + 2 
  \left(
    \stackrel{(1)}{{\cal E}} + \stackrel{(1)}{{\cal P}} 
  \right) 
  \stackrel{(1)}{{\cal U}_{a}} u^{b}
  \nonumber\\
  && \quad
  + \left( \epsilon + p \right) u_{a} \left(
        \stackrel{(2)}{{\cal U}^{b}}
    - 2 {\cal H}^{bc} \stackrel{(1)}{{\cal U}_{c}}
    + 2 {\cal H}^{bc}{\cal H}_{cd} u^{d}
    -   {\cal L}^{bd} u_{d}
  \right)
  \nonumber\\
  && \quad
  + 2 \left( \epsilon + p \right)
  \stackrel{(1)}{{\cal U}_{a}} \left(
    \stackrel{(1)}{{\cal U}^{b}} 
    - {\cal H}^{bc} u_{c}
  \right)
  + \left( \epsilon + p \right) \stackrel{(2)}{{\cal U}_{a}} u^{b}
  +   \stackrel{(2)}{{\cal P}} \delta_{a}^{\;\;b},
  \label{eq:second-energy-momentum-tensor-perfect-gauge-inv}
\end{eqnarray}
where we defined
\begin{eqnarray}
  \stackrel{(1)}{{\cal U}^{a}} := g^{ab} \stackrel{(1)}{{\cal U}_{b}},
  \quad
  \stackrel{(2)}{{\cal U}^{a}} := g^{ab} \stackrel{(2)}{{\cal U}_{b}}.
  \label{eq:first-second-pert-contravariant-gauge-inv-four-velocity-def}
\end{eqnarray}


We also note that the fluid four-velocities $\bar{u}_{a}$ and
$u_{a}$ should satisfy the normalization conditions of the
four-velocity 
\begin{eqnarray}
  \bar{g}^{ab}\bar{u}_{a}\bar{u}_{a} = g^{ab}u_{a}u_{b} = -1.
\end{eqnarray}
These normalization conditions yield
\begin{eqnarray}
  \label{eq:kouchan-17.30}
  u^{a} \stackrel{(1)}{{\cal U}_{a}}
  &=& \frac{1}{2} {\cal H}_{ab} u^{a} u^{b}
  , \\
  \label{eq:kouchan-17.31}
  u^{a} \stackrel{(2)}{{\cal U}_{a}}
  &=&
  -             g_{cb} \left(
      \stackrel{(1)}{{\cal U}^{b}}
    - {\cal H}^{db} u_{d}
  \right)
  \left(
      \stackrel{(1)}{{\cal U}^{c}}
    - {\cal H}^{ac} u_{a}
  \right)
  + \frac{1}{2} {\cal L}_{ab} u^{a} u^{b}.
\end{eqnarray}


We have to emphasize that the perturbative expressions
(\ref{eq:first-energy-momentum-tensor-perfect-decomposed}) and
(\ref{eq:second-energy-momentum-tensor-perfect-decomposed}) are
not definitions but the results which are derived from the
definitions (\ref{eq:kouchan-16.13})--(\ref{eq:kouchan-16.18}).
These are natural results from general formulae
(\ref{eq:matter-gauge-inv-decomp-1.0}) and
(\ref{eq:matter-gauge-inv-decomp-2.0}).
However, these results imply that the framework developed in
KN2003 and KN2005 does work in the case of the perturbations of
the energy momentum tensor of a perfect fluid.


\subsubsection{Perturbations of the continuity equation}
\label{sec:Perfect-fluid-continuity-equation-generic}


Here, we consider the perturbations of the continuity equation
for the perfect fluid which is derived from 
$\bar{u}^{a}\bar{\nabla}_{b}{}^{(p)}\!\bar{T}_{a}^{\;\;b} = 0$.
This equation yields 
\begin{eqnarray}
  \label{eq:continuity-equation-of-perfect-fluid}
  &&
  \bar{C}_{0}^{(p)} := \bar{u}^{a}\bar{\nabla}_{a}\bar{\epsilon}
  + (\bar{\epsilon} + \bar{p}) \bar{\theta} = 0
  .
\end{eqnarray}
The energy density $\bar{\epsilon}$ and the pressure $\bar{p}$
are expanded as Eqs.~(\ref{eq:energy-density-expansion}),
(\ref{eq:pressure-expansion}), respectively.
Further, as shown in Appendix
\ref{sec:Perturbations-of-accerelation-expansion-shear-and-rotation},
the four-velocity $\bar{u}^{a}$ and the expansion $\bar{\theta}$
associated with the four-velocity $\bar{u}_{a}$ are expanded as
Eq.~(\ref{eq:contravariant-four-velocity-expansion}) and
(\ref{eq:theta-expansion}).
Through these equations, we obtain the perturbative expansion of
the continuity equation as
\begin{eqnarray}
  \bar{C}_{0}^{(p)}
  =
  C_{0}^{(p)}
  + \lambda {}^{(1)}\!C_{0}^{(p)}
  + \frac{1}{2} \lambda^{2} {}^{(2)}\!C_{0}^{(p)}
  + O(\lambda^{3})
  = 
  0,
  \label{eq:continuity-perfect-expansion}
\end{eqnarray}
where the continuity equation of each order is given as follows: 
\begin{eqnarray}
  \label{eq:background-continuity-perfect}
  C_{0}^{(p)} &:=& 
  u^{a} \nabla_{a}\epsilon + \left(\epsilon + p\right) \theta = 0, \\
  \label{eq:first-order-continuity-perfect}
  {}^{(1)}\!C_{0}^{(p)} &:=&
  u^{a} \nabla_{a}\stackrel{(1)}{\epsilon}
  + \stackrel{(1)}{(u^{a})} \nabla_{a}\epsilon 
  + \left(\stackrel{(1)}{\epsilon} + \stackrel{(1)}{p} \right) \theta
  + \stackrel{(1)}{\theta} \left(\epsilon + p\right) = 0, \\
  \label{eq:second-order-continuity-perfect}
  {}^{(2)}\!C_{0}^{(p)} &:=&
  u^{a} \nabla_{a}\stackrel{(2)}{\epsilon}
  + 2 \stackrel{(1)}{(u^{a})} \nabla_{a}\stackrel{(1)}{\epsilon}
  + \stackrel{(2)}{(u^{a})} \nabla_{a}\epsilon 
  \nonumber\\
  && \quad
  + \theta \left(
    \stackrel{(2)}{\epsilon} + \stackrel{(2)}{p}
  \right)
  + 2 \stackrel{(1)}{\theta} 
  \left(\stackrel{(1)}{\epsilon} + \stackrel{(1)}{p} \right)
  + \stackrel{(2)}{\theta} \left(\epsilon + p\right)
  = 0.
\end{eqnarray}


The gauge-invariant variables for each order perturbations of
$\bar{\epsilon}$, $\bar{p}$, $\bar{u}^{a}$, and $\bar{\theta}$
are also given by
Eqs.~(\ref{eq:kouchan-16.13})--(\ref{eq:kouchan-16.18}),
(\ref{eq:kouchan-16.40}), (\ref{eq:kouchan-16.41}),
(\ref{eq:first-theta-gauge-inv})--(\ref{eq:kouchan-17.342}).
In terms of these gauge-invariant variables, the first- and the
second-order perturbations
(\ref{eq:first-order-continuity-perfect}) and
(\ref{eq:second-order-continuity-perfect}) of the continuity
equation are given in the gauge-invariant form. 
First, we derive the gauge-invariant expression of the
first-order perturbation
(\ref{eq:first-order-continuity-perfect}) of the continuity
equation 
(\ref{eq:continuity-equation-of-perfect-fluid}). 
Substituting Eqs.~(\ref{eq:kouchan-16.13}),
(\ref{eq:kouchan-16.40}), and (\ref{eq:first-theta-gauge-inv}) 
into Eq.~(\ref{eq:first-order-continuity-perfect}), we can
decompose the first-order perturbation ${}^{(1)}\!C_{0}^{(p)}$
into the gauge-invariant and gauge-variant parts as 
\begin{eqnarray}
  {}^{(1)}\!C_{0}^{(p)}
  &=&
  {}^{(1)}\!{\cal C}_{0}^{(p)}
  + {\pounds}_{X}C_{0}^{(p)},
  \label{eq:first-order-continuity-perfect-decomp}
\end{eqnarray}
where
\begin{eqnarray}
  {}^{(1)}\!{\cal C}_{0}^{(p)} &:=&
  u^{a}\nabla_{a}\stackrel{(1)}{{\cal E}}
  + \left(
      \stackrel{(1)}{{\cal U}^{a}} 
    - {\cal H}^{ab} u_{b}
  \right)
  \nabla_{a}\epsilon 
  + \left(
      \stackrel{(1)}{{\cal E}}
    + \stackrel{(1)}{{\cal P}}
  \right) \theta
  + \left(\epsilon + p\right) \stackrel{(1)}{\Theta}
  .
  \label{eq:first-order-continuity-perfect-gauge-inv}
\end{eqnarray}
We note that
Eq.~(\ref{eq:first-order-continuity-perfect-decomp}) has the
same form as Eq.~(\ref{eq:matter-gauge-inv-decomp-1.0}). 
By virtue of the background continuity equation
(\ref{eq:background-continuity-perfect}), the first-order
perturbation (\ref{eq:first-order-continuity-perfect}) of the
continuity equation is given in a gauge-invariant form:
\begin{eqnarray}
  {}^{(1)}\!{\cal C}_{0}^{(p)} = 0.
  \label{eq:first-order-continuity-perfect-gauge-inv-equation}
\end{eqnarray}
Further, through the definitions
(\ref{eq:kouchan-16.13})--(\ref{eq:kouchan-16.18}) and the
decomposition formulae (\ref{eq:kouchan-16.40}),
(\ref{eq:kouchan-16.41}), (\ref{eq:first-theta-gauge-inv}), and 
(\ref{eq:second-theta-gauge-inv}), the second-order perturbation 
${}^{(2)}\!C_{0}^{(p)}$ of the continuity equation defined by
(\ref{eq:second-order-continuity-perfect}) is decomposed into
the form 
\begin{eqnarray}
  {}^{(2)}\!C_{0}^{(p)}
  &=&
  {}^{(2)}\!{\cal C}_{0}^{(p)}
  + 2 {\pounds}_{X}{}^{(1)}\!C_{0}^{(p)}
  + \left( {\pounds}_{Y} - {\pounds}_{X}^{2} \right)C_{0}^{(p)}, 
  \label{eq:second-order-continuity-perfect-decomp}
\end{eqnarray}
where
\begin{eqnarray}
  {}^{(2)}\!{\cal C}_{0}^{(p)}
  &=&
      u^{a}\nabla_{a}\stackrel{(2)}{{\cal E}} 
  + \left(
        \stackrel{(2)}{{\cal U}^{a}}
    - 2 {\cal H}^{ab} \stackrel{(1)}{{\cal U}_{b}}
    + 2 {\cal H}^{ac} {\cal H}_{cb} u^{b}
    -   {\cal L}^{ab} u_{b}
  \right)
  \nabla_{a}\epsilon 
  \nonumber\\
  && \quad
  +   \theta \left(
      \stackrel{(2)}{{\cal E}} 
    + \stackrel{(2)}{{\cal P}}
  \right)
  +   \stackrel{(2)}{\Theta} \left(\epsilon + p\right)
  + 2 \left(
    \stackrel{(1)}{{\cal U}^{a}} - {\cal H}^{ab} u_{b}
  \right)
  \nabla_{a}\stackrel{(1)}{{\cal E}} 
  \nonumber\\
  && \quad
  + 2 \stackrel{(1)}{\Theta}
  \left(
    \stackrel{(1)}{{\cal E}} + \stackrel{(1)}{{\cal P}}
  \right)
  \label{eq:second-order-continuity-perfect-gauge-inv}
  .
\end{eqnarray}
We also note that
Eq.~(\ref{eq:second-order-continuity-perfect-decomp}) has the
same form as Eq.~(\ref{eq:matter-gauge-inv-decomp-2.0}).
Through the background equation
(\ref{eq:background-continuity-perfect}) and the first-order
perturbation (\ref{eq:first-order-continuity-perfect}) of the
continuity equation, the second-order perturbation
(\ref{eq:second-order-continuity-perfect}) of the continuity
equation is given in the gauge-invariant form: 
\begin{eqnarray}
  {}^{(2)}\!{\cal C}_{0}^{(p)} = 0.
  \label{eq:second-order-continuity-perfect-gauge-inv-equation}
\end{eqnarray}
Thus, we have obtained the gauge-invariant form of the first-
and the second-order perturbations of the continuity equation
for a perfect fluid through the lower order equations without
any gauge fixing.


\subsubsection{Perturbations of the Euler equation}
\label{sec:Perfect-fluid-Euler-equation-generic}


Here, we consider the perturbations of the Euler equations.
The component of 
$\bar{\nabla}_{b}{}^{(p)}\!\bar{T}_{a}^{\;\;b} = 0$ orthogonal
to $\bar{u}^{a}$ gives the Euler equation 
\begin{eqnarray}
  &&
  \bar{C}_{b}^{(p)}
  :=
  (\bar{\epsilon} + \bar{p}) \bar{a}_{b}
  + \bar{g}^{ac} \bar{q}_{bc} \bar{\nabla}_{a}\bar{p} 
  = 0.
  \label{eq:Euler-equation-of-perfect-fluid}
\end{eqnarray}
Here, the three-metric $\bar{q}_{ab}$, the acceleration vector
$\bar{a}_{b}$ associated with the four-velocity $\bar{u}_{a}$
are defined by Eqs.~(\ref{eq:three-metric-definition-full}),
(\ref{eq:acceleration-def-full}), and (\ref{eq:kouchan-17.89})
in Appendix
\ref{sec:Perturbations-of-accerelation-expansion-shear-and-rotation}.
Substituting the perturbative expansions
(\ref{eq:energy-density-expansion}),
(\ref{eq:pressure-expansion}),
(\ref{eq:acceleration-expansion}),
(\ref{eq:three-metric-expansion}), and 
(\ref{eq:inverse-metric-expansion}) into the Euler equation
(\ref{eq:Euler-equation-of-perfect-fluid}), we obtain the
expansion form of Eq.~(\ref{eq:Euler-equation-of-perfect-fluid})
as 
\begin{eqnarray}
  \bar{C}_{b}^{(p)}
  =:
  C_{b}^{(p)}
  + \lambda {}^{(1)}\!C_{b}^{(p)}
  + \frac{1}{2} \lambda^{2} {}^{(2)}\!C_{b}^{(p)}
  + O(\lambda^{3})
  = 0 .
\end{eqnarray}
Then, we obtain the Euler equation of each order:
\begin{eqnarray}
  C_{b}^{(p)}
  &:=&
  \left( \epsilon + p \right) a_{b} + g^{ac} q_{bc} \nabla_{a}p
  =
  0
  \label{eq:background-Euler-perfect}
  , \\
  {}^{(1)}\!C_{b}^{(p)}
  &:=&
  \left( \epsilon + p \right) \stackrel{(1)}{(a_{b})}
  + \left( \stackrel{(1)}{\epsilon} + \stackrel{(1)}{p} \right) a_{b}
  + g^{ac} q_{bc} \nabla_{a} \stackrel{(1)}{p}
  \nonumber\\
  &&
  + g^{ac} \stackrel{(1)}{(q_{bc})} \nabla_{a}p
  - h^{ac} q_{bc} \nabla_{a}p
  =
  0
  \label{eq:first-order-Euler-perfect}
  , \\
  {}^{(2)}\!C_{b}^{(p)}
  &:=&
  \left( \epsilon + p \right) \stackrel{(2)}{(a_{b})}
  + 2
  \left(
    \stackrel{(1)}{\epsilon} + \stackrel{(1)}{p}
  \right)
  \stackrel{(1)}{(a_{b})}
  +
  \left(
    \stackrel{(2)}{\epsilon} + \stackrel{(2)}{p}
  \right)
  a_{b}
  + g^{ac} q_{bc} \nabla_{a} \stackrel{(2)}{p}
  \nonumber\\
  &&
  + 2 g^{ac} \stackrel{(1)}{(q_{bc})} \nabla_{a}\stackrel{(1)}{p}
  + g^{ac} \stackrel{(2)}{(q_{bc})} \nabla_{a}p
  - 2 h^{ac} q_{bc} \nabla_{a}\stackrel{(1)}{p}
  \nonumber\\
  &&
  - 2 h^{ac} \stackrel{(1)}{(q_{bc})} \nabla_{a}p
  + \left( 2 h^{ad}h_{d}^{\;\;c} - l^{ac}\right) q_{bc} \nabla_{a}p
  =
  0
  .
  \label{eq:second-order-Euler-perfect}
\end{eqnarray}


As in the case of the continuity equation, the first- and the
second-order perturbations (\ref{eq:first-order-Euler-perfect}) and
(\ref{eq:second-order-Euler-perfect}) of the Euler equation
should be given in the gauge-invariant form. 
First, we consider the gauge-invariant expression of the
first-order perturbation (\ref{eq:first-order-Euler-perfect}) of
the Euler equation. 
Substituting Eqs.~(\ref{eq:linear-metric-decomp}),
(\ref{eq:kouchan-16.13}),
(\ref{eq:first-three-metric-gauge-inv}), and
(\ref{eq:first-acceleration-gauge-inv}), 
we obtain the expression of ${}^{(1)}\!C_{b}^{(p)}$ as 
\begin{eqnarray}
  {}^{(1)}\!C_{b}^{(p)}
  &=&
  {}^{(1)}\!{\cal C}_{b}^{(p)}
  + {\pounds}_{X}C_{b}^{(p)}
  ,
  \label{eq:first-order-Euler-perfect-decomp}
\end{eqnarray}
where
\begin{eqnarray}
  {}^{(1)}\!{\cal C}_{b}^{(p)}
  &:=&
    \left( \epsilon + p \right) \stackrel{(1)}{{\cal A}_{b}}
  + \left( \stackrel{(1)}{{\cal E}} + \stackrel{(1)}{{\cal P}}\right) a_{b}
  + g^{ac} q_{bc} \nabla_{a}\stackrel{(1)}{{\cal P}}
  \nonumber\\
  &&
  + g^{ac} \stackrel{(1)}{{\cal Q}_{bc}} \nabla_{a}p
  - g^{ad} g^{ce} {\cal H}_{de} q_{bc} \nabla_{a}p
  \label{eq:first-order-Euler-perfect-gauge-inv}
  ,
\end{eqnarray}
The equation (\ref{eq:first-order-Euler-perfect-decomp}) has the
same form as Eq.~(\ref{eq:matter-gauge-inv-decomp-1.0}).
By virtue of the background Euler equation
(\ref{eq:background-Euler-perfect}), the first-order
perturbation (\ref{eq:first-order-Euler-perfect}) of the Euler
equation is given in a gauge-invariant form: 
\begin{eqnarray}
  {}^{(1)}\!{\cal C}_{b}^{(p)} = 0.
  \label{eq:first-order-Euler-perfect-gauge-inv-equation}
\end{eqnarray}
Second, through the definitions (\ref{eq:kouchan-16.13}),
(\ref{eq:kouchan-16.17}) of gauge invariant variables, and the
decomposition formulae (\ref{eq:linear-metric-decomp}),
(\ref{eq:second-metric-decomp}), 
(\ref{eq:first-three-metric-gauge-inv}),
(\ref{eq:second-three-metric-gauge-inv}),
(\ref{eq:first-acceleration-gauge-inv}), and
(\ref{eq:second-acceleration-gauge-inv}) of gauge-invariant
variables, the second-order perturbation
${}^{(2)}\!C_{b}^{(p)}$ defined by
(\ref{eq:second-order-Euler-perfect}) is decomposed in the form
\begin{eqnarray}
  {}^{(2)}\!C_{b}^{(p)}
  &=&
  {}^{(2)}\!{\cal C}_{b}^{(p)}
  + 2 {\pounds}_{X}{}^{(1)}\!C_{b}^{(p)}
  +   \left(
    {\pounds}_{Y} - {\pounds}_{X}^{2}
  \right)C_{b}^{(p)}
  \label{eq:second-order-Euler-perfect-decomp}
  ,
\end{eqnarray}
where
\begin{eqnarray}
  {}^{(2)}\!{\cal C}_{b}^{(p)}
  &:=&
    \left( \epsilon + p \right) \stackrel{(2)}{{\cal A}_{b}}
  + \left(\stackrel{(2)}{{\cal E}} + \stackrel{(2)}{{\cal P}}\right) a_{b}
  + g^{ac} q_{bc} \nabla_{a}\stackrel{(2)}{{\cal P}}
  + g^{ac} \stackrel{(2)}{{\cal Q}_{bc}} \nabla_{a}p
  \nonumber\\
  &&
  - g^{ad} g^{ce} {\cal L}_{de} q_{bc} \nabla_{a}p
  + 2 \left(
      \stackrel{(1)}{{\cal E}}
    + \stackrel{(1)}{{\cal P}}
  \right) \stackrel{(1)}{{\cal A}_{b}}
  + 2 g^{ac} \stackrel{(1)}{{\cal Q}_{bc}} \nabla_{a}\stackrel{(1)}{{\cal P}}
  - 2 {\cal H}^{ac} q_{bc} \nabla_{a}\stackrel{(1)}{{\cal P}}
  \nonumber\\
  && 
  - 2 {\cal H}^{ac} \stackrel{(1)}{{\cal Q}_{bc}} \nabla_{a}p
  + 2 {\cal H}^{ad} g^{ce} {\cal H}_{de} q_{bc} \nabla_{a}p
  .
  \label{eq:second-order-Euler-perfect-gauge-inv}
\end{eqnarray}
The equation (\ref{eq:second-order-Euler-perfect-decomp}) has
the same form as Eq.~(\ref{eq:matter-gauge-inv-decomp-2.0}).
Through the background equation
(\ref{eq:background-Euler-perfect}) and the first-order
perturbation (\ref{eq:first-order-Euler-perfect}) of the
Euler equation, the second-order perturbation
(\ref{eq:second-order-Euler-perfect}) of the Euler
equation is given in the gauge-invariant form: 
\begin{eqnarray}
  {}^{(2)}\!{\cal C}_{b}^{(p)} = 0.
  \label{eq:second-order-Euler-perfect-gauge-inv-eqution}
\end{eqnarray}
Thus, we have obtained the gauge-invariant form of the first-
and the second-order perturbations of the Euler equation for a
perfect fluid without any gauge fixing.


\subsection{Imperfect fluid}
\label{sec:imperfect-fluid-generic}


In this subsection, we consider the perturbative expressions of
the energy momentum tensor and the equations of motion for an
imperfect fluid which includes important effects in the recent
cosmology.
We derive the gauge-invariant part of the perturbations of the
energy-momentum tensor (\ref{eq:imperfect-fluid-ene-mom-full-1})
for an imperfect fluid in
\S\ref{sec:Imerfect-fluid-ene-mon-generic}.
Then, we derive the gauge-invariant continuity equation for an
imperfect fluid
\S\ref{sec:Imerfect-fluid-continuity-equation-generic}.
Further, we derive the generalized Navier-Stokes equation in
\S\ref{sec:Imerfect-fluid-Navier-Stokes-equation-generic}, which
corresponds to the Euler equation for a perfect fluid.


\subsubsection{Energy momentum tensor}
\label{sec:Imerfect-fluid-ene-mon-generic}


To derive the perturbations of the energy momentum tensor for an
imperfect fluid, it is convenient to consider the perturbation
of the contravariant energy flux $\bar{q}^{a}:=\bar{g}^{ab}q_{b}$:
\begin{eqnarray}
  \label{eq:kouchan-16.72-2}
  \bar{q}^{a} 
  &=:& 
  q^{a}
  + \lambda \stackrel{(1)}{(q^{a})}
  + \frac{1}{2} \lambda^{2} \stackrel{(2)}{(q^{a})}
  + O(\lambda^{3})
  .
\end{eqnarray}
The perturbations $\stackrel{(1)}{(q^{a})}$ and
$\stackrel{(2)}{(q^{a})}$ are given by the same procedure as the
derivations of the perturbations of $\bar{u}^{a}$ in Appendix
\ref{sec:acceleration-appendix} and these are decomposed into
gauge-invariant and gauge-variant parts:
\begin{eqnarray}
  \stackrel{(1)}{(q^{a})}
  &=&
  \stackrel{(1)}{{\cal Q}^{a}}
  - {\cal H}^{ab} q_{b}
  + {\pounds}_{X}q^{a}
  ,
  \label{eq:kouchan-16.115}
  \\
  \stackrel{(2)}{(q^{a})}
  &=&
    \stackrel{(2)}{{\cal Q}^{a}}
  - q_{b} {\cal L}^{ab}
  - 2 {\cal H}^{ab} \stackrel{(1)}{{\cal Q}_{b}}
  + 2 {\cal H}^{ac} {\cal H}_{cd} q^{d}
  + 2 {\pounds}_{X}\stackrel{(1)}{(q^{a})}
  + {\pounds}_{Y}q^{a}
  - {\pounds}_{X}^{2}q^{a}
  ,
  \label{eq:kouchan-16.119}
\end{eqnarray}
where we defined
\begin{eqnarray}
  \label{eq:kouchan-16.114}
  \stackrel{(1)}{{\cal Q}^{a}} := g^{ab} \stackrel{(1)}{{\cal Q}_{b}},
  \quad
  \stackrel{(2)}{{\cal Q}^{a}} := g^{ac} \stackrel{(2)}{{\cal Q}_{c}}
  .
\end{eqnarray}
Further, $\bar{\pi}_{a}^{\;\;b}:=\bar{g}^{bc}\bar{\pi}_{ac}$ is
expanded as 
\begin{eqnarray}
  \label{eq:kouchan-16.74-2}
  \bar{\pi}_{a}^{\;\;b} 
  &=:& 
  \pi_{a}^{\;\;b}
  + \lambda \stackrel{(1)}{(\pi_{a}^{\;\;b})}
  + \frac{1}{2} \lambda^{2} \stackrel{(2)}{(\pi_{a}^{\;\;b})}
  + O(\lambda^{3})
  ,
\end{eqnarray}
and the similar procedure to decompose the perturbations of
$\bar{u}^{a}$ into the gauge-invariant and the gauge-variant
parts yields
\begin{eqnarray}
  \stackrel{(1)}{(\pi_{a}^{\;\;b})}
  &=& 
    \stackrel{(1)}{\Pi_{a}^{\;\;b}}
  - {\cal H}^{bc} \pi_{ac}
  + {\pounds}_{X}\pi_{a}^{\;\;b},
  \label{eq:kouchan-16.127}
  \\
  \stackrel{(2)}{(\pi_{a}^{\;\;b})}
  &=&
  \stackrel{(2)}{\Pi_{a}^{\;\;b}}
  - 2 {\cal H}^{cb} \stackrel{(1)}{\Pi_{ca}}
  - \pi_{ac} {\cal L}^{cb}
  + 2 {\cal H}^{cd} {\cal H}_{d}^{\;\;b} \pi_{ac}
  \nonumber\\
  && \quad
  + 2 {\pounds}_{X}\stackrel{(1)}{(\pi_{a}^{\;\;b})}
  + {\pounds}_{Y}\pi_{a}^{\;\;b}
  - {\pounds}_{X}^{2}\pi_{a}^{\;\;b}
  ,
  \label{eq:kouchan-16.138}
\end{eqnarray}
where we defined 
\begin{eqnarray}
  \label{eq:kouchan-16.128}
  \stackrel{(1)}{\Pi_{a}^{\;\;b}} := g^{bc} \stackrel{(1)}{\Pi_{ac}},
  \quad
  \stackrel{(2)}{\Pi_{a}^{\;\;b}} := g^{cb} \stackrel{(2)}{\Pi_{ca}}.
\end{eqnarray}
The perturbative expansions of the traceless property [the last 
equation in Eqs.(\ref{eq:kouchan-16.59})] of the anisotropic
stress $\bar{\pi}_{a}^{\;\;b}$ are given in gauge-invariant
forms as
\begin{eqnarray}
  \pi_{a}^{\;\;a}
  = 
  0
  ,
  \quad
  \stackrel{(1)}{\Pi_{a}^{\;\;a}}
  = {\cal H}^{ac} \pi_{ac}
  ,
  \quad
  \stackrel{(2)}{\Pi_{a}^{\;\;a}}
  =
      \pi_{ac} {\cal L}^{ca}
  + 2 {\cal H}^{ca} \stackrel{(1)}{\Pi_{ca}}
  - 2 {\cal H}^{cd} {\cal H}_{d}^{\;\;a} \pi_{ac}
  \label{eq:kouchan-16.132}
  ,
\end{eqnarray}
where we used
$\pi_{a}^{\;\;a}=\stackrel{(1)}{(\pi_{a}^{\;\;b})}=\stackrel{(2)}{(\pi_{a}^{\;\;b})}=0$.


The orthogonal condition (\ref{eq:kouchan-16.58}) of the energy
flux $\bar{q}_{a}$ to the four-velocity $\bar{u}_{a}$
is also expanded perturbatively through
Eqs.~(\ref{eq:kouchan-16.72}) and
(\ref{eq:contravariant-four-velocity-expansion}).
The perturbations of the orthogonal condition
(\ref{eq:kouchan-16.58}) are given in the gauge-invariant form
as follows:
\begin{eqnarray}
  \label{kouchan-16.77}
  u^{a} q_{a} &=& 0
  , \\
  u^{a} \stackrel{(1)}{{\cal Q}_{a}}
  &=&
  - q_{a} \stackrel{(1)}{{\cal U}^{a}}
  + q_{a} {\cal H}^{ab} u_{b},
  \label{eq:kouchan-16.95-4}
  \\
  u^{a} \stackrel{(2)}{{\cal Q}_{a}}
  &=&
  - q_{a} \stackrel{(2)}{{\cal U}^{a}}
  + 2 q_{a} {\cal H}^{ab} \stackrel{(1)}{{\cal U}_{b}}
  - 2 q_{a} {\cal H}^{ac}{\cal H}_{cb} u^{b}
  +   q_{a} {\cal L}^{ab} u_{b}
  \nonumber\\
  &&
  - 2 \stackrel{(1)}{{\cal Q}_{a}} \stackrel{(1)}{{\cal U}^{a}}
  + 2 \stackrel{(1)}{{\cal Q}_{a}} {\cal H}^{ab} u_{b}
  .
  \label{eq:kouchan-16.95-6}
\end{eqnarray}
Here, we have to emphasize that we did not fix any gauge choice.
The perturbations of the orthogonal conditions
(\ref{eq:kouchan-16.58}) are also decomposed into the 
gauge-invariant and the gauge-variant parts as the formulae
(\ref{eq:matter-gauge-inv-decomp-1.0}) and
(\ref{eq:matter-gauge-inv-decomp-2.0}). 
Since the gauge-variant parts of the perturbations of
Eq.~(\ref{eq:kouchan-16.58}) are given by the Lie derivative of
its background value and the first-order perturbation of
Eq.~(\ref{eq:kouchan-16.58}), the perturbations of
Eq.~(\ref{eq:kouchan-16.58}) are necessarily given in the
gauge-invariant form through the lower order perturbations of
Eq.~(\ref{eq:kouchan-16.58}).


Similarly, the orthogonal condition [the second equation in
Eqs.~(\ref{eq:kouchan-16.59})] of the anisotropic stress
$\bar{\pi}_{ab}$ to the four-velocity $\bar{u}^{a}$ are also 
perturbatively expanded through
Eqs.~(\ref{eq:contravariant-four-velocity-expansion}) and
(\ref{eq:kouchan-16.74}) and these are given by
\begin{eqnarray}
  \label{eq:kouchan-16.81}
  u^{a}\pi_{ab} &=& 0
  , \\
  u^{a} \stackrel{(1)}{\Pi_{ab}}
  &=&
  - \stackrel{(1)}{{\cal U}^{a}} \pi_{ab} 
  + {\cal H}^{ac} u_{c} \pi_{ab}
  \label{eq:kouchan-16.95-8}
  , \\
  u^{a} \stackrel{(2)}{\Pi_{ab}}
  &=&
  -   \stackrel{(2)}{{\cal U}^{a}} \pi_{ab}
  + 2 {\cal H}^{ac} \stackrel{(1)}{{\cal U}_{c}} \pi_{ab}
  - 2 {\cal H}^{ac}{\cal H}_{cd} u^{d} \pi_{ab}
  +   {\cal L}^{ac} u_{c} \pi_{ab}
  \nonumber\\
  &&
  - 2 \stackrel{(1)}{\Pi_{ab}}
  \left(
    \stackrel{(1)}{{\cal U}^{a}} - {\cal H}^{ab} u_{b}
  \right)
  \label{eq:kouchan-16.95-13}
  ,
\end{eqnarray}
where we have used the background orthogonal condition
(\ref{eq:kouchan-16.81}) for the anisotropic stress $u_{a}$ and
its first-order perturbation.
Eqs.~(\ref{eq:kouchan-16.95-8}) and (\ref{eq:kouchan-16.95-13})
are gauge-invariant.
This is due to the fact that the gauge-variant parts of the
perturbations of $\bar{u}^{a} \bar{\pi}_{ab}$ are given by the
Lie derivative of the lower order perturbations of
$\bar{u}^{a}\bar{\pi}_{ab}$ as
Eqs.~(\ref{eq:matter-gauge-inv-decomp-1.0}) and
(\ref{eq:matter-gauge-inv-decomp-2.0}) and these gauge-variant
parts vanish as in the case of the perturbations of the
orthogonal condition $\bar{u}^{a}\bar{q}_{a}=0$.


These orthogonal conditions
(\ref{kouchan-16.77})--(\ref{eq:kouchan-16.95-13}) for the
perturbations of the energy flux and the anisotropic stress are
necessary to specify the independent components of the
gauge-invariant parts of the perturbations of the energy flux
and the anisotropic stress.


Now, we consider the first- and the second-order perturbations
of the imperfect part ${}^{(i)}\!\bar{T}_{a}^{\;\;b}$ of the
energy momentum tensor for an imperfect fluid.
Through Eqs.~(\ref{eq:four-velocity-expansion}),
(\ref{eq:contravariant-four-velocity-expansion}),
(\ref{eq:kouchan-16.72}), (\ref{eq:kouchan-16.72-2}), and
(\ref{eq:kouchan-16.74-2}), the perturbative expansion of the
imperfect part (\ref{eq:imperfect-part-ene-mom-full}) of the
energy momentum tensor is expanded as 
\begin{eqnarray}
  \label{eq:energy-momentum-tensor-expansion-imperfect-part}
  {}^{(i)}\!\bar{T}_{a}^{\;\;b}
  &=&
  {}^{(i)}\!T_{a}^{\;\;b}
  + \lambda \stackrel{(1)}{{}^{(i)}\!T_{a}^{\;\;b}}
  + \frac{1}{2} \lambda^{2} \stackrel{(2)}{{}^{(i)}\!T_{a}^{\;\;b}}
  + O(\lambda^{3})
  ,
\end{eqnarray}
where
\begin{eqnarray}
  {}^{(i)}\!T_{a}^{\;\;b}
  &=&
    q_{a}u^{b}
  + u_{a}q^{b}
  + \pi_{a}^{\;\;b}
  \label{eq:kouchan-16.146}
  , \\
  \stackrel{(1)}{{}^{(i)}\!T_{a}^{\;\;b}}
  &=&
    q_{a} \stackrel{(1)}{(u^{b})} 
  + \stackrel{(1)}{(q_{a})} u^{b}
  + u_{a} \stackrel{(1)}{(q^{b})}
  + \stackrel{(1)}{(u_{a})} q^{b}
  + \stackrel{(1)}{(\pi_{a}^{\;\;b})}
  \label{eq:kouchan-16.147}
  , \\
  \stackrel{(2)}{{}^{(i)}\!T_{a}^{\;\;b}}
  &=&
      q_{a} \stackrel{(2)}{(u^{b})}
  + 2 \stackrel{(1)}{(q_{a})} \stackrel{(1)}{(u^{b})}
  +   \stackrel{(2)}{(q_{a})} u^{b} 
  +   u_{a} \stackrel{(2)}{(q^{b})}
  + 2 \stackrel{(1)}{(u_{a})} \stackrel{(1)}{(q^{b})}
  +   \stackrel{(2)}{(u_{a})} q^{b}
  \nonumber\\
  && \quad
  +   \stackrel{(2)}{(\pi_{a}^{\;\;b})}
  \label{eq:kouchan-16.148}
  .
\end{eqnarray}


Substituting (\ref{eq:kouchan-16.13}), (\ref{eq:kouchan-16.40}),
(\ref{eq:kouchan-16.92}), (\ref{eq:kouchan-16.115}), and
(\ref{eq:kouchan-16.127}), into (\ref{eq:kouchan-16.147}), the
first-order perturbation of the imperfect part of the energy
momentum tensor for an imperfect fluid is decomposed as 
\begin{eqnarray}
  \stackrel{(1)}{{}^{(i)}\!T_{a}^{\;\;b}}
  &=&
  \stackrel{(1)}{{}^{(i)}\!{\cal T}_{a}^{\;\;b}}
  + {\pounds}_{X}{}^{(i)}T_{a}^{\;\;b}
  ,
  \label{eq:kouchan-16.149}
\end{eqnarray}
where
\begin{eqnarray}
  \stackrel{(1)}{{}^{(i)}\!{\cal T}_{a}^{\;\;b}}
  &=&
    q_{a} \stackrel{(1)}{{\cal U}^{b}} 
  + \stackrel{(1)}{{\cal U}_{a}} q^{b}
  + \stackrel{(1)}{{\cal Q}_{a}} u^{b}
  + u_{a} \stackrel{(1)}{{\cal Q}^{b}}
  - 2 u_{(a} q_{c)} {\cal H}^{bc}
  + \stackrel{(1)}{\Pi_{a}^{\;\;b}}
  - \pi_{ac} {\cal H}^{bc}
  .
  \label{eq:kouchan-16.149-2}
\end{eqnarray}
Further, through
Eqs.~(\ref{eq:kouchan-16.13})--(\ref{eq:kouchan-16.18}), 
(\ref{eq:kouchan-16.115})--(\ref{eq:kouchan-16.92}),
(\ref{eq:kouchan-16.138}),
(\ref{eq:kouchan-16.40}), and (\ref{eq:kouchan-16.41}),
the second-order perturbation (\ref{eq:kouchan-16.148}) of the 
imperfect part of the energy momentum tensor for an imperfect
fluid is decomposed as
\begin{eqnarray}
  \label{eq:second-energy-momentum-tensor-imperfect-part-decomposed}
  \stackrel{(2)}{{}^{(i)}\!T_{a}^{\;\;b}}
  &=&
  \stackrel{(2)}{{}^{(i)}\!{\cal T}_{a}^{\;\;b}}
  + 2 {\pounds}_{X}\stackrel{(1)}{{}^{(i)}\!T_{a}^{\;\;b}}
  +   \left({\pounds}_{Y} - {\pounds}_{X}^{2}\right){}^{(i)}\!T_{a}^{\;\;b}
  ,
\end{eqnarray}
where
\begin{eqnarray}
  \stackrel{(2)}{{}^{(i)}\!{\cal T}_{a}^{\;\;b}}
  &=&
      q_{a} \stackrel{(2)}{{\cal U}^{b}}
  - 2 q_{a} {\cal H}^{bc} \stackrel{(1)}{{\cal U}_{c}}
  + 2 q_{a} {\cal H}^{bd}{\cal H}_{dc} u^{c}
  -   q_{a} {\cal L}^{bc} u_{c}
  +   \stackrel{(2)}{{\cal Q}_{a}} u^{b} 
  \nonumber\\
  && 
  -   u_{a} q_{c} {\cal L}^{bc}
  - 2 u_{a} {\cal H}^{bc} \stackrel{(1)}{{\cal Q}_{c}}
  + 2 u_{a} {\cal H}^{bc} {\cal H}_{cd} q^{d}
  \nonumber\\
  && 
  +   u_{a} \stackrel{(2)}{{\cal Q}^{b}}
  +   \stackrel{(2)}{{\cal U}_{a}} q^{b}
  +   \stackrel{(2)}{\Pi_{a}^{\;\;b}}
  - 2 {\cal H}^{cb} \stackrel{(1)}{\Pi_{ca}}
  -   \pi_{ac} {\cal L}^{cb}
  + 2 {\cal H}^{cd} {\cal H}_{d}^{\;\;b} \pi_{ac}
  \nonumber\\
  && 
  + 2 \stackrel{(1)}{{\cal Q}_{a}} \stackrel{(1)}{{\cal U}^{b}}
  - 2 {\cal H}^{bc} u_{c} \stackrel{(1)}{{\cal Q}_{a}}
  + 2 \stackrel{(1)}{{\cal Q}^{b}} \stackrel{(1)}{{\cal U}_{a}}
  - 2 {\cal H}^{bc} q_{c} \stackrel{(1)}{{\cal U}_{a}}
  .
  \label{eq:second-energy-momentum-tensor-imperfect-part-gauge-inv}
\end{eqnarray}
Note that Eqs.~(\ref{eq:kouchan-16.149}) and
(\ref{eq:second-energy-momentum-tensor-imperfect-part-decomposed})
have the same form as
Eqs.~(\ref{eq:matter-gauge-inv-decomp-1.0}) and
(\ref{eq:matter-gauge-inv-decomp-2.0}), respectively.


Together with
Eqs.~(\ref{eq:first-energy-momentum-tensor-perfect-decomposed})
and (\ref{eq:second-energy-momentum-tensor-perfect-decomposed})
for the perfect fluid, the total energy momentum tensor of each
order is defined by 
\begin{eqnarray}
  \bar{T}_{a}^{\;\;b}
  &=:&
  T_{a}^{\;\;b}
  + \lambda \stackrel{(1)}{T_{a}^{\;\;b}}
  + \frac{1}{2} \lambda^{2} \stackrel{(2)}{T_{a}^{\;\;b}}
  + O(\lambda^{3})
  \label{eq:perturbative-expression-energy-momentum-tensor-imperfect-total}
  ,
  \\
  T_{a}^{\;\;b}
  &:=& 
  {}^{(p)}\!T_{a}^{\;\;b} 
  +
  {}^{(i)}\!T_{a}^{\;\;b}
  , 
  \quad
  \stackrel{(1)}{T_{a}^{\;\;b}}
  :=
  \stackrel{(1)}{{}^{(p)}\!T_{a}^{\;\;b}}
  +
  \stackrel{(1)}{{}^{(i)}\!T_{a}^{\;\;b}}
  , \quad
  \stackrel{(2)}{T_{a}^{\;\;b}}
  :=
  \stackrel{(2)}{{}^{(p)}\!T_{a}^{\;\;b}}
  +
  \stackrel{(2)}{{}^{(i)}\!T_{a}^{\;\;b}}
  \label{eq:second-order-energy-momentum-tensor-imperfect-total}
  .
\end{eqnarray}
Further, the first- and the second-order perturbations of the
total energy momentum tensor are also decomposed into the
gauge-variant and gauge-invariant parts: 
\begin{eqnarray}
  \stackrel{(1)}{T_{a}^{\;\;b}}
  &=&
  \stackrel{(1)}{{\cal T}_{a}^{\;\;b}} 
  +
  {\pounds}_{X}T_{a}^{\;\;b}
  ,
  \\
  \stackrel{(2)}{T_{a}^{\;\;b}}
  &=&
  \stackrel{(2)}{{\cal T}_{a}^{\;\;b}}
  + 2 {\pounds}_{X}\stackrel{(1)}{T_{a}^{\;\;b}}
  +   \left({\pounds}_{Y} - {\pounds}_{X}^{2}\right) T_{a}^{\;\;b}
  .
\end{eqnarray}
Through
Eqs.~(\ref{eq:first-energy-momentum-tensor-perfect-gauge-inv})
and (\ref{eq:kouchan-16.149-2}), the gauge-invariant part
$\stackrel{(1)}{{\cal T}_{a}^{\;\;b}}$ of the first-order
perturbation of the total energy momentum tensor is given by
\begin{eqnarray}
  \stackrel{(1)}{{\cal T}_{a}^{\;\;b}}
  &:=&
  \stackrel{(1)}{{}^{(p)}\!{\cal T}_{a}^{\;\;b}}
  +
  \stackrel{(1)}{{}^{(i)}\!{\cal T}_{a}^{\;\;b}}
  \\
  &=&
  \left(
    \stackrel{(1)}{{\cal E}}
    + \stackrel{(1)}{{\cal P}}
  \right) u_{a} u^{b}
  + \stackrel{(1)}{{\cal P}} \delta_{a}^{\;\;b}
  + \left( \epsilon + p \right) \left(
    u_{a} \stackrel{(1)}{{\cal U}^{b}} 
    - {\cal H}^{bc} u_{c} u_{a} 
    + \stackrel{(1)}{{\cal U}_{a}} u^{b}
  \right)
  \nonumber\\
  && \quad
  + q_{a} \stackrel{(1)}{{\cal U}^{b}} 
  + \stackrel{(1)}{{\cal U}_{a}} q^{b}
  + \stackrel{(1)}{{\cal Q}_{a}} u^{b}
  + u_{a} \stackrel{(1)}{{\cal Q}^{b}}
  - 2 u_{(a} q_{c)} {\cal H}^{bc}
  \nonumber\\
  && \quad
  + \stackrel{(1)}{\Pi_{a}^{\;\;b}}
  - \pi_{ac} {\cal H}^{bc}
  \label{eq:first-ptotal-ene-mom-imperfect-gauge-inv}
  .
\end{eqnarray}
On the other hand, through
Eqs.~(\ref{eq:second-energy-momentum-tensor-perfect-gauge-inv}),
and
(\ref{eq:second-energy-momentum-tensor-imperfect-part-gauge-inv}),
the gauge-invariant part
$\stackrel{(2)}{{\cal T}_{a}^{\;\;b}}$ of the second-order
perturbation of the total energy momentum tensor is given by
\begin{eqnarray}
  \stackrel{(2)}{{\cal T}_{a}^{\;\;b}} 
  &:=&
  \stackrel{(2)}{{}^{(p)}\!{\cal T}_{a}^{\;\;b}}
  +
  \stackrel{(2)}{{}^{(i)}\!{\cal T}_{a}^{\;\;b}}
  \\
  &=&
  \left( 
    \stackrel{(2)}{{\cal E}} 
    + \stackrel{(2)}{{\cal P}}
  \right) u_{a} u^{b}
  + 2 \left( 
    \stackrel{(1)}{{\cal E}} + \stackrel{(1)}{{\cal P}} 
  \right) 
  u_{a}
  \left(
    \stackrel{(1)}{{\cal U}^{b}} - {\cal H}^{bc} u_{c}
  \right)
  + 2 
  \left(
    \stackrel{(1)}{{\cal E}} + \stackrel{(1)}{{\cal P}} 
  \right) 
  \stackrel{(1)}{{\cal U}_{a}} u^{b}
  \nonumber\\
  && 
  + \left( \epsilon + p \right) u_{a} \left(
        \stackrel{(2)}{{\cal U}^{b}}
    - 2 {\cal H}^{bc} \stackrel{(1)}{{\cal U}_{c}}
    + 2 {\cal H}^{bc}{\cal H}_{cd} u^{d}
    -   {\cal L}^{bd} u_{d}
  \right)
  \nonumber\\
  && 
  + 2 \left( \epsilon + p \right)
  \stackrel{(1)}{{\cal U}_{a}} \left(
    \stackrel{(1)}{{\cal U}^{b}} 
    - {\cal H}^{bc} u_{c}
  \right)
  + \left( \epsilon + p \right) \stackrel{(2)}{{\cal U}_{a}} u^{b}
  +   \stackrel{(2)}{{\cal P}} \delta_{a}^{\;\;b}.
  \nonumber\\
  && 
  +   q_{a} \stackrel{(2)}{{\cal U}^{b}}
  - 2 q_{a} {\cal H}^{bc} \stackrel{(1)}{{\cal U}_{c}}
  + 2 q_{a} {\cal H}^{bd}{\cal H}_{dc} u^{c}
  -   q_{a} {\cal L}^{bc} u_{c}
  +   \stackrel{(2)}{{\cal Q}_{a}} u^{b} 
  \nonumber\\
  && 
  -   u_{a} q_{c} {\cal L}^{bc}
  - 2 u_{a} {\cal H}^{bc} \stackrel{(1)}{{\cal Q}_{c}}
  + 2 u_{a} {\cal H}^{bc} {\cal H}_{cd} q^{d}
  +   u_{a} \stackrel{(2)}{{\cal Q}^{b}}
  +   \stackrel{(2)}{{\cal U}_{a}} q^{b}
  \nonumber\\
  && 
  +   \stackrel{(2)}{\Pi_{a}^{\;\;b}}
  - 2 {\cal H}^{cb} \stackrel{(1)}{\Pi_{ca}}
  -   \pi_{ac} {\cal L}^{cb}
  + 2 {\cal H}^{cd} {\cal H}_{d}^{\;\;b} \pi_{ac}
  \nonumber\\
  && 
  + 2 \stackrel{(1)}{{\cal Q}_{a}} \stackrel{(1)}{{\cal U}^{b}}
  - 2 {\cal H}^{bc} u_{c} \stackrel{(1)}{{\cal Q}_{a}}
  + 2 \stackrel{(1)}{{\cal Q}^{b}} \stackrel{(1)}{{\cal U}_{a}}
  - 2 {\cal H}^{bc} q_{c} \stackrel{(1)}{{\cal U}_{a}}
  .
  \label{eq:second-total-ene-mom-imperfect-gauge-inv}
\end{eqnarray}


\subsubsection{Perturbations of the continuity equation}
\label{sec:Imerfect-fluid-continuity-equation-generic}


The equation of motion $\bar{\nabla}_{b}\bar{T}_{a}^{\;\;b}=0$
is decomposed into the tangential and normal components to the
four-velocity $\bar{u}_{a}$.
The tangential component of
$\bar{\nabla}_{b}\bar{T}_{a}^{\;\;b}=0$ to $\bar{u}_{a}$ is
given by
\begin{eqnarray}
  \label{eq:kouchan-16.69}
  \bar{u}^{b}\bar{\nabla}_{b}\bar{\epsilon}
  + (\bar{\epsilon} + \bar{p}) \bar{\theta}
  + \bar{q}^{b} \bar{a}_{b}
  + \bar{\nabla}_{b}\bar{q}^{b}
  + \bar{\pi}^{ab}\bar{B}_{ab} = 0,
\end{eqnarray}
where we have used Eqs.~(\ref{eq:barab-definition}),
(\ref{eq:barBab-definition}), and (\ref{eq:kouchan-17.89}).
Eq.~(\ref{eq:kouchan-16.69}) is the continuity equation for
an imperfect fluid.
The normal components of $\bar{\nabla}_{b}\bar{T}_{a}^{\;\;b}=0$
to the fluid four-velocity $\bar{u}^{a}$ is discussed in
\S\ref{sec:Imerfect-fluid-Navier-Stokes-equation-generic}.


The first two terms in Eq.~(\ref{eq:kouchan-16.69}) coincide
with $\bar{C}_{0}^{(p)}$ defined by
Eq.~(\ref{eq:continuity-equation-of-perfect-fluid}) and the
perturbative expression of these two terms are given in
\S\ref{sec:Perfect-fluid-continuity-equation-generic}. 
Then, we consider the perturbative expression of the remaining
terms in Eq.~(\ref{eq:kouchan-16.69}): 
\begin{eqnarray}
  \bar{C}_{0}^{(i)}
  &:=&
    \bar{g}^{ab} \left(
    \bar{q}_{a} \bar{a}_{b} + \bar{\nabla}_{a}\bar{q}_{b}
  \right)
  + \bar{g}^{ac} \bar{g}^{bd} \bar{\pi}_{ab}\bar{B}_{cd}.
  \label{eq:kouchan-16.710}
\end{eqnarray}
We expand this $\bar{C}_{0}^{(i)}$ as
\begin{eqnarray}
  \bar{C}_{0}^{(i)} 
  = 
  C_{0}^{(i)}
  + \lambda \stackrel{(1)}{C_{0}^{(i)}}
  + \frac{1}{2} \lambda^{2} \stackrel{(2)}{C_{0}^{(i)}}
  + O(\lambda^{3}).
  \label{eq:kouchan-16.711}
\end{eqnarray}


The first term $\bar{g}^{ab}\bar{q}_{a}\bar{a}_{b}$ in
Eq.~(\ref{eq:kouchan-16.710}) is expanded through 
Eqs.~(\ref{eq:kouchan-16.72}),
(\ref{eq:inverse-metric-expansion}), and
(\ref{eq:acceleration-expansion}).
The last term $\bar{g}^{ac}\bar{g}^{bd}\bar{\pi}_{ab}\bar{B}_{cd}$
in Eq.~(\ref{eq:kouchan-16.710}) is expanded through
Eqs.~(\ref{eq:kouchan-16.74}),
(\ref{eq:inverse-metric-expansion}), and
(\ref{eq:kouchan-17.701}).
Therefore, to derive the explicit expression of $C_{0}^{(i)}$,
$\stackrel{(1)}{C_{0}^{(i)}}$, and
$\stackrel{(2)}{C_{0}^{(i)}}$, we have to derive the
perturbative expression of the term
$\bar{\nabla}_{a}\bar{q}_{b}$. 
As in the case of the perturbations of the tensor $\bar{A}_{ab}$
which is derived in Appendix
\ref{sec:Perturbations-Aab-appendix}, we can derive the
perturbative expansion of the tensor
$\bar{\nabla}_{a}\bar{q}_{b}$ as follows:
\begin{eqnarray}
  \bar{\nabla}_{a}\bar{q}_{b}
  &=&
    \nabla_{a}q_{b}
  + \lambda \left(
      \nabla_{a}\stackrel{(1)}{(q_{b})}
    - H_{ab}^{\;\;\;\;c}\left[h\right] q_{c}
  \right)
  \nonumber\\
  && 
  + \frac{1}{2} \lambda^{2} \left(
      \nabla_{a}\stackrel{(2)}{(q_{b})}
    - 2 H_{ab}^{\;\;\;\;c}\left[h\right] \stackrel{(1)}{(q_{c})}
    - H_{ab}^{\;\;\;\;c}\left[l\right] q_{c}
    + 2 h^{cd} H_{abd}\left[h\right] q_{c}
  \right)
  \nonumber\\
  && 
  + O(\lambda^{3})
  \label{eq:kouchan-16.713}
  .
\end{eqnarray}


Substituting the expansion formulae
(\ref{eq:kouchan-16.72}), (\ref{eq:kouchan-16.74}),
(\ref{eq:kouchan-16.713}), (\ref{eq:inverse-metric-expansion}),
(\ref{eq:acceleration-expansion}), and (\ref{eq:kouchan-17.701}), 
the perturbations $C_{0}^{(i)}$, $\stackrel{(1)}{C_{0}^{(i)}}$,
and $\stackrel{(2)}{C_{0}^{(i)}}$ of $\bar{C}_{0}^{(i)}$ defined
by Eq.~(\ref{eq:kouchan-16.710}) are given by 
\begin{eqnarray}
  C_{0}^{(i)}
  &=&
  q^{b} a_{b} + \nabla_{b}q^{b} + \pi^{ab} B_{ab}
  \label{eq:kouchan-16.716}
  , \\
  \stackrel{(1)}{C_{0}^{(i)}}
  &=&
    q^{b} \stackrel{(1)}{(a_{b})} + \stackrel{(1)}{(q_{b})} a^{b}
  + \nabla^{b}\stackrel{(1)}{(q_{b})}
  - g^{ab} H_{ba}^{\;\;\;\;c}\left[h\right] q_{c}
  - h_{ab} q^{a} a^{b}
  - h_{ab} \nabla^{a}q^{b}
  \nonumber\\
  &&
  + \pi^{ab} \stackrel{(1)}{B_{ab}} + \stackrel{(1)}{(\pi_{ab})} B^{ab}
  - 2 h_{bd} \pi_{a}^{\;\;b} B^{ad}
  \label{eq:kouchan-16.717}
  , \\
  \stackrel{(2)}{C_{0}^{(i)}}
  &=&
      q^{b} \stackrel{(2)}{(a_{b})}
  +   \stackrel{(2)}{(q_{b})} a^{b}
  +   \nabla^{b}\stackrel{(2)}{(q_{b})}
  -   q^{a} a^{b} l_{ab}
  -   \nabla^{a}q^{b} l_{ab}
  \nonumber\\
  &&
  +   \stackrel{(2)}{(\pi_{ab})} B^{ab}
  +   \pi^{cd} \stackrel{(2)}{B_{cd}}
  - 2 l_{bd} \pi^{ab} B_{a}^{\;\;d}
  -   g^{ab} H_{ab}^{\;\;\;\;c}\left[l\right] q_{c}
  \nonumber\\
  && 
  + 2 g^{ab} \stackrel{(1)}{(q_{a})} \stackrel{(1)}{(a_{b})}
  - 2 g^{ab} H_{ab}^{\;\;\;\;c}\left[h\right] \stackrel{(1)}{(q_{c})}
  + 2 g^{ab} h^{cd} H_{abd}\left[h\right] q_{c}
  - 2 h^{ab} q_{a} \stackrel{(1)}{(a_{b})}
  \nonumber\\
  && 
  - 2 h^{ab} \stackrel{(1)}{(q_{a})} a_{b}
  - 2 h^{ab} \nabla_{a}\stackrel{(1)}{(q_{b})}
  + 2 h^{ab} H_{ab}^{\;\;\;\;c}\left[h\right] q_{c}
  + 2 h^{ac} h_{c}^{\;\;b} q_{a} a_{b}
  \nonumber\\
  && 
  + 2 h^{ac} h_{c}^{\;\;b} \nabla_{a}q_{b}
  + 2 g^{ac} g^{bd} \stackrel{(1)}{(\pi_{ab})} \stackrel{(1)}{B_{cd}}
  - 4 h_{b}^{\;\;d} \pi^{cb} \stackrel{(1)}{B_{cd}}
  - 4 h_{d}^{\;\;b} \stackrel{(1)}{(\pi_{ab})} B^{ad}
  \nonumber\\
  && 
  + 4 h_{bf} h^{fd} \pi^{cb} B_{cd}
  + 2 h_{ac} h_{bd} \pi^{ab} B^{cd}
  \label{eq:kouchan-16.718}
  .
\end{eqnarray}


Through Eqs.~
(\ref{eq:linear-metric-decomp}),
(\ref{eq:kouchan-16.92}), 
(\ref{eq:kouchan-16.93}), (\ref{eq:kouchan-17.78}),
(\ref{eq:first-acceleration-gauge-inv}), and
(\ref{eq:first-order-calBab-def}), the first-order perturbation
$\stackrel{(1)}{C_{0}^{(i)}}$ of $\bar{C}_{0}^{(i)}$ is
decomposed as
\begin{eqnarray}
  \stackrel{(1)}{C_{0}^{(i)}}
  &=&
  \stackrel{(1)}{{\cal C}_{0}^{(i)}}
  + {\pounds}_{X}C_{0}^{(i)},
  \label{eq:first-order-continuity-imperfect-part-decmposition}
\end{eqnarray}
where
\begin{eqnarray}
  \stackrel{(1)}{{\cal C}_{0}^{(i)}}
  &:=&
    q^{b} \stackrel{(1)}{{\cal A}_{b}}
  + \stackrel{(1)}{{\cal Q}_{b}} a^{b}
  + \nabla^{b}\stackrel{(1)}{{\cal Q}_{b}}
  - g^{ab} H_{ab}^{\;\;\;\;c}\left[{\cal H}\right] q_{c}
  - {\cal H}_{ab} q^{a} a^{b}
  - {\cal H}_{ab} \nabla^{a}q^{b}
  \nonumber\\
  && 
  + \pi^{ab} \stackrel{(1)}{{\cal B}_{ab}}
  + \stackrel{(1)}{\Pi_{ab}} B^{ab}
  - 2 {\cal H}_{bd} \pi_{a}^{\;\;b} B^{ad}
  \label{eq:kouchan-16.719}
  .
\end{eqnarray}
Similarly, through
Eqs.~
(\ref{eq:linear-metric-decomp}),
(\ref{eq:second-metric-decomp}),
(\ref{eq:kouchan-16.92}),
(\ref{eq:kouchan-16.93}),
(\ref{eq:kouchan-17.78}), (\ref{eq:kouchan-17.79}), 
(\ref{eq:first-acceleration-gauge-inv}),
(\ref{eq:second-acceleration-gauge-inv}),
(\ref{eq:first-order-calBab-def}), and
(\ref{eq:second-order-calBab-def}), 
the second-order perturbation $\stackrel{(2)}{C_{0}^{(i)}}$ is
decomposed as
\begin{eqnarray}
  \stackrel{(2)}{C_{0}^{(i)}}
  &=&
  \stackrel{(2)}{{\cal C}_{0}^{(i)}}
  + 2 {\pounds}_{X}\stackrel{(1)}{C_{0}^{(i)}}
  +   \left(
    {\pounds}_{Y} - {\pounds}_{X}^{2}
  \right)C_{0}^{(i)}
  \label{eq:second-order-continuity-imperfect-part-decmposition}
  ,
\end{eqnarray}
where 
\begin{eqnarray}
  \stackrel{(2)}{{\cal C}_{0}^{(i)}}
  &:=&
      q^{b} \stackrel{(2)}{{\cal A}_{b}}
  +   \stackrel{(2)}{{\cal Q}_{b}} a^{b}
  +   \nabla^{b}\stackrel{(2)}{{\cal Q}_{b}}
  -   q^{a} a^{b} {\cal L}_{ab}
  -   \nabla^{a}q^{b} {\cal L}_{ab}
  \nonumber\\
  && 
  +   \stackrel{(2)}{\Pi_{ab}} B^{ab}
  +   \pi^{cd} \stackrel{(2)}{{\cal B}_{cd}}
  - 2 {\cal L}_{bd} \pi^{ab} B_{a}^{\;\;d}
  -   g^{ab} H_{ab}^{\;\;\;\;c}[{\cal L}] q_{c}
  \nonumber\\
  && 
  + 2 g^{ab} {\cal H}^{cd} H_{abd}\left[{\cal H}\right] q_{c}
  + 2 g^{ab} \stackrel{(1)}{{\cal Q}_{a}} \stackrel{(1)}{{\cal A}_{b}}
  - 2 {\cal H}^{ab} \stackrel{(1)}{{\cal Q}_{a}} a_{b}
  - 2 {\cal H}^{ab} q_{a} \stackrel{(1)}{{\cal A}_{b}}
  \nonumber\\
  && 
  - 2 {\cal H}^{ab} \nabla_{a}\stackrel{(1)}{{\cal Q}_{b}}
  - 2 g^{ab}H_{ab}^{\;\;\;\;c}\left[{\cal H}\right]\stackrel{(1)}{{\cal Q}_{c}}
  + 2 {\cal H}^{ab} H_{ab}^{\;\;\;\;c}\left[{\cal H}\right] q_{c}
  + 2 {\cal H}^{ac} {\cal H}_{cb} q_{a} a^{b}
  \nonumber\\
  && 
  + 2 {\cal H}^{ac} {\cal H}_{ab} \nabla_{c}q^{b}
  + 2 g^{ac} g^{bd} \stackrel{(1)}{\Pi_{ab}} \stackrel{(1)}{{\cal B}_{cd}}
  - 4 {\cal H}_{be} g^{de} \pi^{cb} \stackrel{(1)}{{\cal B}_{cd}}
  - 4 g^{ac} {\cal H}^{bd} \stackrel{(1)}{\Pi_{ab}} B_{cd}
  \nonumber\\
  && 
  + 4 {\cal H}_{bd} {\cal H}^{df} \pi^{cb} B_{cf}
  + 2 {\cal H}^{bd} {\cal H}^{ac} \pi_{ab} B_{cd}
  \label{eq:kouchan-16.722}
  .
\end{eqnarray}
Here again, we have confirmed that the perturbations
$\stackrel{(1)}{C_{0}^{(i)}}$ and $\stackrel{(2)}{C_{0}^{(i)}}$
defined by Eqs.~(\ref{eq:kouchan-16.717}) and
(\ref{eq:kouchan-16.718}) are decomposed into gauge-invariant
and gauge-variant parts as
Eqs.~(\ref{eq:first-order-continuity-imperfect-part-decmposition})
and
(\ref{eq:second-order-continuity-imperfect-part-decmposition}), 
and these have the same forms as
Eqs.~(\ref{eq:matter-gauge-inv-decomp-1.0}) and
(\ref{eq:matter-gauge-inv-decomp-2.0}), respectively.


Hence, through
Eqs.~(\ref{eq:continuity-perfect-expansion}),
(\ref{eq:first-order-continuity-perfect-decomp}),
(\ref{eq:second-order-continuity-perfect-decomp}),
(\ref{eq:kouchan-16.711}), each order perturbation of continuity
equation for an imperfect fluid is given by the gauge-invariant
form as 
\begin{eqnarray}
  C_{0}^{(p)} + C_{0}^{(i)}
  &=& 0
  \label{eq:kouchan-16.726}
  ,\\
  \stackrel{(1)}{C_{0}^{(p)}} + \stackrel{(1)}{C_{0}^{(i)}}
  &=&
  \stackrel{(1)}{{\cal C}_{0}^{(p)}} + \stackrel{(1)}{{\cal C}_{0}^{(i)}}
  + {\pounds}_{X}\left(C_{0}^{(p)} + C_{0}^{(i)}\right)
  \nonumber\\
  &=&
  \stackrel{(1)}{{\cal C}_{0}^{(p)}} + \stackrel{(1)}{{\cal C}_{0}^{(i)}}
  = 0
  \label{eq:kouchan-16.727}
  , \\
  \stackrel{(2)}{C_{0}^{(p)}} + \stackrel{(2)}{C_{0}^{(i)}}
  &=&
  \stackrel{(2)}{{\cal C}_{0}^{(p)}} + \stackrel{(2)}{{\cal C}_{0}^{(i)}}
  + 2 {\pounds}_{X}\left(
    \stackrel{(1)}{C_{0}^{(p)}} + \stackrel{(1)}{C_{0}^{(i)}}
  \right)
  + \left(
    {\pounds}_{Y}
    - {\pounds}_{X}^{2}
  \right)
  \left(
    C_{0}^{(p)} + C_{0}^{(i)}
  \right)
  \nonumber\\
  &=&
  \stackrel{(2)}{{\cal C}_{0}^{(p)}} + \stackrel{(2)}{{\cal C}_{0}^{(i)}}
  =
  0
  ,
  \label{eq:kouchan-16.728}
\end{eqnarray}
where we used the background continuity equation
(\ref{eq:kouchan-16.726}) in the derivation of
Eq.~(\ref{eq:kouchan-16.727}), and used the background continuity
equation (\ref{eq:kouchan-16.726}) and its first-order
perturbation (\ref{eq:kouchan-16.727}) in the derivation of
Eq.~(\ref{eq:kouchan-16.728}).


\subsubsection{Perturbations of the generalized Navier-Stokes equation}
\label{sec:Imerfect-fluid-Navier-Stokes-equation-generic}


The normal component of $\bar{\nabla}_{b}\bar{T}_{a}^{\;\;b}$ to
$\bar{u}_{a}$ for an imperfect fluid
(\ref{eq:imperfect-fluid-ene-mom-full-1}) is given by
\begin{eqnarray}
  \left(\bar{\epsilon}+\bar{p}\right) \bar{a}_{b}
  + \bar{q}_{c}^{\;\;a} \left(
      \bar{\nabla}_{a}\bar{p}
    + \bar{u}^{b}\bar{\nabla}_{b}\bar{q}_{a}
    + \bar{\nabla}_{b}\bar{\pi}_{a}^{\;\;b}
  \right)
  + \bar{q}^{b} \left(
      \bar{q}_{cb} \bar{\theta}
    + \bar{B}_{cb}
  \right)
  = 
  0.
  \label{eq:kouchan-16.71}
\end{eqnarray}
This equation (\ref{eq:kouchan-16.71}) is an extension of the
Euler equation (\ref{eq:Euler-equation-of-perfect-fluid}) for a
perfect fluid to the imperfect fluid case. 
In this paper, we call Eq.~(\ref{eq:kouchan-16.71}) as 
``{\it the generalized Navier-Stokes
  equation}''\cite{Landau-Lifshitz-Fluid}.


As seen in the case of the Euler equation
(\ref{eq:Euler-equation-of-perfect-fluid}) for a perfect fluid,
the first two terms of the generalized Navier-Stokes equation
(\ref{eq:kouchan-16.71}) is given by 
$\bar{C}_{c}^{(p)}$ which is defined in 
Eq.~(\ref{eq:Euler-equation-of-perfect-fluid}).
We call this $\bar{C}_{c}^{(p)}$ as ``{\it the perfect part}''
of the generalized Navier-Stokes equation (\ref{eq:kouchan-16.71}).
Using this perfect part, the equation
(\ref{eq:kouchan-16.71}) is given by
\begin{eqnarray}
  &&
  \bar{C}_{c}^{(p)} + \bar{C}_{c}^{(i)} = 0
  ,
  \label{eq:kouchan-17.761}
\end{eqnarray}
where $\bar{C}_{c}^{(i)}$ is defined by 
\begin{eqnarray}
  \bar{C}_{c}^{(i)} 
  &:=&
    \bar{g}^{ad} \bar{q}_{cd} \left(
      \bar{u}^{b} \bar{\nabla}_{b}\bar{q}_{a}
    + \bar{g}^{be} \bar{\nabla}_{b}\bar{\pi}_{ae}
  \right)
  + \bar{g}^{be} \bar{q}_{e} \left(
      \bar{q}_{bc} \bar{\theta}
    + \bar{B}_{bc}
  \right)
  \label{eq:kouchan-17.764}
  .
\end{eqnarray}
We call this $\bar{C}_{c}^{(i)}$ as ``{\it the imperfect part}''
of the generalized Navier-Stokes equation for an imperfect fluid.
Since the perturbative expansion of the perfect part
$\bar{C}_{c}^{(p)}$ is already given in
\S\ref{sec:Perfect-fluid-Euler-equation-generic}, we may
concentrate on the perturbative expansion of the imperfect part
(\ref{eq:kouchan-17.764}).


As in the cases of the imperfect part of the continuity equation
for an imperfect fluid, the imperfect part of the Navier-Stokes
equation is perturbatively expanded as
\begin{eqnarray}
  \bar{C}_{b}^{(i)}
  &=:&
            C_{b}^{(i)}
  + \lambda \stackrel{(1)}{C_{b}^{(i)}}
  + \frac{1}{2} \lambda^{2} \stackrel{(2)}{C_{b}^{(i)}}
  + O(\lambda^{3}).
\end{eqnarray}
Here, $C_{b}^{(i)}$, $\stackrel{(1)}{C_{b}^{(i)}}$, and
$\stackrel{(2)}{C_{b}^{(i)}}$ are given by the perturbative
expansion of Eq.~(\ref{eq:kouchan-17.764}) through
Eqs.~(\ref{eq:kouchan-16.72}), (\ref{eq:kouchan-16.713}),
(\ref{eq:three-metric-expansion}),
(\ref{eq:contravariant-four-velocity-expansion}),
(\ref{eq:inverse-metric-expansion}),
(\ref{eq:kouchan-17.701}), (\ref{eq:theta-expansion}), and the
perturbation of the tensor $\bar{\nabla}_{b}\bar{\pi}_{ae}$.
To derive the perturbative expression of
$\bar{\nabla}_{b}\bar{\pi}_{ae}$, the perturbative expansion
(\ref{eq:kouchan-16.74}) of the anisotropic stress and the
perturbation (\ref{eq:perturbative-connection-expansion}) of the
connection $C_{ba}^{c}$ are used.
Further, through Eqs.~(\ref{eq:linear-metric-decomp}),
(\ref{eq:second-metric-decomp}),
(\ref{eq:kouchan-16.92}), (\ref{eq:kouchan-16.93}),
(\ref{eq:first-three-metric-gauge-inv}),
(\ref{eq:second-three-metric-gauge-inv}),
(\ref{eq:kouchan-17.78}), (\ref{eq:kouchan-17.79}),
(\ref{eq:first-order-calBab-def}),
(\ref{eq:second-order-calBab-def}),
(\ref{eq:first-theta-gauge-inv}),
(\ref{eq:second-theta-gauge-inv}), we can show that the first-
and the second-order perturbations of the imperfect part
$\bar{C}_{b}^{(i)}$ are also decomposed into the 
gauge-invariant and the gauge-variant parts as
\begin{eqnarray}
  \stackrel{(1)}{C_{b}^{(i)}}
  &=&
      \stackrel{(1)}{{\cal C}_{b}^{(i)}}
  +   {\pounds}_{X}C_{b}^{(i)}
  \label{eq:kouchan-17.907}
  , \\
  \stackrel{(2)}{C_{b}^{(i)}}
  &=&
  \stackrel{(2)}{{\cal C}_{b}^{(i)}}
  +   {\pounds}_{Y}C_{b}^{(i)}
  -   {\pounds}_{X}^{2}C_{b}^{(i)}
  + 2 {\pounds}_{X}\stackrel{(1)}{C_{b}^{(i)}}
  \label{eq:kouchan-17.909}
  .
\end{eqnarray}
These have the same form as
Eqs.~(\ref{eq:matter-gauge-inv-decomp-1.0}) and
(\ref{eq:matter-gauge-inv-decomp-2.0}), respectively.


The derivations of the gauge-invariant part of
$\stackrel{(1)}{{\cal C}_{b}^{(i)}}$ and 
$\stackrel{(2)}{{\cal C}_{b}^{(i)}}$ are straightforward.
Therefore, we just summarize the results, here.
The details of these derivations can be seen in the Appendix in
Ref.~19).
Through
Eqs.~(\ref{eq:background-Euler-perfect})--(\ref{eq:first-order-Euler-perfect-gauge-inv}),
(\ref{eq:second-order-Euler-perfect-decomp}), and
(\ref{eq:second-order-Euler-perfect-gauge-inv}), each order
perturbation of the generalized Navier-Stokes equation for an
imperfect fluid is summarized as follows. 
The background generalized Navier-Stokes equation is given by 
\begin{eqnarray}
  C_{b}^{(p)} + C_{b}^{(i)} 
  &=&
  \left( \epsilon + p \right) a_{b} + g^{ac} q_{bc} \nabla_{a}p
  \nonumber\\
  &&
  + q_{bd} \left(
      u^{c} \nabla_{c}q^{d}
    + \nabla_{c}\pi^{dc}
  \right)
  + q^{c} \left(
      q_{cb} \theta
    + B_{cb}
  \right)
  \\
  &=& 0
  \label{eq:kouchan-16.911}
  .
\end{eqnarray}
The first-order perturbation of the generalized Navier-Stokes
equation is given by 
\begin{eqnarray}
  \stackrel{(1)}{C_{b}^{(p)}} + \stackrel{(1)}{C_{b}^{(i)}}
  &=&
  \stackrel{(1)}{{\cal C}_{b}^{(p)}} + \stackrel{(1)}{{\cal C}_{b}^{(i)}} 
  +   {\pounds}_{X}\left(C_{b}^{(p)} + C_{b}^{(i)}\right),
  \label{eq:kouchan-17.912}
\end{eqnarray}
where
\begin{eqnarray}
  \stackrel{(1)}{{\cal C}_{b}^{(p)}} + \stackrel{(1)}{{\cal C}_{b}^{(i)}} 
  &=&
  \left( \epsilon + p \right) \stackrel{(1)}{{\cal A}_{b}}
  +
  \left(
      \stackrel{(1)}{{\cal E}} + \stackrel{(1)}{{\cal P}}
  \right)
  a_{b}
  + g^{ac} q_{bc} \nabla_{a}\stackrel{(1)}{{\cal P}}
  \nonumber\\
  && 
  + g^{ac} \stackrel{(1)}{{\cal Q}_{bc}} \nabla_{a}p
  - g^{ad} g^{ce} {\cal H}_{de} q_{bc} \nabla_{a}p
  +   q_{bd} u^{c} \nabla_{c}\stackrel{(1)}{{\cal Q}^{d}}
  -   q_{bd} u_{c} H^{dcf}[{\cal H}] q_{f}
  \nonumber\\
  && 
  +   q_{bd} \stackrel{(1)}{{\cal U}^{c}} \nabla_{c}q^{d}
  -   q_{bd} {\cal H}^{ce} u_{e} \nabla_{c}q^{d}
  +   \stackrel{(1)}{{\cal Q}_{bd}} u^{c} \nabla_{c}q^{d}
  -   {\cal H}^{ad} q_{bd} u^{c} \nabla_{c}q_{a}
  \nonumber\\
  && 
  -   {\cal H}^{ad} q_{bd} \nabla^{c}\pi_{ac}
  +   \stackrel{(1)}{{\cal Q}_{bd}} \nabla_{c}\pi^{dc}
  -   q_{bd} {\cal H}_{ce} \nabla^{c}\pi^{de}
  \nonumber\\
  && 
  +   q_{bd} \nabla_{c}\stackrel{(1)}{\Pi^{dc}}
  -   q_{bd} H^{dcf}\left[{\cal H}\right] \pi_{fc}
  -   g^{ad} q_{bd} H_{c}^{\;\;cf}[{\cal H}] \pi_{af}
  \nonumber\\
  && 
  -   {\cal H}^{ce} q_{e} q_{bc} \theta
  +   \stackrel{(1)}{{\cal Q}^{c}} q_{bc} \theta
  +   q^{c} \stackrel{(1)}{{\cal Q}_{bc}} \theta
  +   q^{c} q_{bc} \stackrel{(1)}{\Theta}
  \nonumber\\
  && 
  +   \stackrel{(1)}{{\cal Q}^{c}} B_{cb}
  -   {\cal H}^{ce} q_{e} B_{cb}
  +   q^{c} \stackrel{(1)}{{\cal B}_{cb}}
  \label{eq:kouchan-17.914}
  .
\end{eqnarray}
Then, the first-order perturbation of the generalized
Navier-Stokes equation is given in the gauge-invariant form as 
\begin{eqnarray}
  \stackrel{(1)}{{\cal C}_{b}^{(p)}}
  + \stackrel{(1)}{{\cal C}_{b}^{(i)}} = 0,
  \label{eq:first-order-Navier-Stokes-gauge-inv}
\end{eqnarray}
where we used the background equation (\ref{eq:kouchan-16.911})
of the generalized Navier-Stokes equation.
The second-order perturbation of the generalized Navier-Stokes
equation is given by
\begin{eqnarray}
  \stackrel{(2)}{C_{b}^{(p)}} + \stackrel{(2)}{C_{b}^{(i)}}
  &=&
  \stackrel{(2)}{{\cal C}_{b}^{(p)}} + \stackrel{(2)}{{\cal C}_{b}^{(i)}}
  \nonumber\\
  && 
  + 2 {\pounds}_{X}\left(
    \stackrel{(1)}{C_{b}^{(p)}} + \stackrel{(1)}{C_{b}^{(i)}}
  \right)
  +   \left\{ {\pounds}_{Y}
  -   {\pounds}_{X}^{2} \right\}\left( C_{b}^{(p)} + C_{b}^{(i)} \right)
  \label{eq:kouchan-17.915}
  ,
\end{eqnarray}
where
\begin{eqnarray}
  \stackrel{(2)}{{\cal C}_{b}^{(p)}} + \stackrel{(2)}{{\cal C}_{b}^{(i)}}
  &=&
    \left( \epsilon + p \right) \stackrel{(2)}{{\cal A}_{b}}
  + \left(\stackrel{(2)}{{\cal E}} + \stackrel{(2)}{{\cal P}}\right) a_{b}
  + g^{ac} q_{bc} \nabla_{a}\stackrel{(2)}{{\cal P}}
  + g^{ac} \stackrel{(2)}{{\cal Q}_{bc}} \nabla_{a}p
  \nonumber\\
  && 
  - g^{ad} g^{ce} {\cal L}_{de} q_{bc} \nabla_{a}p
  + 2 \left(
      \stackrel{(1)}{{\cal E}}
    + \stackrel{(1)}{{\cal P}}
  \right) \stackrel{(1)}{{\cal A}_{b}}
  + 2 g^{ac} \stackrel{(1)}{{\cal Q}_{bc}} \nabla_{a}\stackrel{(1)}{{\cal P}}
  \nonumber\\
  && 
  - 2 {\cal H}^{ac} q_{bc} \nabla_{a}\stackrel{(1)}{{\cal P}}
  - 2 {\cal H}^{ac} \stackrel{(1)}{{\cal Q}_{bc}} \nabla_{a}p
  + 2 {\cal H}^{ad} g^{ce} {\cal H}_{de} q_{bc} \nabla_{a}p
  \nonumber\\
  && 
  +   q_{bd} u^{c} \nabla_{c}\stackrel{(2)}{{\cal Q}^{d}}
  -   q_{bd} u_{c} H^{dcf}[{\cal L}] q_{f}
  + 2 q_{bd} u_{c} {\cal H}_{fg} H^{dcg}[{\cal H}] q^{f}
  \nonumber\\
  && 
  +   q_{bd} \stackrel{(2)}{{\cal U}^{c}} \nabla_{c}q^{d}
  -   q_{bd} {\cal L}^{ce} u_{e} \nabla_{c}q^{d}
  +   \stackrel{(2)}{{\cal Q}_{bd}} u^{c} \nabla_{c}q^{d}
  -   {\cal L}^{ad} q_{bd} u^{c} \nabla_{c}q_{a}
  \nonumber\\
  && 
  - 2 q_{bd} u_{c} H^{dcf}[{\cal H}] \stackrel{(1)}{{\cal Q}_{f}}
  + 2 q_{bd} \stackrel{(1)}{{\cal U}^{c}} \nabla_{c}\stackrel{(1)}{{\cal Q}^{d}}
  - 2 q_{bd} \stackrel{(1)}{{\cal U}_{c}} H^{dcf}[{\cal H}] q_{f}
  \nonumber\\
  && 
  - 2 q_{bd} {\cal H}^{ce} u_{e} \nabla_{c}\stackrel{(1)}{{\cal Q}^{d}}
  + 2 q_{bd} {\cal H}_{ce} u^{e} H^{dcf}[{\cal H}] q_{f}
  - 2 q_{bd} {\cal H}^{ce} \stackrel{(1)}{{\cal U}_{e}} \nabla_{c}q^{d}
  \nonumber\\
  && 
  + 2 q_{bd} {\cal H}_{fg} {\cal H}^{cf} u^{g} \nabla_{c}q^{d}
  + 2 \stackrel{(1)}{{\cal Q}_{bd}} u^{c} \nabla_{c}\stackrel{(1)}{{\cal Q}^{d}}
  - 2 \stackrel{(1)}{{\cal Q}_{bd}} u_{c} H^{dcf}[{\cal H}] q_{f}
  \nonumber\\
  && 
  + 2 \stackrel{(1)}{{\cal Q}_{bd}} \stackrel{(1)}{{\cal U}^{c}} \nabla_{c}q^{d}
  - 2 \stackrel{(1)}{{\cal Q}_{bd}} {\cal H}^{ce} u_{e} \nabla_{c}q^{d}
  - 2 {\cal H}^{ad} q_{bd} u^{c} \nabla_{c}\stackrel{(1)}{{\cal Q}_{a}}
  \nonumber\\
  && 
  + 2 {\cal H}^{ad} q_{bd} u^{c} H_{acf}[{\cal H}] q^{f}
  - 2 {\cal H}^{ad} q_{bd} \stackrel{(1)}{{\cal U}^{c}} \nabla_{c}q_{a}
  + 2 {\cal H}^{ad} q_{bd} {\cal H}^{ce} u_{e} \nabla_{c}q_{a}
  \nonumber\\
  && 
  - 2 {\cal H}^{ad} \stackrel{(1)}{{\cal Q}_{bd}} u^{c} \nabla_{c}q_{a}
  + 2 {\cal H}_{af} {\cal H}^{fd} q_{bd} u^{c} \nabla_{c}q^{a}
  -   {\cal L}^{ad} q_{bd} \nabla^{c}\pi_{ac}
  \nonumber\\
  && 
  +   \stackrel{(2)}{{\cal Q}_{bd}} \nabla_{c}\pi^{dc}
  -   q_{bd} {\cal L}_{ce} \nabla^{c}\pi^{de}
  +   q_{bd} \nabla_{c}\stackrel{(2)}{\Pi^{dc}}
  \nonumber\\
  && 
  -   q_{bd} H^{dcf}[{\cal L}] \pi_{fc}
  + 2 g^{ad} q_{bd} {\cal H}_{fg} H_{ac}^{\;\;\;\;\;\;g}[{\cal H}] \pi^{fc}
  -   g^{ad} q_{bd} H_{c}^{\;\;cf}[{\cal L}] \pi_{af}
  \nonumber\\
  && 
  + 2 q_{bd} {\cal H}_{fg} H_{c}^{\;\;cg}[{\cal H}] \pi^{df}
  - 2 {\cal H}^{ad} \stackrel{(1)}{{\cal Q}_{bd}} \nabla^{c}\pi_{ac}
  + 2 {\cal H}^{ad} q_{bd} {\cal H}^{ce} \nabla_{c}\pi_{ae}
  \nonumber\\
  && 
  - 2 {\cal H}^{ad} q_{bd} \nabla^{c}\stackrel{(1)}{(\Pi_{ac})}
  + 2 {\cal H}^{ad} q_{bd} H_{a}^{\;\;cf}[{\cal H}] \pi_{fc}
  + 2 {\cal H}^{ad} q_{bd} H_{c}^{\;\;cf}[{\cal H}] \pi_{af}
  \nonumber\\
  && 
  - 2 \stackrel{(1)}{{\cal Q}_{bd}} {\cal H}_{ce} \nabla^{c}\pi^{de}
  + 2 \stackrel{(1)}{{\cal Q}_{bd}} \nabla_{c}\stackrel{(1)}{\Pi^{dc}}
  - 2 \stackrel{(1)}{{\cal Q}_{bd}} H^{dcf}[{\cal H}] \pi_{fc}
  \nonumber\\
  && 
  - 2 g^{ad} \stackrel{(1)}{{\cal Q}_{bd}} H_{c}^{\;\;cf}[{\cal H}] \pi_{af}
  - 2 q_{bd} {\cal H}_{ce} \nabla^{c}\stackrel{(1)}{\Pi^{de}}
  + 2 q_{bd} {\cal H}_{ap} {\cal H}^{pd} \nabla_{c}\pi^{ac}
  \nonumber\\
  && 
  + 2 q_{bd} {\cal H}^{cq} {\cal H}_{qp} \nabla_{c}\pi^{dp}
  + 2 g^{ad} q_{bd} {\cal H}^{ce} H_{ac}^{\;\;\;\;f}[{\cal H}] \pi_{fe}
  + 2 q_{bd} {\cal H}^{ce} H_{ecf}[{\cal H}] \pi^{df}
  \nonumber\\
  && 
  - 2 q_{bd} H^{dcf}[{\cal H}] \stackrel{(1)}{\Pi_{fc}}
  - 2 g^{ad} q_{bd} H_{c}^{\;\;cf}[{\cal H}] \stackrel{(1)}{\Pi_{af}}
  -   {\cal L}^{ce} q_{e} q_{bc} \theta
  \nonumber\\
  && 
  +   \stackrel{(2)}{{\cal Q}^{c}} q_{bc} \theta
  +   q^{c} \stackrel{(2)}{{\cal Q}_{bc}} \theta
  +   q^{c} q_{bc} \stackrel{(2)}{\Theta}
  \nonumber\\
  && 
  - 2 {\cal H}^{ce} \stackrel{(1)}{{\cal Q}_{e}} q_{bc} \theta
  + 2 {\cal H}^{cf} {\cal H}_{fe} q^{e} q_{bc} \theta
  - 2 {\cal H}^{ce} q_{e} \stackrel{(1)}{{\cal Q}_{bc}} \theta
  \nonumber\\
  && 
  - 2 {\cal H}^{ce} q_{e} q_{bc} \stackrel{(1)}{\Theta}
  + 2 \stackrel{(1)}{{\cal Q}^{c}} \stackrel{(1)}{{\cal Q}_{bc}} \theta
  + 2 \stackrel{(1)}{{\cal Q}^{c}} q_{bc} \stackrel{(1)}{\Theta}
  \nonumber\\
  && 
  + 2 q^{c} \stackrel{(1)}{{\cal Q}_{bc}} \stackrel{(1)}{\Theta}
  +   \stackrel{(2)}{{\cal Q}^{c}} B_{cb}
  -   {\cal L}^{ce} q_{e} B_{cb}
  \nonumber\\
  && 
  +   q^{c} \stackrel{(2)\;\;\;\;}{{\cal B}_{cb}}
  - 2 {\cal H}^{ce} \stackrel{(1)}{{\cal Q}_{e}} B_{cb}
  - 2 {\cal H}^{ce} q_{e} \stackrel{(1)}{{\cal B}_{cb}}
  \nonumber\\
  && 
  + 2 \stackrel{(1)}{{\cal Q}^{c}} \stackrel{(1)}{{\cal B}_{cb}}
  + 2 {\cal H}^{cd} {\cal H}_{de} q^{e} B_{cb}
  \label{eq:kouchan-17.916}
  .
\end{eqnarray}
Through Eqs.~(\ref{eq:kouchan-16.911}) and
(\ref{eq:first-order-Navier-Stokes-gauge-inv}), the second-order
perturbation of the generalized Navier-Stokes equation is given
in the gauge-invariant form as
\begin{eqnarray}
  \stackrel{(2)}{{\cal C}_{b}^{(p)}}
  + \stackrel{(2)}{{\cal C}_{b}^{(i)}} = 0.
  \label{eq:second-order-Navier-Stokes-gauge-inv}
\end{eqnarray}


\subsection{Scalar fluid}
\label{sec:scalar-field-generic}


Here, we consider the perturbations of the energy momentum
tensor (\ref{eq:MFB-6.2-again}) and the Klein-Gordon equation of
the single scalar field $\bar{\varphi}$.
Since the perturbative expression of the energy momentum tensor
of the single scalar field is already discussed in KN2007, we
just summarize the formulae for the energy momentum tensor in
\S\ref{sec:scalar-field-generic-ene-mom}.
After that, we consider the perturbations of the Klein-Gordon
equation in \S\ref{sec:scalar-field-generic-Klein-Gordon}.


\subsubsection{Energy momentum tensor}
\label{sec:scalar-field-generic-ene-mom}


Through the perturbative expansions
(\ref{eq:scalar-field-expansion-second-order}) and
(\ref{eq:inverse-metric-expansion}) of the scalar 
field $\bar{\varphi}$ and the metric, the energy momentum tensor 
(\ref{eq:MFB-6.2-again}) is also expanded as 
\begin{eqnarray}
  \bar{T}_{a}^{\;\;b} = T_{a}^{\;\;b} 
  + \lambda {}^{(1)}\!\left(T_{a}^{\;\;b}\right)
  + \frac{1}{2} \lambda^{2} {}^{(2)}\!\left(T_{a}^{\;\;b}\right)
  + O(\lambda^{3}).
\end{eqnarray}
The background energy momentum tensor $T_{a}^{\;\;b}$ is given
by
\begin{eqnarray}
  \label{eq:background-energy-momentum-scalar}
  T_{a}^{\;\;b}
  &:=& 
  \nabla_{a}\varphi g^{bc} \nabla_{c}\varphi 
  - \frac{1}{2} \delta_{a}^{\;\;b}
  \left(
    \nabla_{c}\varphi\nabla^{c}\varphi
    + 2 V\left( \varphi \right)
  \right)
  .
\end{eqnarray}
Further, through the decompositions
(\ref{eq:linear-metric-decomp}),
(\ref{eq:second-metric-decomp}), (\ref{eq:varphi-1-def}), and
(\ref{eq:varphi-2-def}) of the perturbations of the metric and
the scalar field, the perturbations of the energy momentum
tensor ${}^{(1)}\!\left(T_{a}^{\;\;b}\right)$ and
${}^{(2)}\!\left(T_{a}^{\;\;b}\right)$ are also decomposed into
the gauge-invariant and the gauge-variant parts as 
\begin{eqnarray}
  {}^{(1)}\!\left(T_{a}^{\;\;b}\right)
  &=:&
  {}^{(1)}\!{\cal T}_{a}^{\;\;b} + {\pounds}_{X}T_{a}^{\;\;b}
  \label{eq:first-order-energy-momentum-scalar-decomp}
  , \\
  {}^{(2)}\!\left(T_{a}^{\;\;b}\right)
  &=:&
  {}^{(2)}\!{\cal T}_{a}^{\;\;b}
  + 2 {\pounds}_{X}{}^{(1)}\!\left(T_{a}^{\;\;b}\right)
  + \left( {\pounds}_{Y} - {\pounds}_{X}^{2}\right) T_{a}^{\;\;b},
  \label{eq:second-order-energy-momentum-scalar-decomp}
\end{eqnarray}
where the gauge-invariant parts ${}^{(1)}\!{\cal T}_{a}^{\;\;b}$
and ${}^{(2)}\!{\cal T}_{a}^{\;\;b}$ of the first and the
second order are given by 
\begin{eqnarray}
  {}^{(1)}\!{\cal T}_{a}^{\;\;b}
  &:=& 
  \nabla_{a}\varphi \nabla^{b}\varphi_{1} 
  - \nabla_{a}\varphi {\cal H}^{bc} \nabla_{c}\varphi 
  + \nabla_{a}\varphi_{1} \nabla^{b} \varphi 
  \nonumber\\
  && \quad
  - \frac{1}{2} \delta_{a}^{\;\;b}
  \left(
    \nabla_{c}\varphi\nabla^{c}\varphi_{1}
    - \nabla_{c}\varphi {\cal H}^{dc} \nabla_{d} \varphi 
    + \nabla_{c}\varphi_{1}\nabla^{c} \varphi
    + 2 \varphi_{1} \frac{\partial V}{\partial\varphi}
  \right)
  \label{eq:first-order-energy-momentum-scalar-gauge-inv}
  , \\
  {}^{(2)}\!{\cal T}_{a}^{\;\;b}
  &:=&
  \nabla_{a}\varphi \nabla^{b}\varphi_{2} 
  - 2 \nabla_{a}\varphi {\cal H}^{bc} \nabla_{c}\varphi_{1} 
  + 2 \nabla_{a}\varphi {\cal H}^{bd}{\cal H}_{dc} \nabla^{c}\varphi
  - \nabla_{a}\varphi g^{bd} {\cal L}_{dc} \nabla^{c}\varphi
  \nonumber\\
  && \quad
  + 2 \nabla_{a}\varphi_{1} \nabla^{b}\varphi_{1}
  - 2 \nabla_{a}\varphi_{1} {\cal H}^{bc} \nabla_{c}\varphi 
  + \nabla_{a}\varphi_{2} \nabla^{b}\varphi  
  \nonumber\\
  && \quad
  - \frac{1}{2} \delta_{a}^{\;\;b}
  \left(
    \nabla_{c}\varphi\nabla^{c}\varphi_{2}
    - 2 \nabla_{c}\varphi {\cal H}^{dc} \nabla_{d}\varphi_{1} 
    + 2 \nabla^{c}\varphi {\cal H}^{de}{\cal H}_{ec} \nabla_{d}\varphi 
  \right.
  \nonumber\\
  && \quad\quad\quad\quad\quad
  \left.
    - \nabla^{c}\varphi {\cal L}_{dc}\nabla^{d}\varphi 
    + 2 \nabla_{c}\varphi_{1}\nabla^{c}\varphi_{1} 
    - 2 \nabla_{c}\varphi_{1} {\cal H}^{dc} \nabla_{d} \varphi 
  \right.
  \nonumber\\
  && \quad\quad\quad\quad\quad
  \left.
    + \nabla_{c}\varphi_{2}\nabla^{c} \varphi 
    + 2 \varphi_{2}\frac{\partial V}{\partial\varphi}
    + 2 \varphi_{1}^{2}\frac{\partial^{2}V}{\partial\varphi^{2}}
  \right)
  \label{eq:second-order-energy-momentum-scalar-gauge-inv}
  .
\end{eqnarray}
We also note that in these derivations, we did not used the
homogeneous condition for the background scalar field
$\varphi$ and the decompositions
(\ref{eq:first-order-energy-momentum-scalar-decomp}) and
(\ref{eq:second-order-energy-momentum-scalar-decomp}) coincide
with the formulae (\ref{eq:matter-gauge-inv-decomp-1.0}) and
(\ref{eq:matter-gauge-inv-decomp-2.0}), respectively.


\subsubsection{Klein-Gordon equation}
\label{sec:scalar-field-generic-Klein-Gordon}


Here, we consider the perturbation of the Klein-Gordon equation 
\begin{eqnarray}
  \bar{C}_{(K)} := \bar{\nabla}^{a}\bar{\nabla}_{a}\bar{\varphi}
  - \frac{\partial V}{\partial\bar{\varphi}}(\bar{\varphi}) = 0.
  \label{eq:Klein-Gordon-equation}
\end{eqnarray}
Through the perturbative expansions
(\ref{eq:scalar-field-expansion-second-order}) and
(\ref{eq:metric-expansion}), the Klein-Gordon equation
(\ref{eq:Klein-Gordon-equation}) is expanded as
\begin{eqnarray}
  \bar{C}_{(K)}
  =:
  C_{(K)}
  + \lambda \stackrel{(1)}{C_{(K)}}
  + \frac{1}{2} \lambda^{2} \stackrel{(2)}{C_{(K)}}
  + O(\lambda^{3}),
\end{eqnarray}
where
\begin{eqnarray}
  \label{eq:Klein-Gordon-eq-background}
  C_{(K)}
  &:=&
  \nabla_{a}\nabla^{a}\varphi
  - \frac{\partial V}{\partial\bar{\varphi}}(\varphi)
  = 0
  , \\
  \label{eq:Klein-Gordon-eq-first-order}
  \stackrel{(1)}{C_{(K)}}
  &:=&
  \nabla^{a}\nabla_{a}\hat{\varphi}_{1}
  - g^{ab} H_{ab}^{\;\;\;\;c}\left[h\right] \nabla_{c}\varphi
  - h^{ab} \nabla_{a}\nabla_{b}\varphi
  - \hat{\varphi}_{1} \frac{\partial^{2}V}{\partial\bar{\varphi}^{2}}(\varphi)
  = 0
  , \\
  \label{eq:Klein-Gordon-eq-second-order}
  \stackrel{(2)}{C_{(K)}}
  &:=&
  \nabla^{a}\nabla_{a}\hat{\varphi}_{2}
  - 2 h^{ab} \nabla_{a}\nabla_{b}\hat{\varphi}_{1}
  - 2 g^{ab} H_{ab}^{\;\;\;\;c}\left[h\right] \nabla_{c}\hat{\varphi}_{1}
  + 2 h^{ab} H_{ab}^{\;\;\;\;c}\left[h\right] \nabla_{c}\varphi
  \nonumber\\
  && 
  -   g^{ab} \left(
    H_{ab}^{\;\;\;\;c}\left[l\right]
    - 2 h^{cd} H_{abd}\left[h\right]
  \right) \nabla_{c}\varphi
  + 2 h^{ae}h_{e}^{\;\;b} \nabla_{a}\nabla_{b}\varphi
  -   l^{ab} \nabla_{a}\nabla_{b}\varphi
  \nonumber\\
  && 
  -
  \hat{\varphi}_{2}
  \frac{\partial^{2}V}{\partial\bar{\varphi}^{2}}(\varphi)
  -
  \left(
    \hat{\varphi}_{1}
  \right)^{2}
  \frac{\partial^{3}V}{\partial\bar{\varphi}^{3}}(\varphi)
  = 0
  .
\end{eqnarray}
Here, to derive the perturbative expansion of the kinetic terms
of the Klein-Gordon equation (\ref{eq:Klein-Gordon-equation}),
we use the connection
(\ref{eq:perturbative-connection-expansion}) between the
covariant derivatives $\bar{\nabla}_{a}$ and $\nabla_{a}$ as in
the case of the perturbative expansion of the tensor
$\bar{A}_{ab}$ in Appendix \ref{sec:Perturbations-Aab-appendix}.


The first- and the second-order perturbations
$\stackrel{(1)}{C_{(K)}}$ and $\stackrel{(2)}{C_{(K)}}$ of the
Klein-Gordon equation are also decomposed into the
gauge-invariant and the gauge-variant parts. 
Through Eqs.~(\ref{eq:linear-metric-decomp}),
(\ref{eq:second-metric-decomp}), (\ref{eq:varphi-1-def}),
(\ref{eq:varphi-2-def}), (\ref{eq:kouchan-17.78}), and 
(\ref{eq:kouchan-17.79}), the first- and the second-order
perturbations (\ref{eq:Klein-Gordon-eq-first-order}) and
(\ref{eq:Klein-Gordon-eq-second-order}) of the Klein-Gordon 
equation are decomposed as
\begin{eqnarray}
  \label{eq:Klein-Gordon-eq-first-decomp}
  \stackrel{(1)}{C_{(K)}}
  &=:&
  \stackrel{(1)}{{\cal C}_{(K)}}
  + {\pounds}_{X}C_{(K)},
  \\
  \label{eq:Klein-Gordon-eq-second-decomp}
  \stackrel{(2)}{C_{(K)}}
  &=:&
  \stackrel{(2)}{{\cal C}_{(K)}}
  + 2 {\pounds}_{X}\stackrel{(1)}{C_{(K)}}
  + \left( {\pounds}_{Y} - {\pounds}_{X}^{2} \right) C_{(K)}
  ,
\end{eqnarray}
where
\begin{eqnarray}
  \label{eq:Klein-Gordon-eq-first-gauge-inv-def}
  \stackrel{(1)}{{\cal C}_{(K)}}
  &:=&
  \nabla^{a}\nabla_{a}\varphi_{1}
  - H_{a}^{\;\;ac}[{\cal H}]\nabla_{c}\varphi
  - {\cal H}^{ab} \nabla_{a}\nabla_{b}\varphi
  - \varphi_{1} \frac{\partial^{2}V}{\partial\bar{\varphi}^{2}}(\varphi)
  , \\
  \label{eq:Klein-Gordon-eq-second-gauge-inv-def}
  \stackrel{(2)}{{\cal C}_{(K)}}
  &:=&
      \nabla^{a}\nabla_{a}\varphi_{2} 
  -   H_{a}^{\;\;ac}[{\cal L}] \nabla_{c}\varphi
  + 2 H_{a}^{\;\;ad}[{\cal H}] {\cal H}_{cd} \nabla^{c}\varphi
  - 2 H_{a}^{\;\;ac}[{\cal H}] \nabla_{c}\varphi_{1}
  \nonumber\\
  && \quad
  + 2 {\cal H}^{ab} H_{ab}^{\;\;\;\;c}[{\cal H}] \nabla_{c}\varphi
  -   {\cal L}^{ab} \nabla_{a}\nabla_{b}\varphi
  + 2 {\cal H}^{a}_{\;\;d} {\cal H}^{db} \nabla_{a}\nabla_{b}\varphi
  - 2 {\cal H}^{ab} \nabla_{a}\nabla_{b}\varphi_{1}
  \nonumber\\
  && \quad
  -   \varphi_{2} \frac{\partial^{2}V}{\partial\bar{\varphi}^{2}}(\varphi)
  -   (\varphi_{1})^{2}\frac{\partial^{3}V}{\partial\bar{\varphi}^{3}}(\varphi)
  .
\end{eqnarray}
Here, we note that Eqs.~(\ref{eq:Klein-Gordon-eq-first-decomp})
and (\ref{eq:Klein-Gordon-eq-second-decomp}) have the same form
as Eqs.~(\ref{eq:matter-gauge-inv-decomp-1.0}) and
(\ref{eq:matter-gauge-inv-decomp-2.0}), respectively.


By virtue of the background Klein-Gordon equation
(\ref{eq:Klein-Gordon-eq-background}) and the first-order
perturbation (\ref{eq:Klein-Gordon-eq-first-order}) of the
Klein-Gordon equation, the first- and the second-order
perturbation of the Klein-Gordon equation are given in terms of
gauge-invariant variables: 
\begin{eqnarray}
  \label{eq:Klein-Gordon-eq-first-second-gauge-inv}
   \stackrel{(1)}{{\cal C}_{(K)}} = 0,
  \quad
   \stackrel{(2)}{{\cal C}_{(K)}} = 0.
\end{eqnarray}


\section{Energy momentum tensors and equations of motion in
  cosmological situations} 
\label{sec:Explicit}


Here, we derive the explicit components of the energy momentum
tensors and the equations of motion for a perfect fluid, an
imperfect fluid, and a scalar field.
In the derivation of these equations, we do not use any
information of the Einstein equations to guarantee the validity
of the formulae to wide applications.


\subsection{Perfect fluid} 


\subsubsection{Energy momentum tensor} 


As shown in \S\ref{sec:Perfect-fluid-ene-mon-generic}, 
the gauge-invariant parts of the first- and the second-order
perturbations of the energy momentum tensor for a perfect
fluid are given by
Eqs.~(\ref{eq:first-energy-momentum-tensor-perfect-gauge-inv})
and (\ref{eq:second-energy-momentum-tensor-perfect-gauge-inv}).
From Eqs.~(\ref{eq:components-calHab}),
(\ref{eq:kouchan-17.378}), and (\ref{eq:kouchan-17.380}),
the components of the first-order perturbation of the
energy-momentum tensor of the single perfect fluid are given by
\begin{eqnarray}
  \stackrel{(1)}{{}^{(p)}\!{\cal T}_{\eta}^{\;\;\eta}}
  &=&
  - \stackrel{(1)}{{\cal E}}
  , \quad
  \stackrel{(1)}{{}^{(p)}\!{\cal T}_{\eta}^{\;\;i}}
  =
  - \left( \epsilon + p \right) \left(
      D^{i} \stackrel{(1)}{v} 
    + \stackrel{(1)}{{\cal V}^{i}}
    - \stackrel{(1)}{\nu^{i}}
  \right)
  \label{eq:kouchan-19.23}
  , \\
  \stackrel{(1)}{{}^{(p)}\!{\cal T}_{i}^{\;\;\eta}}
  &=&
  \left( \epsilon + p \right) \left(
      D_{i}\stackrel{(1)}{v} 
    + \stackrel{(1)}{{\cal V}_{i}}
  \right)
  , \quad
  \stackrel{(1)}{{}^{(p)}\!{\cal T}_{i}^{\;\;j}}
  =
  \stackrel{(1)}{{\cal P}} \delta_{i}^{\;\;j}
  \label{eq:kouchan-19.25}
  .
\end{eqnarray}
Further, from the components of the first- and the second-order
perturbations of the fluid four-velocity which are given by
Eqs.~(\ref{eq:kouchan-17.378})--(\ref{eq:kouchan-17.399}), 
the components (\ref{eq:components-calHab}) and
(\ref{eq:components-calLab}) of the first- and the second-order
metric perturbations, the components of the second-order
perturbation of the energy-momentum tensor of the single perfect
fluid are given by  
\begin{eqnarray}
  \stackrel{(2)}{{}^{(p)}\!{\cal T}_{\eta}^{\;\;\eta}}
  &=&
  -   \stackrel{(2)}{{\cal E}} 
  - 2 \left( \epsilon + p \right) \left(
      D_{i}\stackrel{(1)}{v} 
    + \stackrel{(1)}{{\cal V}_{i}}
  \right)
  \left(
      D^{i} \stackrel{(1)}{v} 
    + \stackrel{(1)}{{\cal V}^{i}}
    - \stackrel{(1)}{\nu^{i}}
  \right)
  \label{eq:kouchan-19.34}
  , \\
  \stackrel{(2)}{{}^{(p)}\!{\cal T}_{i}^{\;\;\eta}}
  &=&
    2 
  \left(
    \stackrel{(1)}{{\cal E}} + \stackrel{(1)}{{\cal P}} 
  \right) 
  \left(
      D_{i}\stackrel{(1)}{v}
    + \stackrel{(1)}{{\cal V}_{i}}
  \right)
  \nonumber\\
  && 
  + \left( \epsilon + p \right) \left(
      D_{i}\stackrel{(2)}{v}
    + \stackrel{(2)}{{\cal V}_{i}}
    - 2 \stackrel{(1)}{\Phi} D_{i}\stackrel{(1)}{v}
    - 2 \stackrel{(1)}{\Phi} \stackrel{(1)}{{\cal V}_{i}}
  \right)
  \label{eq:kouchan-19.36}
  , \\
  \stackrel{(2)}{{}^{(p)}\!{\cal T}_{\eta}^{\;\;i}}
  &=&
  - 2 \left( 
    \stackrel{(1)}{{\cal E}} + \stackrel{(1)}{{\cal P}} 
  \right) 
  \left(
      D^{i}\stackrel{(1)}{v}
    + \stackrel{(1)}{{\cal V}^{i}}
    - \stackrel{(1)}{\nu^{i}}
  \right)
  \nonumber\\
  && 
  + \left( \epsilon + p \right) \left\{
    -   D^{i}\stackrel{(2)}{v}
    -   \stackrel{(2)}{{\cal V}^{i}}
    +   \stackrel{(2)}{\nu^{i}}
    - 2 \stackrel{(1)}{\Phi} \left(
        D^{i}\stackrel{(1)}{v}
      + \stackrel{(1)}{{\cal V}^{i}}
    \right)
  \right.
  \nonumber\\
  && \quad\quad\quad\quad\quad
  \left.
    + 2 \left(
      - 2 \stackrel{(1)}{\Psi} \gamma^{ij}
      +   \stackrel{(1)}{\chi^{ij}}
    \right) 
    \left(
        D_{j}\stackrel{(1)}{v}
      + \stackrel{(1)}{{\cal V}_{j}}
      - \stackrel{(1)}{\nu_{j}}
    \right)
  \right\}
  \label{eq:kouchan-19.35}
  , \\
  \stackrel{(2)}{{}^{(p)}\!{\cal T}_{i}^{\;\;j}}
  &=&
    2 \left( \epsilon + p \right) \left(
      D_{i}\stackrel{(1)}{v}
    + \stackrel{(1)}{{\cal V}_{i}}
  \right)
  \left(
      D^{j}\stackrel{(1)}{v}
    + \stackrel{(1)}{{\cal V}^{j}}
    - \stackrel{(1)}{\nu^{j}}
  \right)
  +   \stackrel{(2)}{{\cal P}} \delta_{i}^{\;\;j}
  \label{eq:kouchan-19.37}
  .
\end{eqnarray}
In Eqs.~(\ref{eq:kouchan-19.34})--(\ref{eq:kouchan-19.37}), 
the vector- and the tensor-modes of the first-order
perturbations are included into our consideration, which are
ignored in KN2007.


\subsubsection{Perturbative continuity equation of a perfect fluid}


Now, we derive the explicit expression of the first- and the 
second-order perturbations
(\ref{eq:first-order-continuity-perfect-gauge-inv-equation}) and
(\ref{eq:second-order-continuity-perfect-gauge-inv-equation}) of
the continuity equation for a perfect fluid in the
cosmological situation.
Before the derivation of the perturbations, we first show the
background continuity equation
(\ref{eq:background-continuity-perfect}) in the cosmological 
situation, which is given by 
\begin{eqnarray}
  a C_{0}^{(p)}
  &=&
  \partial_{\eta}\epsilon
  +
  3 {\cal H} \left(\epsilon + p\right) = 0
  \label{eq:kouchan-17.638}
  .
\end{eqnarray}
Next, we consider the first-order perturbation
(\ref{eq:first-order-continuity-perfect-gauge-inv}) of the
continuity equation:
\begin{eqnarray}
  a {}^{(1)}\!{\cal C}_{0}^{(p)}
  &=&
      \partial_{\eta}\stackrel{(1)}{{\cal E}}
  + 3 {\cal H} \left(
      \stackrel{(1)}{{\cal E}}
    + \stackrel{(1)}{{\cal P}}
  \right)
  + \left(\epsilon + p\right) \left(
                  \Delta\stackrel{(1)}{v} 
    -          3  \partial_{\eta}\stackrel{(1)}{\Psi}
  \right)
  = 0
  \label{eq:kouchan-17.640}
  ,
\end{eqnarray}
where we have used Eqs.~(\ref{eq:components-calHab}),
(\ref{eq:kouchan-17.378}), (\ref{eq:kouchan-17.380}),
(\ref{eq:kouchan-17.498}), (\ref{eq:kouchan-17.501}),
$\epsilon=\epsilon(\eta)$, and (\ref{eq:kouchan-17.638}).


On the other hand, the explicit expression of the second-order
perturbation
(\ref{eq:second-order-continuity-perfect-gauge-inv}) in the
cosmological situation is given as follows.
Through Eqs.~(\ref{eq:components-calHab}),
(\ref{eq:components-calLab}),
(\ref{eq:kouchan-17.378})--(\ref{eq:kouchan-17.399}),
(\ref{eq:kouchan-17.498})--(\ref{eq:kouchan-17.505}),
$\epsilon=\epsilon(\eta)$, (\ref{eq:kouchan-17.638}), and
(\ref{eq:kouchan-17.640}), the second-order perturbation
${}^{(2)}\!{\cal C}_{0}^{(p)}$ of the continuity equation is
given by 
\begin{eqnarray}
  a {}^{(2)}\!{\cal C}_{0}^{(p)}
  &=&
      \partial_{\eta}\stackrel{(2)}{{\cal E}} 
  + 3 {\cal H} \left(
      \stackrel{(2)}{{\cal E}} 
    + \stackrel{(2)}{{\cal P}}
  \right)
  + \left(
         \Delta \stackrel{(2)}{v}
    -  3 \partial_{\eta}\stackrel{(2)}{\Psi}
  \right)
  \left(\epsilon + p\right)
  - \Xi_{0}^{(p)}
  = 0,
  \label{eq:kouchan-17.644}
\end{eqnarray}
where
\begin{eqnarray}
  \Xi_{0}^{(p)}
  &:=&
  2 \left[
       6 \stackrel{(1)}{\Psi} \partial_{\eta}\stackrel{(1)}{\Psi}
    -  \left(
      2 \stackrel{(1)}{\Psi} + \stackrel{(1)}{\Phi}
    \right) \Delta\stackrel{(1)}{v}
    -    \stackrel{(1)}{\nu^{k}} D_{k}\stackrel{(1)}{\Psi}
  \right.
  \nonumber\\
  && \quad\quad
  \left.
    + \left(
        D^{k}\stackrel{(1)}{v}
      + \stackrel{(1)}{{\cal V}^{k}}
    \right)
    \left\{
        D_{k}\left(\stackrel{(1)}{\Psi} - \stackrel{(1)}{\Phi}\right)
      - \partial_{\eta}\left(
          D_{k}\stackrel{(1)}{v}
        + \stackrel{(1)}{{\cal V}_{k}}
      \right)
    \right\}
  \right.
  \nonumber\\
  && \quad\quad
  \left.
    + \stackrel{(1)}{\chi^{ik}}
    \left\{
        D_{i}\left(
            D_{k}\stackrel{(1)}{v} 
          + \stackrel{(1)}{{\cal V}_{k}}
          - \stackrel{(1)}{\nu_{k}}
      \right)
      + \frac{1}{2} \partial_{\eta}\stackrel{(1)}{\chi_{ik}}
    \right\}
  \right]
  \left(\epsilon + p\right)
  \nonumber\\
  && 
  - 2 \left(
      D^{i}\stackrel{(1)}{v}
    + \stackrel{(1)}{{\cal V}^{i}}
    - \stackrel{(1)}{\nu^{i}}
  \right)
  D_{i}\stackrel{(1)}{{\cal E}} 
  + 2 \left(
               3  \partial_{\eta}\stackrel{(1)}{\Psi}
    -             \Delta\stackrel{(1)}{v} 
  \right)
  \left(
    \stackrel{(1)}{{\cal E}} + \stackrel{(1)}{{\cal P}}
  \right)
  .
  \label{eq:kouchan-17.645-2}
\end{eqnarray}


\subsubsection{Perturbation of the Euler equation}
\label{perturbation-of-Euler-equation-for-a-perfect-fluid}


Here, we consider the explicit expressions of the perturbations 
(\ref{eq:first-order-Euler-perfect-gauge-inv-equation}) and
(\ref{eq:second-order-Euler-perfect-gauge-inv-eqution}) of the
Euler equation for the perfect fluid.


First, we consider the background Euler equation
(\ref{eq:background-Euler-perfect}) in the cosmological
situation. 
In this case, the integral curve of the four-velocity
$u^{a}=g^{ab}u_{a}$, whose component is given by
Eq.~(\ref{eq:kouchan-17.378}), is a geodesic. 
Then, we obtain 
\begin{eqnarray}
  \label{eq:kouchan-17.670}
  a_{b} = 0.
\end{eqnarray}
Through the background Euler equation
(\ref{eq:background-Euler-perfect}), this yields 
\begin{eqnarray}
  \label{eq:kouchan-17.671}
   q_{b}^{\;\;a} \nabla_{a}p = 0.
\end{eqnarray}
This is supported by the homogeneous spatial distribution of the
background pressure.
Therefore, the background Euler equation
(\ref{eq:background-Euler-perfect}) is trivial due to the facts
that the matter distribution is spatially homogeneous.


Next, we consider the components of the gauge-invariant
first-order perturbation
(\ref{eq:first-order-Euler-perfect-gauge-inv-equation}) of the
Euler equation.
Since the background fluid four-velocity $u_{a}$ is tangent to
geodesics, i.e., $a_{b}=0$, we can easily see that the
$\eta$-component $u^{b} {}^{(1)}\!{\cal C}_{b}^{(p)}$ of
(\ref{eq:first-order-Euler-perfect-gauge-inv}) is trivial due
to the equations $u^{b}q_{bc}=0$, Eqs.~(\ref{eq:kouchan-17.517-2}),
(\ref{eq:kouchan-17.438}), and (\ref{eq:kouchan-17.671}).
On the other hand, the spatial component ($i$-component) of
Eq.~(\ref{eq:first-order-Euler-perfect-gauge-inv}) gives the
Euler equation for the perfect fluid:
\begin{eqnarray}
  {}^{(1)}\!{\cal C}_{i}^{(p)}
  &=&
  \left( \epsilon + p \right) \left\{
    \left(
      \partial_{\eta} + {\cal H}
    \right) \left(
      D_{i}\stackrel{(1)}{v} 
      + \stackrel{(1)}{{\cal V}_{i}}
    \right)
    + D_{i}\stackrel{(1)}{\Phi}
  \right\}
  \nonumber\\
  &&
  + D_{i}\stackrel{(1)}{{\cal P}}
  + \partial_{\eta}p \left(
      D_{i}\stackrel{(1)}{v}
    + \stackrel{(1)}{{\cal V}_{i}}
  \right)
  = 0
  ,
  \label{eq:kouchan-17.674}
\end{eqnarray}
where we have used Eqs.~(\ref{eq:components-calHab}),
(\ref{eq:kouchan-17.517-2}), and (\ref{eq:kouchan-17.439}).
This equation can be decomposed into scalar- and vector-parts 
\begin{eqnarray}
  \left( \epsilon + p \right) \left\{
    \left(
      \partial_{\eta} + {\cal H} 
    \right) D_{i}\stackrel{(1)}{v} 
    + D_{i}\stackrel{(1)}{\Phi}
  \right\}
  + D_{i}\stackrel{(1)}{{\cal P}}
  + \partial_{\eta}p D_{i}\stackrel{(1)}{v}
  &=&
  0
  \label{eq:kouchan-17.676}
  , \\
  \left( \epsilon + p \right) \left(
    \partial_{\eta}
    + {\cal H} 
  \right)\stackrel{(1)}{{\cal V}_{i}}
  + \partial_{\eta}p \stackrel{(1)}{{\cal V}_{i}}
  &=&
  0
  .
  \label{eq:kouchan-17.677}
\end{eqnarray}


Since the background value of the acceleration $a_{b}$ vanishes,
the $\eta$-component of ${}^{(2)}\!{\cal C}_{b}^{(p)}$ in
Eq.~(\ref{eq:second-order-Euler-perfect-gauge-inv}) is given by 
\begin{eqnarray}
  {}^{(2)}\!{\cal C}_{\eta}^{(p)}
  &=&
  2 {}^{(1)}\!{\cal C}_{i}^{(p)}
  \left(
           \stackrel{(1)}{\nu^{i}}
    -       D^{i}\stackrel{(1)}{v}
    - \stackrel{(1)}{{\cal V}^{i}}
  \right)
  = 0
  ,
  \label{eq:kouchan-17.679}
\end{eqnarray}
where we have used Eqs.~(\ref{eq:kouchan-17.517-2}),
(\ref{eq:kouchan-17.517-6}), (\ref{eq:kouchan-17.438}),
(\ref{eq:kouchan-17.446}), $q_{\eta c}=0$, $p=p(\eta)$, and
(\ref{eq:kouchan-17.674}).
On the other hand, through Eq.~(\ref{eq:components-calHab}),
(\ref{eq:kouchan-17.517-7}), (\ref{eq:kouchan-17.517-8}),
(\ref{eq:kouchan-17.439}), and (\ref{eq:kouchan-17.447}),
the spatial component of the second-order perturbation
(\ref{eq:second-order-Euler-perfect-gauge-inv}) of the Euler
equation is given by 
\begin{eqnarray}
  {}^{(2)}\!{\cal C}_{i}^{(p)}
  &=&
  \left( \epsilon + p \right) \left\{
    \left(
      \partial_{\eta} + {\cal H}
    \right)
    \left(
      D_{i}\stackrel{(2)}{v}
      + \stackrel{(2)}{{\cal V}_{i}}
    \right)
    + D_{i}\stackrel{(2)}{\Phi}
  \right\}
  \nonumber\\
  &&
  + D_{i}\stackrel{(2)}{{\cal P}}
  + \partial_{\eta}p \left(
      D_{i}\stackrel{(2)}{v}
    + \stackrel{(2)}{{\cal V}_{i}}
  \right)
  - \Xi_{i}^{(p)}
  \nonumber\\
  &=&
  0
  ,
  \label{eq:kouchan-17.680}
\end{eqnarray}
where $\Xi_{i}^{(p)}$ is the collection of the quadratic terms
of the linear-order perturbations defined by 
\begin{eqnarray}
  \Xi_{i}^{(p)}
  &:=&
  - 2 \stackrel{(1)}{\Phi}
  D_{i}\left\{
      \stackrel{(1)}{{\cal P}}
    - \left( \epsilon + p \right) \stackrel{(1)}{\Phi}
  \right\}
  \nonumber\\
  &&
  - 2 \left( \epsilon + p \right) \left(
      \stackrel{(1)}{\nu^{j}}
    - D^{j}\stackrel{(1)}{v} 
    - \stackrel{(1)}{{\cal V}^{j}}
  \right)
  \left\{
      D_{i}\stackrel{(1)}{\nu_{j}}
    - D_{j}\left(
        D_{i}\stackrel{(1)}{v} 
      + \stackrel{(1)}{{\cal V}_{i}}
    \right)
  \right\}
  \nonumber\\
  &&
  - 2 \left(
      \stackrel{(1)}{{\cal E}}
    + \stackrel{(1)}{{\cal P}}
  \right)
  \left\{
      D_{i}\stackrel{(1)}{\Phi}
    + \partial_{\eta}\left(
      D_{i}\stackrel{(1)}{v} 
      + \stackrel{(1)}{{\cal V}_{i}}
    \right)
    + {\cal H} \left(
      D_{i}\stackrel{(1)}{v}
      + \stackrel{(1)}{{\cal V}_{i}}
    \right)
  \right\}
  \nonumber\\
  &&
  - 2 \left(
      D_{i}\stackrel{(1)}{v}
    + \stackrel{(1)}{{\cal V}_{i}}
  \right) \partial_{\eta}\stackrel{(1)}{{\cal P}}
  \label{eq:kouchan-17.682}
  .
\end{eqnarray}
As in the case of the first-order perturbations of the Euler
equation, the equation (\ref{eq:kouchan-17.680}) is decomposed
into the scalar- and the vector-parts as  
\begin{eqnarray}
  &&
  \left( \epsilon + p \right) \left\{
    \left( \partial_{\eta} + {\cal H} \right) D_{i}\stackrel{(2)}{v}
    + D_{i}\stackrel{(2)}{\Phi}
  \right\}
  + D_{i}\stackrel{(2)}{{\cal P}}
  + \partial_{\eta}p D_{i}\stackrel{(2)}{v}
  =
  D_{i} \Delta^{-1} D^{j}\Xi_{j}^{(p)} 
  \label{eq:kouchan-17.684}
  , \\
  &&
  \left( \epsilon + p \right) \left(
    \partial_{\eta} + {\cal H}
  \right)
    \stackrel{(2)}{{\cal V}_{i}}
  + \partial_{\eta}p \stackrel{(2)}{{\cal V}_{i}}
  = \Xi_{i}^{(p)} - D_{i} \Delta^{-1} D^{j}\Xi_{j}^{(p)}
  \label{eq:kouchan-17.685}
  .
\end{eqnarray}


\subsection{Imperfect fluid}

\subsubsection{Energy momentum tensor} 
\label{sec:Energy-momentum-tensor-imperfect}


Here, we consider the components of the each order perturbation
of the energy momentum tensor for an imperfect fluid in the
context of cosmological perturbations.
In the context of the cosmological perturbations, the background
spacetime is homogeneous and isotropic and the energy momentum
tensor on the background spacetime is described by a perfect
fluid.
In other words, we can choose the four-velocity $u_{a}$ of the
dominant fluid so that the energy flux of the fluid vanishes,
\begin{eqnarray}
  q_{a} = 0.
  \label{eq:background-energy-flux-vanishes}
\end{eqnarray}
Further, at the background level, we may neglect the anisotropic
stress
\begin{eqnarray}
  \pi_{ab} = 0.
  \label{eq:background-anisotropic-stress-vanishes}
\end{eqnarray}
Then, the background energy momentum tensor is given by
(\ref{eq:background-energy-momentum-tensor-perfect}). 
Choosing the coordinate system as
Eqs.~(\ref{eq:kouchan-17.378}), the components of the projection
operator $q_{a}^{\;\;b}:=g^{bc}q_{ac}$ is given through
Eq.~(\ref{eq:background-three-metric-components}) and the
background energy momentum tensor is given by
Eq.~(\ref{eq:background-energy-momentum-tensor-perfect}).


Next, we consider the components of the gauge-invariant part of
the first-order perturbation of the energy momentum tensor for
imperfect fluid.
In the situation where
Eqs.~(\ref{eq:background-energy-flux-vanishes}) and
(\ref{eq:background-anisotropic-stress-vanishes}) are satisfied, 
the properties (\ref{eq:kouchan-16.95-4}) and
(\ref{eq:kouchan-16.95-6}) give the $\eta$-component of the
first- and the second-order perturbations of the energy flux
are given by  
\begin{eqnarray}
  \stackrel{(1)}{\hat{\cal Q}_{\eta}}
  = 0
  , \quad
  \stackrel{(2)}{\hat{\cal Q}_{\eta}}
  =
  - 2 \stackrel{(1)}{\hat{\cal Q}_{i}} \left(
      D^{i}\stackrel{(1)}{v}
    + \stackrel{(1)}{{\cal V}^{i}}
    - \stackrel{(1)}{\nu^{i}}
  \right)
  \label{eq:gauge-inv-energy-flux-orthogonal-each-cosmo-component}
  ,
\end{eqnarray}
where we defined 
$\stackrel{(1)}{{\cal Q}_{a}}=:a\stackrel{(1)}{\hat{\cal Q}_{a}}$,
$\stackrel{(2)}{{\cal Q}_{a}}=:a\stackrel{(2)}{\hat{\cal Q}_{a}}$,
and used Eqs.~(\ref{eq:components-calHab}) and
(\ref{eq:kouchan-17.380}).
In the same situation, Eq.~(\ref{eq:kouchan-16.95-8})
and (\ref{eq:kouchan-16.95-13}) are given by 
\begin{eqnarray}
  \stackrel{(1)}{\hat{\Pi}_{\eta\eta}} = 0
  = \stackrel{(1)}{\hat{\Pi}_{i\eta}}
  , \quad
  \stackrel{(2)}{\hat{\Pi}_{\eta\eta}} = 0
  , \quad
  \stackrel{(2)}{\hat{\Pi}_{\eta i}}
  =
  - 2 \stackrel{(1)}{\hat{\Pi}_{ji}}
  \left(
        D^{j}\stackrel{(1)}{v} 
      + \stackrel{(1)}{{\cal V}^{j}}
      - \stackrel{(1)}{\nu^{j}}
  \right)
  \label{eq:gauge-inv-anisotropic-stress-orthogonal-each-cosmo-component}
\end{eqnarray}
through Eqs.~(\ref{eq:components-calHab}) and
(\ref{eq:kouchan-17.380}), where we defined 
$\stackrel{(1)}{\Pi_{ab}}=:a^{2}\stackrel{(1)}{\hat{\Pi}_{ab}}$
and 
$\stackrel{(2)}{\Pi_{ab}}=:a^{2}\stackrel{(2)}{\hat{\Pi}_{ab}}$.
Further, the traceless properties (\ref{eq:kouchan-16.132}) of
the gauge-invariant parts of the first- and the second-order
perturbations of the anisotropic stress are given by
\begin{eqnarray}
  \gamma^{ji} \stackrel{(1)}{\hat{\Pi}_{ji}}
  = 0
  , \quad
  \gamma^{ji} \stackrel{(2)}{\hat{\Pi}_{ji}}
  =
  2 \stackrel{(1)}{\chi^{ki}}\stackrel{(1)}{\hat{\Pi}_{ki}}
  .
  \label{eq:traceless-Piab-gauge-inv-each-cosmo}
\end{eqnarray}


The components of the gauge-invariant part
(\ref{eq:first-ptotal-ene-mom-imperfect-gauge-inv}) of the
first-order perturbation of the total energy momentum tensor for
an imperfect fluid are summarized as   
\begin{eqnarray}
  &&
  \stackrel{(1)}{{\cal T}_{\eta}^{\;\;\eta}}
  =
  - \stackrel{(1)}{{\cal E}}
  , \quad
  \stackrel{(1)}{{\cal T}_{i}^{\;\;\eta}} 
  =
  \left( \epsilon + p \right) \left(
      D_{i}\stackrel{(1)}{v}
    + \stackrel{(1)}{{\cal V}_{i}}
  \right)
  + \stackrel{(1)}{\hat{\cal Q}_{i}}
  , \nonumber\\
  &&
  \stackrel{(1)}{{\cal T}_{\eta}^{\;\;i}} 
  =
  - \left( \epsilon + p \right) \left(
      D^{i}\stackrel{(1)}{v} 
    + \stackrel{(1)}{{\cal V}^{i}} 
    - \stackrel{(1)}{\nu^{i}}
  \right)
  - \stackrel{(1)}{\hat{\cal Q}^{i}}
  , \quad
  \stackrel{(1)}{{\cal T}_{i}^{\;\;j}} 
  =
    \stackrel{(1)}{{\cal P}} \gamma_{i}^{\;\;j}
  + \stackrel{(1)}{\hat{\Pi}_{i}^{\;\;j}}
  ,
\end{eqnarray}
where we have defined
$\stackrel{(1)}{\hat{\cal Q}^{i}}:=\gamma^{ij}\stackrel{(1)}{\hat{\cal Q}_{j}}$
and
$\stackrel{(1)}{\hat{\Pi}_{i}^{\;\;j}}:=\gamma^{jk}\stackrel{(1)}{\hat{\Pi}_{ik}}$.
On the other hand, in the same situation, the components of the
gauge-invariant part
(\ref{eq:second-total-ene-mom-imperfect-gauge-inv}) of the
second-order perturbation of the total energy momentum tensor
for an imperfect fluid are summarized as 
\begin{eqnarray}
  \stackrel{(2)}{{\cal T}_{\eta}^{\;\;\eta}} 
  &=&
  - \stackrel{(2)}{{\cal E}} 
  - 2 \left( \epsilon + p \right) \left(
      D^{i}\stackrel{(1)}{v}
    + \stackrel{(1)}{{\cal V}^{i}}
  \right)
  \left(
      D_{i}\stackrel{(1)}{v}
    +   \stackrel{(1)}{{\cal V}_{i}}
    -   \stackrel{(1)}{\nu_{i}}
  \right)
  \nonumber\\
  && 
  - 2 \stackrel{(1)}{\hat{\cal Q}_{i}} \left(
      2 D^{i}\stackrel{(1)}{v}
    + 2 \stackrel{(1)}{{\cal V}^{i}}
    -   \stackrel{(1)}{\nu^{i}}
  \right)
  , \\
  \stackrel{(2)}{{\cal T}_{i}^{\;\;\eta}} 
  &=&
  2 
  \left(
    \stackrel{(1)}{{\cal E}} + \stackrel{(1)}{{\cal P}} 
  \right) 
  \left(
      D_{i}\stackrel{(1)}{v}
    + \stackrel{(1)}{{\cal V}_{i}}
  \right)
  + \left( \epsilon + p \right) \left\{
        D_{i}\stackrel{(2)}{v}
    +   \stackrel{(2)}{{\cal V}_{i}}
    - 2 \stackrel{(1)}{\Phi} \left(
      D_{i}\stackrel{(1)}{v}
      + \stackrel{(1)}{{\cal V}_{i}}
    \right)
  \right\}
  \nonumber\\
  &&
  +   \stackrel{(2)}{\hat{\cal Q}_{i}}
  - 2 \stackrel{(1)}{\Phi} \stackrel{(1)}{\hat{\cal Q}_{i}}
  + 2 \stackrel{(1)}{\hat{\Pi}_{ij}}
  \left(
      D^{j}\stackrel{(1)}{v} 
    + \stackrel{(1)}{{\cal V}^{j}}
  \right)
  , \\
  \stackrel{(2)}{{\cal T}_{\eta}^{\;\;i}}
  &=&
  2 \left( 
    \stackrel{(1)}{{\cal E}} + \stackrel{(1)}{{\cal P}} 
  \right) 
  \left(
     \stackrel{(1)}{\nu^{i}}
    - D^{i}\stackrel{(1)}{v}
    - \stackrel{(1)}{{\cal V}^{i}}
  \right)
  \nonumber\\
  && 
  + \left( \epsilon + p \right) \left\{
    -   D^{i}\stackrel{(2)}{v}
    -   \stackrel{(2)}{{\cal V}^{i}}
    +   \stackrel{(2)}{\nu^{i}}
    - 2 \stackrel{(1)}{\Phi} \left(
      D^{i}\stackrel{(1)}{v}
      + \stackrel{(1)}{{\cal V}^{i}}
    \right)
  \right.
  \nonumber\\
  && \quad\quad\quad\quad\quad\quad
  \left.
    + 2 \left(
      - 2 \stackrel{(1)}{\Psi} \gamma^{il}
      +   \stackrel{(1)}{\chi^{il}}
    \right)
    \left(
        D_{l}\stackrel{(1)}{v}
      + \stackrel{(1)}{{\cal V}_{l}}
      - \stackrel{(1)}{\nu_{l}}
    \right)
  \right\}
  \nonumber\\
  &&
  -   \stackrel{(2)}{\hat{\cal Q}^{i}}
  - 2 \left(
    \stackrel{(1)}{\Phi} + 2 \stackrel{(1)}{\Psi} 
  \right) \stackrel{(1)}{\hat{\cal Q}^{i}}
  + 2 \stackrel{(1)}{\chi^{il}} \stackrel{(1)}{\hat{\cal Q}_{l}}
  - 2 \stackrel{(1)}{\hat{\Pi}^{ki}} \left(
      D_{k}\stackrel{(1)}{v} 
    + \stackrel{(1)}{{\cal V}_{k}}
    - \stackrel{(1)}{\nu_{k}}
  \right)
  , \\
  \stackrel{(2)}{{\cal T}_{i}^{\;\;j}} 
  &=&
  2 \left( \epsilon + p \right) \left(
      D_{i}\stackrel{(1)}{v}
    + \stackrel{(1)}{{\cal V}_{i}}
  \right)
  \left(
      D^{j}\stackrel{(1)}{v} 
    + \stackrel{(1)}{{\cal V}^{j}} 
    - \stackrel{(1)}{\nu^{j}}
  \right)
  +   \stackrel{(2)}{{\cal P}} \delta_{i}^{\;\;j}
  \nonumber\\
  && \quad
  + 2 \stackrel{(1)}{\hat{\cal Q}_{i}} \left(
      D^{j}\stackrel{(1)}{v}
    + \stackrel{(1)}{{\cal V}^{j}}
    - \stackrel{(1)}{\nu^{j}}
  \right)
  + 2 \stackrel{(1)}{\hat{\cal Q}^{j}}
  \left(
      D_{i}\stackrel{(1)}{v}
    + \stackrel{(1)}{{\cal V}_{i}}
  \right)
  \nonumber\\
  && 
  +   \stackrel{(2)}{\hat{\Pi}_{i}^{\;\;j}}
  + 4 \stackrel{(1)}{\Psi} \stackrel{(1)}{\hat{\Pi}_{i}^{\;\;j}}
  - 2 \stackrel{(1)}{\chi^{jm}} \stackrel{(1)}{\hat{\Pi}_{im}}
  ,
\end{eqnarray}
where we have used
Eqs.~(\ref{eq:components-calHab}), (\ref{eq:components-calLab}),
(\ref{eq:gauge-inv-energy-flux-orthogonal-each-cosmo-component}),
(\ref{eq:kouchan-17.378})--(\ref{eq:kouchan-17.399}), 
and defined
$\stackrel{(2)}{\hat{\cal Q}^{i}}:=\gamma^{ij}\stackrel{(2)}{\hat{\cal Q}_{j}}$
and
$\stackrel{(2)}{\hat{\Pi}_{i}^{\;\;j}}:=\gamma^{jk}\stackrel{(2)}{\hat{\Pi}_{ik}}$.


\subsubsection{Perturbative continuity equation of an imperfect fluid}


Here, we derive the explicit expression of each order
perturbation of the continuity equation
(\ref{eq:kouchan-16.726}), (\ref{eq:kouchan-16.727}), and
(\ref{eq:kouchan-16.728}) for an imperfect fluid in terms of
the components of the gauge-invariant variables.


Since we consider the situation where
Eqs.~(\ref{eq:background-energy-flux-vanishes}) and
(\ref{eq:background-anisotropic-stress-vanishes}) are satisfied,
the background continuity equation coincides with
Eq.(\ref{eq:kouchan-17.638}) for a perfect fluid.
In the same situation, the gauge-invariant expression
(\ref{eq:kouchan-16.727}) of the first-order perturbation of the
continuity equation for an imperfect fluid is given by
\begin{eqnarray}
  a \left(
    \stackrel{(1)}{{\cal C}_{0}^{(p)}} + \stackrel{(1)}{{\cal C}_{0}^{(i)}}
  \right)
  &=&
    \partial_{\eta}\stackrel{(1)}{{\cal E}}
  + 3 {\cal H} \left(
      \stackrel{(1)}{{\cal E}}
    + \stackrel{(1)}{{\cal P}}
  \right)
  + \left(\epsilon + p\right) \left(
                  \Delta\stackrel{(1)}{v} 
    -          3  \partial_{\eta}\stackrel{(1)}{\Psi}
  \right)
  \nonumber\\
  &&
  + D^{j}\stackrel{(1)}{\hat{\cal Q}_{j}} = 0
  \label{eq:kouchan-16.749}
  ,
\end{eqnarray}
where we have used
Eqs.~(\ref{eq:kouchan-17.451}) and
(\ref{eq:gauge-inv-energy-flux-orthogonal-each-cosmo-component}).


Next, we consider the explicit expression of the second-order
perturbation (\ref{eq:kouchan-16.728}) of the continuity 
equation for an imperfect fluid in the situation where
Eqs.~(\ref{eq:kouchan-17.670}),
(\ref{eq:background-energy-flux-vanishes}), and  
(\ref{eq:background-anisotropic-stress-vanishes}) are satisfied. 
In this situation, the gauge-invariant imperfect part
(\ref{eq:kouchan-16.722}) of the second-order perturbation of
the continuity equation is given by
\begin{eqnarray}
  a \stackrel{(2)}{{\cal C}_{0}^{(i)}}
  &=&
      D^{i}\stackrel{(2)}{\hat{\cal Q}_{i}}
  + 4 \stackrel{(1)}{\Psi} D^{i}\stackrel{(1)}{\hat{\cal Q}_{i}}
  - 2 \stackrel{(1)}{\chi^{ik}} D_{i}\stackrel{(1)}{\hat{\cal Q}_{k}}
  + 2 \partial_{\eta}\stackrel{(1)}{\hat{\cal Q}_{i}} \left(
      D^{i}\stackrel{(1)}{v}
    + \stackrel{(1)}{{\cal V}^{i}}
  \right)
  \nonumber\\
  && \quad\quad
  + 2 \stackrel{(1)}{\hat{\cal Q}_{i}} \left\{
        D^{i}\left( 2 \stackrel{(1)}{\Phi} - \stackrel{(1)}{\Psi} \right)
    + 2 \left(\partial_{\eta} + 2 {\cal H} \right)
    \left(
        D^{i}\stackrel{(1)}{v} 
      + \stackrel{(1)}{{\cal V}^{i}}
    \right)
  \right\}
  \nonumber\\
  && \quad\quad
  + 2 \stackrel{(1)}{\hat{\Pi}_{ik}} \left(
    D^{i}\left(
        D^{k}\stackrel{(1)}{v} 
      + \stackrel{(1)}{{\cal V}^{k}}
      - \stackrel{(1)}{\nu^{k}}
    \right)
    + \frac{1}{2} \partial_{\eta}\stackrel{(1)}{\chi^{ik}}
  \right)
  ,
\end{eqnarray}
where we have used
Eqs.~(\ref{eq:components-calHab}),
(\ref{eq:gauge-inv-energy-flux-orthogonal-each-cosmo-component})--(\ref{eq:traceless-Piab-gauge-inv-each-cosmo}),  
(\ref{eq:kouchan-17.438}),
(\ref{eq:kouchan-17.439}), and
(\ref{eq:kouchan-17.451})--(\ref{eq:kouchan-17.462}).
Together with the perfect part (\ref{eq:kouchan-17.644}) of the
continuity equation, the explicit form of the second-order
perturbation (\ref{eq:kouchan-16.728}) of the continuity
equation for an imperfect fluid is given by
\begin{eqnarray}
  a \left(
    \stackrel{(2)}{{\cal C}_{0}^{(p)}} + \stackrel{(2)}{{\cal C}_{0}^{(i)}}
  \right)
  &=&
    \partial_{\eta}\stackrel{(2)}{{\cal E}} 
  + 3 {\cal H} \left(
      \stackrel{(2)}{{\cal E}} 
    + \stackrel{(2)}{{\cal P}}
  \right)
  + \left(
         \Delta \stackrel{(2)}{v}
    -  3 \partial_{\eta}\stackrel{(2)}{\Psi}
  \right)
  \left(\epsilon + p\right)
  \nonumber\\
  &&
  +   D^{i}\stackrel{(2)}{\hat{\cal Q}_{i}}
  -
  \Xi_{0}^{(p+i)}
  ,
  \label{eq:kouchan-16.754}
\end{eqnarray}
where we defined
\begin{eqnarray}
  \Xi_{0}^{(p+i)} &:=& 
  - 2 \left[
    \left(
        2 \stackrel{(1)}{\Psi}
      +   \stackrel{(1)}{\Phi}
    \right)\Delta\stackrel{(1)}{v} 
    +    \stackrel{(1)}{\nu^{k}} D_{k}\stackrel{(1)}{\Psi}
    -  6 \stackrel{(1)}{\Psi} \partial_{\eta}\stackrel{(1)}{\Psi}
  \right.
  \nonumber\\
  && \quad\quad
  \left.
    + \left(
        D^{k}\stackrel{(1)}{v}
      + \stackrel{(1)}{{\cal V}^{k}}
    \right)
    \left\{
        D_{k}\left(\stackrel{(1)}{\Phi} - \stackrel{(1)}{\Psi}\right)
      + \partial_{\eta}\left(
          D_{k}\stackrel{(1)}{v}
        + \stackrel{(1)}{{\cal V}_{k}}
      \right)
    \right\}
  \right.
  \nonumber\\
  && \quad\quad
  \left.
    + \stackrel{(1)}{\chi^{ik}}
    \left\{
        D_{i}\left(
          \stackrel{(1)}{\nu_{k}}
          - D_{k}\stackrel{(1)}{v} 
          - \stackrel{(1)}{{\cal V}_{k}}
        \right)
      - \frac{1}{2} \partial_{\eta}\stackrel{(1)}{\chi_{ik}}
    \right\}
  \right]
  \left(\epsilon + p\right)
  \nonumber\\
  && 
  - 2 \left(
      D^{i}\stackrel{(1)}{v}
    + \stackrel{(1)}{{\cal V}^{i}}
    - \stackrel{(1)}{\nu^{i}}
  \right)
  D_{i}\stackrel{(1)}{{\cal E}} 
  - 2 \left(
                  \Delta\stackrel{(1)}{v} 
    -          3  \partial_{\eta}\stackrel{(1)}{\Psi}
  \right)
  \left(
    \stackrel{(1)}{{\cal E}} + \stackrel{(1)}{{\cal P}}
  \right)
  \nonumber\\
  && 
  - 2 D_{i}\stackrel{(1)}{\hat{\cal Q}_{k}} \left(
      2 \stackrel{(1)}{\Psi} \gamma^{ik}
    -   \stackrel{(1)}{\chi^{ik}}
  \right)
  - 2 \partial_{\eta}\stackrel{(1)}{\hat{\cal Q}_{i}} \left(
      D^{i}\stackrel{(1)}{v}
    + \stackrel{(1)}{{\cal V}^{i}}
  \right)
  \nonumber\\
  && 
  - 4 \stackrel{(1)}{\hat{\cal Q}_{i}} \left\{
      D^{i}\left(
        \stackrel{(1)}{\Phi}
        - \frac{1}{2} \stackrel{(1)}{\Psi}
      \right)
    +  \left(
      \partial_{\eta} + 2 {\cal H}
    \right)
    \left(
      D^{i}\stackrel{(1)}{v} 
      +  \stackrel{(1)}{{\cal V}^{i}}
    \right)
  \right\}
  \nonumber\\
  && 
  - 2 \stackrel{(1)}{\hat{\Pi}_{ik}} \left\{
    D^{i}\left(
        D^{k}\stackrel{(1)}{v} 
      + \stackrel{(1)}{{\cal V}^{k}}
      - \stackrel{(1)}{\nu^{k}}
    \right)
    + \frac{1}{2} \partial_{\eta}\stackrel{(1)}{\chi^{ik}}
  \right\}
  .
  \label{eq:kouchan-16.753}
\end{eqnarray}


\subsubsection{Perturbations of the generalized Navier-Stokes equation} 
\label{sec:perturbation-Generalized-Navier-Stokes-equation}


Here, we derive the explicit expression of each order
perturbation (\ref{eq:kouchan-16.911}),
(\ref{eq:first-order-Navier-Stokes-gauge-inv}), and
(\ref{eq:second-order-Navier-Stokes-gauge-inv}) of the
generalized Navier-Stokes equation in terms 
of the components of the gauge-invariant variables.


First, in the situation where
Eqs.~(\ref{eq:background-energy-flux-vanishes}) and
(\ref{eq:background-anisotropic-stress-vanishes}) are satisfied,
the background generalized Navier-Stokes equation
(\ref{eq:kouchan-16.911}) coincides with the background Euler
equation (\ref{eq:background-Euler-perfect}) for a perfect fluid. 
Therefore, we obtain Eq.~(\ref{eq:kouchan-17.670}), again due to
Eqs.~(\ref{eq:background-energy-flux-vanishes}) and
(\ref{eq:background-anisotropic-stress-vanishes}).
Hence, the background generalized Navier-Stokes equation
(\ref{eq:kouchan-16.911}) is trivial as mentioned above.


Next, we consider the first-order perturbation
(\ref{eq:first-order-Navier-Stokes-gauge-inv}) of the
generalized Navier-Stokes equation.
As seen in \S \ref{sec:Energy-momentum-tensor-imperfect},
through
Eqs.~(\ref{eq:kouchan-17.670}),
(\ref{eq:gauge-inv-energy-flux-orthogonal-each-cosmo-component})--(\ref{eq:traceless-Piab-gauge-inv-each-cosmo}),
(\ref{eq:background-three-metric-components}),
(\ref{eq:kouchan-17.517-2}), and (\ref{eq:kouchan-17.451}),
we can easily see that the $\eta$-component of the first-order
perturbation (\ref{eq:kouchan-17.914}) is trivial.
On the other hand, in the same situation, the spatial component of
Eq.~(\ref{eq:kouchan-17.914}) is given by 
\begin{eqnarray}
  \stackrel{(1)}{{\cal C}_{i}^{(p)}} + \stackrel{(1)}{{\cal C}_{i}^{(i)}} 
  &=&
  \left( \epsilon + p \right) \left(
    \left(\partial_{\eta} + {\cal H}\right)
    \left(
      D_{i}\stackrel{(1)}{v} 
      + \stackrel{(1)}{{\cal V}_{i}}
    \right)
    + D_{i}\stackrel{(1)}{\Phi}
  \right)
  \nonumber\\
  &&
  +   D_{i}\stackrel{(1)}{{\cal P}}
  +   \partial_{\eta}p \left(
    D_{i}\stackrel{(1)}{v}
    +   \stackrel{(1)}{{\cal V}_{i}}
  \right)
  +   \left(
    \partial_{\eta} + 4 {\cal H}
  \right)
  \stackrel{(1)}{\hat{\cal Q}_{i}}
  +   D^{k}\stackrel{(1)}{\hat{\Pi}_{ik}}
  \label{eq:kouchan-17.927}
  \\
  &=&
  0
  \label{eq:kouchan-17.927-2}
  ,
\end{eqnarray}
where we have used Eqs.~(\ref{eq:components-calHab}),
(\ref{eq:gauge-inv-energy-flux-orthogonal-each-cosmo-component}),
(\ref{eq:gauge-inv-anisotropic-stress-orthogonal-each-cosmo-component}),
(\ref{eq:kouchan-17.378}),
(\ref{eq:background-three-metric-components})--(\ref{eq:kouchan-17.517-5}),
(\ref{eq:kouchan-17.439}), (\ref{eq:kouchan-17.451}), and
(\ref{eq:kouchan-17.498}). 
The scalar-part of the generalized Navier-Stokes equation
(\ref{eq:kouchan-17.927-2}) is given by 
\begin{eqnarray}
  \left( \epsilon + p \right) \left\{
    \left(
      \partial_{\eta} + {\cal H}
    \right) D_{i}\stackrel{(1)}{v} 
    + D_{i}\stackrel{(1)}{\Phi}
  \right\}
  +   D_{i}\stackrel{(1)}{{\cal P}}
  +   \partial_{\eta}p D_{i}\stackrel{(1)}{v}
  &&
  \nonumber\\
  + D_{i}\Delta^{-1} D^{j}\left[ 
    \left(
      \partial_{\eta} + 4 {\cal H}
    \right)
    \stackrel{(1)}{\hat{\cal Q}_{j}}
    +   D^{k}\stackrel{(1)}{\hat{\Pi}_{jk}}
  \right]
  &=&
  0
  .
  \label{eq:kouchan-17.929-2}
\end{eqnarray}
Subtracting Eq.~(\ref{eq:kouchan-17.929-2}) from
Eq.~(\ref{eq:kouchan-17.927-2}), we obtain the vector-part of
the first-order perturbation (\ref{eq:kouchan-17.927-2}) of the
generalized Navier-Stokes equation:
\begin{eqnarray}
  &&
      \left( \epsilon + p \right) \left(
      \partial_{\eta}
    + {\cal H}
  \right) \stackrel{(1)}{{\cal V}_{i}}
  +   \partial_{\eta}p  \stackrel{(1)}{{\cal V}_{i}}
  +   \left(
    \partial_{\eta} + 4 {\cal H}
  \right)\stackrel{(1)}{\hat{\cal Q}_{i}}
  +   D^{k}\stackrel{(1)}{\hat{\Pi}_{ik}}
  \nonumber\\
  && \quad\quad
  - D_{i}\Delta^{-1} D^{j}\left[ 
    \left(
      \partial_{\eta} + 4 {\cal H}
    \right) \stackrel{(1)}{\hat{\cal Q}_{j}}
    +   D^{k}\stackrel{(1)}{\hat{\Pi}_{jk}}
  \right]
  =
  0
  .
  \label{eq:kouchan-17.930-2}
\end{eqnarray}


Finally, we consider the explicit expression of the second-order 
perturbation of the generalized Navier-Stokes equation
given by (\ref{eq:second-order-Navier-Stokes-gauge-inv}). 
In the situation where Eqs.~(\ref{eq:kouchan-17.670}),
(\ref{eq:background-energy-flux-vanishes}) and 
(\ref{eq:background-anisotropic-stress-vanishes}) are satisfied,
the $\eta$-component of this generalized Navier-Stokes equation
(\ref{eq:kouchan-17.916}) is trivial due to the $i$-component of
the first-order perturbation (\ref{eq:kouchan-17.927-2}).
Actually, through
Eqs.~(\ref{eq:gauge-inv-energy-flux-orthogonal-each-cosmo-component})--(\ref{eq:traceless-Piab-gauge-inv-each-cosmo}),
(\ref{eq:kouchan-17.517-2})--(\ref{eq:kouchan-17.517-8}),
(\ref{eq:kouchan-17.438})--(\ref{eq:kouchan-17.447}), 
(\ref{eq:kouchan-17.459})--(\ref{eq:kouchan-17.462}), and 
(\ref{eq:kouchan-17.498}),
the $\eta$-component of Eq.~(\ref{eq:kouchan-17.916}) is given
by 
\begin{eqnarray}
  \stackrel{(2)}{{\cal C}_{\eta}^{(p)}} + \stackrel{(2)}{{\cal C}_{\eta}^{(i)}}
  &=&
    2 \left[
      \stackrel{(1)}{{\cal C}_{i}^{(p)}} + \stackrel{(1)}{{\cal C}_{i}^{(i)}} 
  \right]
  \left(
           \stackrel{(1)}{\nu^{i}}
    -       D^{i}\stackrel{(1)}{v}
    - \stackrel{(1)}{{\cal V}^{i}}
  \right)
  = 0
  \label{eq:kouchan-17.932-4}
  .
\end{eqnarray}
On the other hand, the $i$-component of the second-order
perturbation (\ref{eq:kouchan-17.916}) of the generalized
Navier-Stokes equation is given by 
\begin{eqnarray}
  \stackrel{(2)}{{\cal C}_{i}^{(p)}} + \stackrel{(2)}{{\cal C}_{i}^{(i)}}
  &=&
  \left( \epsilon + p \right) \left\{
    \left(\partial_{\eta} + {\cal H}\right)
    \left(
      D_{i}\stackrel{(2)}{v}
      + \stackrel{(2)}{{\cal V}_{i}}
    \right)
    + D_{i}\stackrel{(2)}{\Phi}
  \right\}
  +    D_{i}\stackrel{(2)}{{\cal P}}
  \nonumber\\
  &&
  +    \partial_{\eta}p 
  \left(
        D_{i}\stackrel{(2)}{v}
    +   \stackrel{(2)}{{\cal V}_{i}}
  \right)
  +    \left(
    \partial_{\eta} + 4 {\cal H}
  \right) \stackrel{(2)}{\hat{\cal Q}_{i}}
  +    D^{k}\stackrel{(2)}{\hat{\Pi}_{ik}}
  -
  \Xi_{i}^{(p+i)}
  \nonumber\\
  &=& 0
  ,
  \label{eq:kouchan-17.932-5}
\end{eqnarray}
where
\begin{eqnarray}
  \Xi_{i}^{(p+i)}
  &:=&
     2 \left( \epsilon + p \right) \stackrel{(1)}{\Phi} D_{i}\stackrel{(1)}{\Phi}
  \nonumber\\
  &&
  -  2 \left( \epsilon + p \right) \left(
      \stackrel{(1)}{\nu^{j}}
    - D^{j}\stackrel{(1)}{v} 
    - \stackrel{(1)}{{\cal V}^{j}}
  \right)
  \left\{
      D_{i}\stackrel{(1)}{\nu_{j}}
    - D_{j}\left(
      D_{i}\stackrel{(1)}{v} 
      + \stackrel{(1)}{{\cal V}_{i}}
    \right)
  \right\}
  \nonumber\\
  &&
  -  2 \left(
      \stackrel{(1)}{{\cal E}}
    + \stackrel{(1)}{{\cal P}}
  \right)
  \left\{
    \left(
      \partial_{\eta} + {\cal H}
    \right)
    \left(
      D_{i}\stackrel{(1)}{v} 
      + \stackrel{(1)}{{\cal V}_{i}}
    \right)
    + D_{i}\stackrel{(1)}{\Phi}
  \right\}
  \nonumber\\
  &&
  -  2 \left(
      D_{i}\stackrel{(1)}{v}
    + \stackrel{(1)}{{\cal V}_{i}}
  \right) \partial_{\eta}\stackrel{(1)}{{\cal P}}
  -  2 \stackrel{(1)}{\Phi} D_{i}\stackrel{(1)}{{\cal P}}
  \nonumber\\
  &&
  +  2 \left(
    3 \partial_{\eta}\stackrel{(1)}{\Psi}
    -  \Delta\stackrel{(1)}{v}
  \right) \stackrel{(1)}{\hat{\cal Q}_{i}}
  +  2 \left\{
    D_{i}\stackrel{(1)}{\nu_{j}}
    -  D_{j}\left(
      D_{i}\stackrel{(1)}{v}
      +  \stackrel{(1)}{{\cal V}_{i}}
    \right)
  \right\} \stackrel{(1)}{\hat{\cal Q}^{j}}
  \nonumber\\
  &&
  +  2 \left(
       \stackrel{(1)}{\nu_{m}}
    -  D_{m} \stackrel{(1)}{v}
    -  \stackrel{(1)}{{\cal V}_{m}}
  \right) D^{m}\stackrel{(1)}{\hat{\cal Q}_{i}}
  \nonumber\\
  &&
  -  2 \left\{
    \left(
      \stackrel{(1)}{\Phi} + 2 \stackrel{(1)}{\Psi}
    \right) \gamma^{kn}
    -  \stackrel{(1)}{\chi^{nk}}
  \right\} D_{n}\stackrel{(1)}{\hat{\Pi}_{ik}}
  -  2 \left(
    D^{j}\stackrel{(1)}{v}
    + \stackrel{(1)}{{\cal V}^{j}}
  \right) \partial_{\eta}\stackrel{(1)}{\hat{\Pi}_{ij}}
  \nonumber\\
  &&
  +  2 \left\{
    D^{j}\left(\stackrel{(1)}{\Psi}  -  \stackrel{(1)}{\Phi} \right)
    -  \left(\partial_{\eta} + 4 {\cal H} \right) \left(
      D^{j}\stackrel{(1)}{v}
      +  \stackrel{(1)}{{\cal V}^{j}}
    \right)
  \right\} \stackrel{(1)}{\hat{\Pi}_{ij}}
  \nonumber\\
  &&
  +    D_{i}\stackrel{(1)}{\chi^{mk}} \stackrel{(1)}{\hat{\Pi}_{km}}
  \label{eq:kouchan-17.934}
  .
\end{eqnarray}
Equation (\ref{eq:kouchan-17.932-5}) is also decomposed into
the scalar- and the vector-part in the similar form to
Eqs.~(\ref{eq:kouchan-17.929-2}) and (\ref{eq:kouchan-17.930-2})
with an additional source term $-\Xi_{i}^{(p+i)}$.


\subsection{Scalar field}


In the inflationary scenario of the very early universe, scalar
fields play important roles which drive the inflation itself
and generate the seed of the density fluctuations through
their quantum fluctuations.
Keeping in our mind the applications of our framework of the
second-order perturbation theory to this inflationary scenario,
in this subsection, we summarize the explicit expressions of the
perturbative the energy momentum tensor and the Klein-Gordon
equation.


\subsubsection{Energy momentum tensor}


Here, we derive the explicit expression of the components of the
energy momentum tensors
(\ref{eq:background-energy-momentum-scalar}),
(\ref{eq:first-order-energy-momentum-scalar-gauge-inv}),
(\ref{eq:second-order-energy-momentum-scalar-gauge-inv}) for a
single scalar field with the potential $V(\phi)$ in
gauge-invariant form.


In the cosmological situation, we consider the homogeneous
background field:
\begin{eqnarray}
  \varphi = \varphi(\eta).
  \label{eq:kouchan-19.181}
\end{eqnarray}
Through the background metric (\ref{eq:background-metric}), the
components of the background the energy momentum tensor
(\ref{eq:background-energy-momentum-scalar}) are given by 
\begin{eqnarray}
  &&
  T_{\eta}^{\;\;\eta}
  = 
  - \frac{1}{2a^{2}} (\partial_{\eta}\varphi)^{2}
  -             V(\bar{\varphi})
  , \quad
  T_{\eta}^{\;\;i}
  = 
  0
  =
  T_{i}^{\;\;\eta}
  , \\
  &&
  T_{i}^{\;\;j}
  = 
  \frac{1}{2a^{2}}
  \left\{
      (\partial_{\eta}\varphi)^{2}
    - 2 a^{2} V(\bar{\varphi})
  \right\}
  \gamma_{i}^{\;\;j}
  .
\end{eqnarray}
Through the components (\ref{eq:components-calHab}) of the
gauge-invariant part of the first-order metric perturbation, the
components of the first-order perturbation of the
energy-momentum tensor of a single scalar field are given by
\begin{eqnarray}
  {}^{(1)}\!{\cal T}_{\eta}^{\;\;\eta}
  &=& 
  -   \frac{1}{a^{2}} \left\{
      \partial_{\eta}\varphi \partial_{\eta}\varphi_{1} 
    - \stackrel{(1)}{\Phi} (\partial_{\eta}\varphi)^{2}
    + a^{2} \frac{\partial V}{\partial\varphi} \varphi_{1}
  \right\}
  \label{eq:kouchan-19.196}
  , \\
  {}^{(1)}\!{\cal T}_{\eta}^{\;\;i}
  &=&
  \frac{1}{a^{2}} \partial_{\eta}\varphi \left(
      D^{i}\varphi_{1}
    + \stackrel{(1)}{\nu^{i}} \partial_{\eta}\varphi
  \right)
  \label{eq:kouchan-19.197}
  , \\
  {}^{(1)}\!{\cal T}_{i}^{\;\;\eta}
  &=& 
  - \frac{1}{a^{2}} D_{i}\varphi_{1} \partial_{\eta}\varphi 
  \label{eq:kouchan-19.198}
  , \\
  {}^{(1)}\!{\cal T}_{i}^{\;\;j}
  &=& 
  \frac{1}{a^{2}} \gamma_{i}^{\;\;j}
  \left\{
        \partial_{\eta}\varphi \partial_{\eta}\varphi_{1}
    -   \stackrel{(1)}{\Phi} (\partial_{\eta}\varphi)^{2}
    -   a^{2} \frac{\partial V}{\partial\varphi} \varphi_{1}
  \right\}
  .
  \label{eq:kouchan-19.199}
\end{eqnarray}


Finally, we consider the gauge-invariant part
(\ref{eq:second-order-energy-momentum-scalar-gauge-inv}) of the
second-order perturbation of the energy momentum tensor for a
scalar field.
Through the components (\ref{eq:components-calHab}) and
(\ref{eq:components-calLab}) of the gauge-invariant parts of the
first- and the second-order metric perturbations and the 
homogeneous background condition (\ref{eq:kouchan-19.181}), the
components of the second-order perturbation of the
energy-momentum tensor for a single scalar field are given by 
\begin{eqnarray}
  {}^{(2)}\!{\cal T}_{\eta}^{\;\;\eta}
  &=&
  -   \frac{1}{a^{2}} \left\{
        \partial_{\eta}\varphi \partial_{\eta}\varphi_{2}
    -   (\partial_{\eta}\varphi)^{2} \stackrel{(2)}{\Phi}
    +   a^{2} \varphi_{2}\frac{\partial V}{\partial\varphi}
    - 4 \partial_{\eta}\varphi \stackrel{(1)}{\Phi} \partial_{\eta}\varphi_{1}
    + 4 (\partial_{\eta}\varphi)^{2} (\stackrel{(1)}{\Phi})^{2}
  \right.
  \nonumber\\
  && \quad\quad\quad
  \left.
    -   (\partial_{\eta}\varphi)^{2} \stackrel{(1)}{\nu^{i}} \stackrel{(1)}{\nu_{i}}
    +   (\partial_{\eta}\varphi_{1})^{2}
    +   D_{i}\varphi_{1} D^{i}\varphi_{1} 
    +   a^{2} (\varphi_{1})^{2} \frac{\partial^{2}V}{\partial\varphi^{2}}
  \right\}
  \label{eq:kouchan-19.209}
  , \\
  {}^{(2)}\!{\cal T}_{i}^{\;\;\eta}
  &=&
  -   \frac{1}{a^{2}} \left\{
        \partial_{\eta}\varphi \left(
          D_{i}\varphi_{2}
      - 4 D_{i}\varphi_{1} \stackrel{(1)}{\Phi}
    \right)
    + 2 D_{i}\varphi_{1} \partial_{\eta}\varphi_{1}
  \right\}
  \label{eq:kouchan-19.211}
  , \\
  {}^{(2)}\!{\cal T}_{\eta}^{\;\;i}
  &=&
      \frac{1}{a^{2}} \left[
        \partial_{\eta}\varphi D^{i}\varphi_{2}
    + 2 \partial_{\eta}\varphi_{1} D^{i}\varphi_{1}
    + 2 \partial_{\eta}\varphi \left(
        2 \stackrel{(1)}{\nu^{i}} \partial_{\eta}\varphi_{1}
      + 2 \stackrel{(1)}{\Psi} D^{i}\varphi_{1}
      -   \stackrel{(1)}{\chi^{il}} D_{l}\varphi_{1}
    \right)
  \right.
  \nonumber\\
  && \quad\quad
  \left.
    +   (\partial_{\eta}\varphi)^{2} \left(
          \stackrel{(2)}{\nu^{i}}
      - 4 \stackrel{(1)}{\Phi} \stackrel{(1)}{\nu^{i}}
      + 4 \stackrel{(1)}{\Psi} \stackrel{(1)}{\nu^{i}}
      - 2 \stackrel{(1)}{\chi^{ik}} \stackrel{(1)}{\nu_{k}}
    \right)
  \right]
  \label{eq:kouchan-19.210}
  , \\
  {}^{(2)}\!{\cal T}_{i}^{\;\;j}
  &=&
  \frac{2}{a^{2}} \left[
       D_{i}\varphi_{1} D^{j}\varphi_{1}
    +  D_{i}\varphi_{1} \stackrel{(1)}{\nu^{j}} \partial_{\eta}\varphi 
  \right.
  \nonumber\\
  && \quad\quad
  \left.
    + \frac{1}{2} \gamma_{i}^{\;\;j}
    \left\{
      +   \partial_{\eta}\varphi \left(
            \partial_{\eta}\varphi_{2}
        - 4 \stackrel{(1)}{\Phi} \partial_{\eta}\varphi_{1} 
        - 2 \stackrel{(1)}{\nu_{l}} D^{l}\varphi_{1} 
      \right)
    \right.
  \right.
  \nonumber\\
  && \quad\quad\quad\quad\quad\quad
  \left.
    \left.
      +   (\partial_{\eta}\varphi)^{2} \left(
          4 (\stackrel{(1)}{\Phi})^{2}
        -   \stackrel{(1)}{\nu^{l}} \stackrel{(1)}{\nu_{l}}
        -   \stackrel{(2)}{\Phi}
      \right)
      +   (\partial_{\eta}\varphi_{1})^{2}
      -   D_{l}\varphi_{1} D^{l}\varphi_{1} 
    \right.
  \right.
  \nonumber\\
  && \quad\quad\quad\quad\quad\quad
  \left.
    \left.
      -   a^{2} \varphi_{2} \frac{\partial V}{\partial\varphi}
      -   a^{2} (\varphi_{1})^{2} \frac{\partial^{2}V}{\partial\varphi^{2}}
    \right\}
  \right]
  \label{eq:kouchan-19.212}
  .
\end{eqnarray}
In the components
(\ref{eq:kouchan-19.209})--(\ref{eq:kouchan-19.212}), the
vector- and the tensor-modes of the first-order perturbations 
are included into our consideration, which are ignored in
KN2007.


\subsubsection{Perturbative Klein-Gordon equations}


Here, we consider the explicit expression of the gauge-invariant
part of each order perturbation
(\ref{eq:Klein-Gordon-eq-background}) and
(\ref{eq:Klein-Gordon-eq-first-second-gauge-inv}) of the
Klein-Gordon equation in the context of cosmological
perturbations. 
Since the background field $\varphi$ in cosmology is
homogeneous and the background metric is given by
(\ref{eq:background-metric}), the background Klein-Gordon
equation (\ref{eq:Klein-Gordon-eq-background}) yields
\begin{eqnarray}
  - a^{2} C_{(K)}
  &=&
      \partial_{\eta}^{2}\varphi
  + 2 {\cal H} \partial_{\eta}\varphi
  +   a^{2} \frac{\partial V}{\partial\bar{\varphi}}(\varphi)
  = 0
  .
  \label{eq:kouchan-17.806-background-explicit-final}
\end{eqnarray}
On the other hand, the gauge-invariant part
$\stackrel{(1)}{{\cal C}_{(K)}}$ of the first-order perturbation
of $\bar{C}_{(K)}$, which is defined by
Eq.~(\ref{eq:Klein-Gordon-eq-first-gauge-inv-def}), is
explicitly given by
\begin{eqnarray}
  - a^{2} \stackrel{(1)}{{\cal C}_{(K)}}
  &=&
      \partial_{\eta}^{2}\varphi_{1}
  + 2 {\cal H} \partial_{\eta}\varphi_{1}
  -   \Delta\varphi_{1}
  - \left(
        \partial_{\eta}\stackrel{(1)}{\Phi}
    + 3 \partial_{\eta}\stackrel{(1)}{\Psi}
  \right) \partial_{\eta}\varphi
  \nonumber\\
  && \quad
  + 2 a^{2} \stackrel{(1)}{\Phi} \frac{\partial V}{\partial\bar{\varphi}}(\varphi)
  +   a^{2}\varphi_{1} \frac{\partial^{2}V}{\partial\bar{\varphi}^{2}}(\varphi)
  \label{eq:kouchan-17.806-first-explicit}
  = 0
  ,
\end{eqnarray}
where we have used the background metric
(\ref{eq:background-metric}), the components
(\ref{eq:components-calHab}) of the gauge-invariant part of the
linear-order metric perturbation, the homogeneous condition
(\ref{eq:kouchan-19.181}) for the background field $\varphi$,
and the background Klein-Gordon equation
(\ref{eq:kouchan-17.806-background-explicit-final}).


Finally, we consider the explicit expression of the
gauge-invariant second-order perturbation of the Klein-Gordon
equation in Eqs.~(\ref{eq:Klein-Gordon-eq-first-second-gauge-inv}).
The gauge-invariant expression of 
$\stackrel{(2)}{{\cal C}_{(K)}}$ for the Klein-Gordon equation
is defined by (\ref{eq:Klein-Gordon-eq-second-gauge-inv-def}).
Through Eqs.~(\ref{eq:components-calHab}),
(\ref{eq:components-calLab}), (\ref{eq:kouchan-19.181}), and the
background Klein-Gordon equation
(\ref{eq:kouchan-17.806-background-explicit-final}) and its 
first-order perturbation
(\ref{eq:kouchan-17.806-first-explicit}), the explicit
expression of $\stackrel{(2)}{{\cal C}_{(K)}}$ is given by 
\begin{eqnarray}
  - a^{2} \stackrel{(2)}{{\cal C}_{(K)}}
  &=&
       \partial_{\eta}^{2}\varphi_{2} 
  +  2 {\cal H} \partial_{\eta}\varphi_{2} 
  -    \Delta\varphi_{2} 
  -    \left(
    \partial_{\eta}\stackrel{(2)}{\Phi}
    +  3 \partial_{\eta}\stackrel{(2)}{\Psi}
  \right) \partial_{\eta}\varphi
  \nonumber\\
  && 
  +  2 a^{2} \stackrel{(2)}{\Phi} \frac{\partial V}{\partial\bar{\varphi}}(\varphi)
  +    a^{2}\varphi_{2}\frac{\partial^{2}V}{\partial\bar{\varphi}^{2}}(\varphi)
  - \Xi_{(K)}
  = 0
  ,
\end{eqnarray}
where we defined
\begin{eqnarray}
  \Xi_{(K)}
  &:=&
     2 \partial_{\eta}\left(
         \stackrel{(1)}{\Phi}  +  3 \stackrel{(1)}{\Psi}
  \right) \partial_{\eta}\varphi_{1}
  +  2 D^{i}\left(
    \stackrel{(1)}{\Phi} - \stackrel{(1)}{\Psi}
  \right) D_{i}\varphi_{1}
  +  4 \left(
    \stackrel{(1)}{\Phi} + \stackrel{(1)}{\Psi} 
  \right) \Delta\varphi_{1}
  \nonumber\\
  &&
  -  4 a^{2} \stackrel{(1)}{\Phi} \varphi_{1} \frac{\partial^{2}V}{\partial\bar{\varphi}^{2}}(\varphi)
  -    a^{2} (\varphi_{1})^{2}\frac{\partial^{3}V}{\partial\bar{\varphi}^{3}}(\varphi)
  \nonumber\\
  && 
  +  2 \left(
    \partial_{\eta} +  2 {\cal H}
  \right) \stackrel{(1)}{\nu^{i}} D_{i}\varphi_{1}
  +  4 \stackrel{(1)}{\nu^{i}} \partial_{\eta}D_{i}\varphi_{1}
  -  2 \stackrel{(1)}{\chi^{ij}} D_{j}D_{i}\varphi_{1}
  \nonumber\\
  &&
  +  2 \left\{
    - 2 \stackrel{(1)}{\Phi} \partial_{\eta}\stackrel{(1)}{\Phi}
    + 6 \stackrel{(1)}{\Psi} \partial_{\eta}\stackrel{(1)}{\Psi}
    -   D^{i}\left(
      \stackrel{(1)}{\Phi} + \stackrel{(1)}{\Psi}
    \right) \stackrel{(1)}{\nu_{i}}
  \right.
  \nonumber\\
  && \quad\quad
  \left.
    +  \stackrel{(1)}{\nu_{i}} \partial_{\eta}\stackrel{(1)}{\nu^{i}}
    -  \stackrel{(1)}{\chi^{ij}} \left(
      D_{i}\stackrel{(1)}{\nu_{j}}
      - \frac{1}{2} \partial_{\eta}\stackrel{(1)}{\chi_{ij}}
    \right)
  \right\} \partial_{\eta}\varphi
  \nonumber\\
  && 
  -  2 a^{2} \stackrel{(1)}{\nu^{i}} \stackrel{(1)}{\nu_{i}} \frac{\partial V}{\partial\bar{\varphi}}(\varphi)
  .
\end{eqnarray}


\section{Summary and Discussions}
\label{sec:summary}


In summary, we have derived the explicit expressions of the
second-order perturbations of the energy momentum tensor for a
perfect fluid, an imperfect fluid, and a scalar field.
Further, we also derived the explicit expression of the
second-order perturbations the continuity equation and the Euler
equation for a perfect fluid, the continuity equation and the
generalized Navier-Stokes equation for an imperfect fluid, and
the Klein-Gordon equation for a scalar field.
As in KN2007\cite{kouchan-cosmo-second}, we have again confirmed
that the general formulation of the second-order perturbation
theory developed in the papers KN2003\cite{kouchan-gauge-inv}
and KN2005\cite{kouchan-second} does work and it is applicable
to the perturbations of the energy momentum tensors and the
equations of motion for matter fields in the cosmological
perturbation theory.
In the derivations of these equations, we have again seen that 
the decomposition formulae
(\ref{eq:matter-gauge-inv-decomp-1.0}) and
(\ref{eq:matter-gauge-inv-decomp-2.0}) of the perturbations for
the arbitrary fields play crucial roles in the gauge-invariant
perturbation theory.
Since the general relativistic higher-order perturbation theory
requires the long algebraic calculations, in many cases, it is
difficult to have confidence in the resulting long equations.
In spite of this fact, we showed that all perturbative variables
are decomposed into the gauge-invariant and gauge-variant
variables in the forms (\ref{eq:matter-gauge-inv-decomp-1.0})
and (\ref{eq:matter-gauge-inv-decomp-2.0}).
This implies that the decomposition formulae
(\ref{eq:matter-gauge-inv-decomp-1.0}) and
(\ref{eq:matter-gauge-inv-decomp-2.0}) are useful to check
whether the resulting equations are correct or not.
This is the main point of this paper.


As mentioned in Introduction (\S\ref{sec:intro}), there are some
attempts of the derivations of the perturbative expressions of
the evolution equations of matter fields to the
second-order\cite{Noh-Hwang-2004}. 
For example, Noh and Hwang\cite{Noh-Hwang-2004} also summarized
the formulae of the energy momentum tensor, equations of motion
for an imperfect fluid in the cosmological situation up to
second order without any gauge fixing.
However, they implicitly imposed so-called ``normal frame
condition'' in their formulae.
In our notation, this normal frame condition corresponds to
$D_{i}\stackrel{(1)}{v}+\stackrel{(1)}{{\cal V}_{i}}$ $=$
$D_{i}\stackrel{(2)}{v}+\stackrel{(2)}{{\cal V}_{i}}$ $=0$.
Due to this condition, their formulae are not equivalent to
ours.
Further, their formulae include gauge degree of freedom and the
resulting formulae have some complicated forms.
On the other hand, in this paper, all formulae are given in
terms of gauge-invariant variables and there is no gauge
ambiguity in these expressions.
In this sense, the formulae in this paper are irreducible.
We also have to emphasize that we did not fix any gauge when we
derive any perturbative formulae in this paper.
The gauge-invariant treatments of perturbations are
equivalent to the complete gauge-fixing method.
Therefore, we have realized the complete gauge-fixing without
any gauge fixing.
This is an advantage of the gauge-invariant perturbation
theory in this paper.


As emphasized in KN2007, the key point of our procedure is in
the assumption which state that we already know the procedure to
decompose the first-order metric perturbation into the
gauge-invariant and variant parts.
Mathematically speaking, this assumption is expressed by the
statement that
{\it there is a vector field $X^{a}$ which is constructed from
  some components of the linear-order metric perturbation so
  that its gauge transformation rule is given by the second
  equation in Eqs.~(\ref{eq:linear-metric-decomp-gauge-trans})
  and the linear-order metric perturbation is decomposed as
  Eq.~(\ref{eq:linear-metric-decomp}).
}
As shown in KN2007 this assumption is correct at least in the
cosmological perturbation case.
However, even in the cosmological perturbation case, homogeneous
modes of perturbations are excluded from our consideration
because we assumed the existence of the Green functions
$\Delta^{-1}$, $(\Delta+2K)^{-1}$, and $(\Delta+3K)^{-1}$.
These homogeneous modes of the cosmological perturbations
will be some dynamical degrees of freedom in compact universes
and we cannot say that these modes are unphysical.
If we want to include these homogeneous modes into our
considerations, separate treatments will be required. 
Besides this detailed problem to extend the domain of functions
for perturbations to includes homogeneous modes, the above
assumption is correct on some background spacetime other than
the cosmological background spacetime\cite{kouchan-papers}.
Therefore, we propose a conjecture that the above assumption is 
correct on any background spacetime.
To clarify whether this conjecture is true or not, non-local
arguments on the background spacetime will be necessary, since
gauge-invariant variables are non-local one by their definitions.


Even if the assumption is correct on any background spacetime,
the other problem is in the interpretations of the
gauge-invariant variables.
We have commented on the non-uniqueness in the definitions of
the gauge-invariant variables in \S\ref{sec:Second-order-cosmological-perturbatios-metric}. 
This non-uniqueness in the definition of gauge-invariant
variables also leads some ambiguities in the interpretations of 
gauge-invariant variables.
Although the precise interpretations of the gauge-invariant
variables will be accomplished by the clarification of the
relations between gauge-invariant variables and observables in
experiments and observations, some geometrical interpretations
are given through the explicit expressions of the
gauge-invariant variables for the perturbations 
of geometrical quantities like the acceleration $\bar{a}_{a}$,
the expansion $\bar{\theta}$, the shear $\bar{\sigma}_{ab}$, and
the rotation $\bar{\omega}_{ab}$. 
For example, it is well-known that the expressions
(\ref{eq:kouchan-17.438}) and (\ref{eq:kouchan-17.439}) of the
first-order perturbation of the acceleration vector give the
clear interpretation of the scalar-mode $\stackrel{(1)}{\Phi}$
of the first-order metric perturbation.
If $\bar{u}^{a}$ is a tangent to a geodesic even in the
first-order perturbation, Eq.~(\ref{eq:kouchan-17.439}) gives
the equation of motion of a free-falling object in expanding
universe and we can regard the scalar function
$\stackrel{(1)}{\Phi}$ as the Newton gravitational potential.
Further, the similar interpretation for $\stackrel{(2)}{\Phi}$
is also possible through Eq.~(\ref{eq:kouchan-17.447}) though
the quadratic terms of the linear-order perturbations in
Eq.~(\ref{eq:kouchan-17.447}) are difficult to interpret.  
Similarly, $\stackrel{(1)}{\Psi}$ and $\stackrel{(2)}{\Psi}$
contribute to the expansion of the first- and the second-order
perturbations of the expansion as in
Eqs.~(\ref{eq:kouchan-17.501}) and (\ref{eq:kouchan-17.505}),
respectively, and the vector- and tensor-modes
$\stackrel{(1)}{\nu_{i}}$, $\stackrel{(2)}{\nu_{i}}$,
$\stackrel{(1)}{\chi_{ij}}$, and $\stackrel{(2)}{\chi_{ij}}$
contribute to the perturbations of the shear tensor through
Eq.~(\ref{eq:shear-explicit-first-order-2}) and
(\ref{eq:shear-explicit-second-order-3}). 
As seen in these equations, the contributions of the second-order 
gauge-invariant variables $\stackrel{(2)}{\Phi}$,
$\stackrel{(2)}{\Psi}$, $\stackrel{(2)}{\nu_{i}}$, and
$\stackrel{(2)}{\chi_{ij}}$ to the gauge-invariant part of the
second-order perturbations of the geometrical quantities
$\bar{\theta}$, $\bar{\sigma}_{ab}$, $\bar{\omega}_{ab}$ are
similar to the contribution of the first-order gauge-invariant
variables $\stackrel{(1)}{\Phi}$, $\stackrel{(1)}{\Psi}$,
$\stackrel{(1)}{\nu_{i}}$, and $\stackrel{(1)}{\chi_{ij}}$ of
the first-order metric perturbation to the gauge-invariant part
of the first-order perturbations of these geometrical
quantities. 
Therefore, we may regard the gauge-invariant variables
$\stackrel{(2)}{\Phi}$, $\stackrel{(2)}{\Psi}$,
$\stackrel{(2)}{\nu_{i}}$, and $\stackrel{(2)}{\chi_{ij}}$ for
the second-order metric perturbation are natural extensions of
the gauge-invariant variables $\stackrel{(1)}{\Phi}$,
$\stackrel{(1)}{\Psi}$, $\stackrel{(1)}{\nu_{i}}$, and
$\stackrel{(1)}{\chi_{ij}}$ for the first-order perturbations to
the second-order one and the geometrical interpretations of these
second-order variables will be similar to those of the first-order
one.


In all derivations of the perturbative expressions of the energy
momentum tensors and the equations of motion in
\S\ref{sec:Generic-form-of-perturbations-of-energy-momentum-tensors-and-equations-of-motion},
we did not use any information of the Einstein equations nor
equations of state for the matter fields. 
This implies that the formulae in
\S\ref{sec:Generic-form-of-perturbations-of-energy-momentum-tensors-and-equations-of-motion} 
are valid for the very wide class of perturbation theories of
gravitational field.
However, we have to emphasize that the equations of motion
derived in \S\ref{sec:Explicit} are not able to solve by
themselves.
These equations include metric perturbations.
Therefore, to solve these equations, we have to use the Einstein
equations.
Further, we have treated the anisotropic stress and the energy
flux in the case of an imperfect fluid as independent variables
of other perturbative variables.
If we specify the micro-physics, these variables are related to
the other perturbative
variables\cite{Kodama-Sasaki-1984,anisotropic-stress-in-cosmology}. 
Therefore, in general, it is meaningless to try to discuss the
solutions to the equations derived in \S\ref{sec:Explicit} by
themselves.
To discuss the solutions to them, we have to treat the Einstein
equations and we have to specify the interactions between matter
fields at micro- or macro-physical level.


Finally, we can show that these equations of motion are not
independent of the perturbative Einstein
equations in the case where the spacetime is filled with a
single matter field\cite{kouchan-in-prep}. 
This is a natural result because the Einstein equation includes
the equations of motion for the matter field through the Bianchi 
identity at least in the case of the single matter content. 
Therefore, we may concentrate on the Einstein equations when
we solve the system of the single matter field.
On the other hand, when we consider the multi-fluids or
multi-fields system, we will have to specify the interactions
between these matter fields and to treat the equations of motion
derived in this paper, seriously.
Therefore, in the realistic situations of cosmology, the
formulae summarized in this paper will play crucial role.


\section*{Acknowledgements}


The author acknowledges Prof. J.~Nester, Prof. C.M.~Chen,
Prof. Galsov and their colleagues for their hospitality during
his visit to the National Central University in Taiwan and
Prof. K.W.~Ng, and Prof. H.L.~Yu, and their colleagues for their
hospitality during his visit to the Academia Scinica in Taiwan.
In particular, the author also thanks to Dr. M.~Narita for his 
kind many suggestions in Taiwan.
The author also acknowledge to Dr. K.~Ichiki for valuable
information.
He also thanks to Prof. J.~Yokoyama and Prof. M.~Kasai for their
valuable comments and their encouragements and to
Prof. H.~Kodama and A.~Ishibashi for their hospitality during
his visit to KEK.
Finally, the author deeply thanks members of Division of
Theoretical Astronomy at NAOJ and his family for their
continuous encouragement.


\appendix


\section{Perturbations of geometrical quantities} 
\label{sec:Perturbations-of-accerelation-expansion-shear-and-rotation} 


The explicit expressions of the perturbative equations of motion
for fluids in 
\S\ref{sec:Generic-form-of-perturbations-of-energy-momentum-tensors-and-equations-of-motion}
are given by the perturbative expansions of the fluid component
$\bar{\epsilon}$, $\bar{p}$, $\bar{u}_{a}$ $\bar{q}_{a}$, and
$\bar{\pi}_{ab}$ as
Eqs.~(\ref{eq:energy-density-expansion})--(\ref{eq:four-velocity-expansion}),
(\ref{eq:kouchan-16.72})--(\ref{eq:kouchan-16.74}) together
with the perturbative expansion (\ref{eq:metric-expansion}) of
the metric perturbation. 
Further, to derive the perturbations of the equations of motion
for matter fields, it is convenient to consider the perturbation
of the tensor $\bar{\nabla}_{a}\bar{u}_{b}$.
From the components of the tensor $\bar{\nabla}_{a}\bar{u}_{b}$,
we define the acceleration, the expansion, the shear, and the
rotation associated with the four-velocity $\bar{u}_{a}$. 
The acceleration $\bar{a}_{b}$ are defined by
\begin{eqnarray}
  \label{eq:barab-definition}
 \bar{a}_{b}
 := \bar{u}^{a}\bar{\nabla}_{a}\bar{u}_{b}
\end{eqnarray}
and the expansion, the shear, and the rotation are defined as
the trace, the traceless symmetric part, the antisymmetric part
of the tensor
\begin{eqnarray}
  \label{eq:barBab-definition}
  \bar{B}_{ab}
  :=
  \bar{q}_{a}^{\;\;c}\bar{q}_{b}^{\;\;d}\bar{\nabla}_{c}\bar{u}_{d}
  =
  \bar{\nabla}_{a}\bar{u}_{b} + \bar{u}_{a}\bar{a}_{b},
\end{eqnarray}
respectively.
In this Appendix, we show the perturbations of the tensor field
$\bar{\nabla}_{a}\bar{u}_{b}$ and the acceleration, 
expansion, shear, and rotation.
All of the perturbations of these geometrical quantities are
also decomposed into the gauge-invariant and the gauge-variant
parts as Eqs.~(\ref{eq:matter-gauge-inv-decomp-1.0}) and
(\ref{eq:matter-gauge-inv-decomp-2.0}).


\subsection{Components of the gauge-invariant three-metric} 
\label{sec:Perturbations-three-metric-appendix}


Before discussing the components of the perturbations of tensor
$\bar{B}_{ab}$, we first derive the components of the
perturbation of the three-metric $\bar{q}_{ab}$, which is
defined by 
\begin{eqnarray}
  \label{eq:three-metric-definition-full}
  \bar{q}_{ab} := \bar{g}_{ab} + \bar{u}_{a}\bar{u}_{b}
\end{eqnarray}
and it is expanded as
\begin{eqnarray}
  \bar{q}_{bc}
  &=&
  q_{bc}
  + \lambda \stackrel{(1)}{(q_{bc})}
  + \frac{1}{2} \lambda^{2} \stackrel{(2)}{(q_{bc})}
  + O(\lambda^{3}).
  \label{eq:three-metric-expansion}
\end{eqnarray}
From the perturbative expansions
(\ref{eq:metric-expansion}), (\ref{eq:four-velocity-expansion}),
each order perturbation of the three metric is summarized as 
\begin{eqnarray}
  \label{eq:kouchan-17.345}
  q_{ab} &:=& g_{ab} + u_{a} u_{b}
  , \\
  \stackrel{(1)}{(q_{ab})}
  &:=&
    h_{ab}
  + u_{a} \stackrel{(1)}{(u_{b})}
  + \stackrel{(1)}{(u_{a})} u_{b}
  \label{eq:kouchan-17.346}
  , \\
  \stackrel{(2)}{(q_{ab})}
  &:=&
    l_{ab}
  + u_{a} \stackrel{(2)}{(u_{b})}
  + 2 \stackrel{(1)}{(u_{a})} \stackrel{(1)}{(u_{b})}
  + \stackrel{(2)}{(u_{a})} u_{b}
  .
  \label{eq:kouchan-17.347}
\end{eqnarray}
Further, substituting the decompositions
(\ref{eq:linear-metric-decomp}),
(\ref{eq:second-metric-decomp}) of the first- and the
second-order metric perturbations and the definitions of the
gauge-invariant variables for the first- and the second-order
perturbations of the fluid four-velocity in
Eqs.~(\ref{eq:kouchan-16.13}) and (\ref{eq:kouchan-16.18}) into
Eqs.~(\ref{eq:kouchan-17.346}) and (\ref{eq:kouchan-17.347}), we
can show that each order perturbation of $\bar{q}_{ab}$ is
decomposed into the gauge-invariant and gauge-variant parts as
\begin{eqnarray}
  \stackrel{(1)}{(q_{bc})}
  &=&
  \stackrel{(1)}{{\cal Q}}_{bc} + {\pounds}_{X}q_{bc}
  \label{eq:first-three-metric-gauge-inv}
  , \\
  \stackrel{(2)}{(q_{bc})}
  &=&
  \stackrel{(2)}{{\cal Q}}_{bc}
  + 2 {\pounds}_{X}\stackrel{(1)}{(q_{bc})}
  + \left\{
      {\pounds}_{Y}
    - {\pounds}_{X}^{2}
  \right\}q_{bc}
  \label{eq:second-three-metric-gauge-inv}
  .
\end{eqnarray}
Here, the gauge-invariant parts $\stackrel{(1)}{{\cal Q}_{ab}}$
and $\stackrel{(2)}{{\cal Q}_{ab}}$ are defined by 
\begin{eqnarray}
  \label{eq:kouchan-17.350}
  \stackrel{(1)}{{\cal Q}_{ab}}
  &:=&
    {\cal H}_{ab}
  + u_{a} \stackrel{(1)}{{\cal U}_{b}}
  + \stackrel{(1)}{{\cal U}_{a}} u_{b}
  , \\
  \label{eq:kouchan-17.352}
  \stackrel{(2)}{{\cal Q}_{ab}}
  &:=&
    {\cal L}_{ab}
  + u_{a} \stackrel{(2)\;\;}{{\cal U}_{b}}
  + \stackrel{(2)\;\;}{{\cal U}_{a}} u_{b}
  + 2 \stackrel{(1)}{{\cal U}_{a}} \stackrel{(1)}{{\cal U}_{b}}
  .
\end{eqnarray}


To derive the components of the gauge-invariant part
(\ref{eq:kouchan-17.350}) and (\ref{eq:kouchan-17.352}) of the
perturbative three-metric, the components of the gauge-invariant
part of the fluid four-velocity are necessary. 
Components of the background and the first-order perturbations
of the four-velocity are summarized as 
\begin{eqnarray}
  \label{eq:kouchan-17.378}
  u_{a} &=& - a (d\eta)_{a}, \\
  \label{eq:kouchan-17.380}
  \stackrel{(1)}{{\cal U}_{a}}
  &=&
  - a \stackrel{(1)}{\Phi} (d\eta)_{a}
  + a \left(
    D_{i} \stackrel{(1)}{v} 
    + 
    \stackrel{(1)}{{\cal V}_{i}}
  \right) (dx^{i})_{a}
  , \quad
  D^{i}\stackrel{(1)}{{\cal V}_{i}} = 0
  , \\
  \label{eq:kouchan-17.399}
  \stackrel{(2)}{{\cal U}_{a}} 
  &=&
  \stackrel{(2)}{{\cal U}_{\eta}} (d\eta)_{a}
  + a \left(
    D_{i} \stackrel{(2)}{v} 
    + 
    \stackrel{(2)}{{\cal V}_{i}}
  \right) (dx^{i})_{a}
  , \quad
  D^{i}\stackrel{(2)}{{\cal V}_{i}} = 0
  ,
\end{eqnarray}
where the $\eta$-component of the first-order perturbation
$\stackrel{(1)}{{\cal U}_{a}}$ is determined by
Eq.~(\ref{eq:kouchan-17.30}) and the $\eta$-component of the
gauge-invariant part $\stackrel{(2)}{{\cal U}_{a}}$ in
Eq.~(\ref{eq:kouchan-17.399}) is given by 
\begin{eqnarray}
  \label{eq:kouchan-17.398}
  \stackrel{(2)}{{\cal U}_{\eta}} 
  &=&
  a \left\{
        \left(\stackrel{(1)}{\Phi}\right)^{2}
    -   \stackrel{(2)}{\Phi}
    - \left(
        D_{i}\stackrel{(1)}{v}
      + \stackrel{(1)}{{\cal V}_{i}}
      - \stackrel{(1)}{\nu_{i}}
    \right)
    \left(
        D^{i}\stackrel{(1)}{v}
      + \stackrel{(1)}{{\cal V}^{i}}
      - \stackrel{(1)}{\nu^{i}}
    \right)
  \right\}
\end{eqnarray}
through Eq.~(\ref{eq:kouchan-17.31}),
where we have used the components (\ref{eq:components-calHab})
and (\ref{eq:components-calLab}) of the first- and the
second-order metric perturbations.


The components of the background three-metric $q_{ab}$ defined
by Eq.~(\ref{eq:kouchan-17.345}) is given by 
\begin{eqnarray}
  \label{eq:background-three-metric-components}
  q_{ab}
  =
  a^{2} \gamma_{ij} (dx^{i})_{a}(dx^{j})_{b}
  =:
  a^{2} \gamma_{ab}
  .
\end{eqnarray}
Through Eqs.~(\ref{eq:components-calHab}),
(\ref{eq:kouchan-17.378}), and (\ref{eq:kouchan-17.380}), the
components of the gauge-invariant part (\ref{eq:kouchan-17.350})
of the first-order perturbation of the three-metric, which is 
defined by (\ref{eq:kouchan-17.350}), are given by 
\begin{eqnarray}
  \label{eq:kouchan-17.517-2}
  \stackrel{(1)}{{\cal Q}_{\eta\eta}}
  &=& 
  0
  , \quad
  \stackrel{(1)}{{\cal Q}_{i\eta}}
  =
  \stackrel{(1)}{{\cal Q}_{\eta i}}
  =
  a^{2} \left(
      \stackrel{(1)}{\nu_{i}}
    - D_{i}\stackrel{(1)}{v}
    - \stackrel{(1)}{{\cal V}_{i}}
  \right)
  , \\
  \stackrel{(1)}{{\cal Q}_{ij}}
  &=&
  {\cal H}_{ij}
  =
  a^{2} \left(
    - 2 \stackrel{(1)}{\Psi} \gamma_{ij} + \stackrel{(1)}{\chi_{ij}}
  \right)
  \label{eq:kouchan-17.517-5}
  .
\end{eqnarray}
Similarly, through Eqs.~(\ref{eq:components-calLab}),
(\ref{eq:kouchan-17.378})--(\ref{eq:kouchan-17.398}), the
components of the gauge-invariant part of the second-order
perturbation of the three-metric, which is defined by
Eq.~(\ref{eq:kouchan-17.352}), are summarized as
\begin{eqnarray}
  \label{eq:kouchan-17.517-6}
  \stackrel{(2)}{{\cal Q}_{\eta\eta}}
  &=&
  2 a^{2} \left(
      D_{i}\stackrel{(1)}{v}
    + \stackrel{(1)}{{\cal V}_{i}}
    - \stackrel{(1)}{\nu_{i}}
  \right)
  \left(
      D^{i}\stackrel{(1)}{v}
    + \stackrel{(1)}{{\cal V}^{i}}
    - \stackrel{(1)}{\nu^{i}}
  \right)
  , \\
  \label{eq:kouchan-17.517-7}
  \stackrel{(2)}{{\cal Q}_{i\eta}}
  &=&
  \stackrel{(2)}{{\cal Q}_{\eta i}}
  =
  a^{2} \left\{
      \stackrel{(2)}{\nu_{i}}
    - D_{i}\stackrel{(2)}{v}
    - \stackrel{(2)}{{\cal V}_{i}}
    - 2 \stackrel{(1)}{\Phi}
    \left(
        D_{i}\stackrel{(1)}{v}
      + \stackrel{(1)}{{\cal V}_{i}}
    \right)
  \right\}
  , \\
  \label{eq:kouchan-17.517-8}
  \stackrel{(2)}{{\cal Q}_{ij}}
  &=&
  a^{2} \left\{
    - 2 \stackrel{(2)}{\Psi} \gamma_{ij} + \stackrel{(2)}{\chi_{ij}}
    + 2 \left(
        D_{i}\stackrel{(1)}{v}
      + \stackrel{(1)}{{\cal V}_{i}}
    \right)
    \left(
        D_{j}\stackrel{(1)}{v}
      + \stackrel{(1)}{{\cal V}_{j}}
    \right)
  \right\}
  .
\end{eqnarray}


\subsection{Perturbation of the tensor
  $\bar{A}_{ab}=\bar{\nabla}_{a}\bar{u}_{b}$} 
\label{sec:Perturbations-Aab-appendix}


Here, we consider the perturbation of the tensor defined by 
\begin{eqnarray}
  \label{eq:kouchan-17.93}
  \bar{A}_{ab} := \bar{\nabla}_{a}\bar{u}_{b}.
\end{eqnarray}
The covariant derivative $\bar{\nabla}_{a}$ associated with the
metric $\bar{g}_{ab}$ on the physical spacetime is related to 
the covariant derivative $\nabla_{a}$ associated with the
background metric $g_{ab}$ as
\begin{eqnarray}
  \label{eq:kouchan-17.94}
  \bar{A}_{ab}
  &=&
  \nabla_{a}\bar{u}_{b} - C_{ba}^{c}\bar{u}_{c}
  ,
\end{eqnarray}
where the connection $C_{ba}^{c}$ is given by 
\begin{eqnarray}
  C^{c}_{\;\;ab}
  &=&
  \frac{1}{2} \bar{g}^{cd}
  \left(
      \nabla_{a}\bar{g}_{db}
    + \nabla_{b}\bar{g}_{da}
    - \nabla_{d}\bar{g}_{ab}
  \right)
  \\
  &=:&
  \lambda H_{ab}^{\;\;\;\;c}\left[h\right]
  + \frac{1}{2}\lambda^{2} \left(
    H_{ab}^{\;\;\;\;c}\left[l\right]
    - 2 h^{cd} H_{abd}\left[h\right]
  \right)
  + O(\lambda^{3}),
  \label{eq:perturbative-connection-expansion}
\end{eqnarray}
where we defined 
\begin{eqnarray}
  H_{abc}\left[t\right]
  :=
  \frac{1}{2} \left(
      \nabla_{a}t_{cb}
    + \nabla_{b}t_{ca}
    - \nabla_{c}t_{ab}
  \right)
  , 
  \quad
  H_{ab}^{\;\;\;\;c}\left[t\right] := g^{cd} H_{abd}\left[t\right]
  \label{eq:Habc-def-general}
\end{eqnarray}
for any tensor $t_{ab}$ of the second rank. 
Through Eqs.~(\ref{eq:four-velocity-expansion}) and
(\ref{eq:perturbative-connection-expansion}), the tensor
$\bar{A}_{ab}$ is expanded as 
\begin{eqnarray}
  \bar{A}_{ab}
  &=&
  \nabla_{a}u_{b}
  + \lambda \left( 
    \nabla_{a}\stackrel{(1)}{(u_{b})}
    - H_{ba}^{\;\;\;\;c}\left[h\right] u_{c}
  \right)
  \nonumber\\
  &&
  + \frac{1}{2} \lambda^{2} \left\{
    \nabla_{a}\stackrel{(2)}{(u_{b})}
    - 2 H_{ab}^{\;\;\;\;c}\left[h\right] \stackrel{(1)}{(u_{c})}
    - \left(
      H_{ab}^{\;\;\;\;c}\left[l\right]
      - 2 h^{cd} H_{abd}\left[h\right]
    \right) u_{c}
  \right\}
  \nonumber\\
  &&
  + O(\lambda^{3})
  \label{eq:kouchan-17.98-0}
  \\
  &=:&
  A_{ab}
  + \lambda \stackrel{(1)}{A_{ab}}
  + \frac{1}{2} \lambda^{2} \stackrel{(2)}{A_{ab}}
  + O(\lambda^{3}).
  \label{eq:kouchan-17.98}
\end{eqnarray}


Further, through Eqs.~(\ref{eq:linear-metric-decomp}) and
(\ref{eq:second-metric-decomp}), the first- and the
second-order perturbations of the connection
(\ref{eq:perturbative-connection-expansion}) are given by  
\begin{eqnarray}
  H_{ab}^{\;\;\;\;c}\left[h\right]
  &=&
  H_{ab}^{\;\;\;\;c}\left[{\cal H}\right]
  + \nabla_{a}\nabla_{b}X^{c}
  + R_{aeb}^{\;\;\;\;\;\;c}X^{e}
  ,
  \label{eq:kouchan-17.78}
  \\
  H_{ab}^{\;\;\;\;c}\left[l\right]
  - 2 h^{cd} H_{abd}\left[h\right]
  &=& 
  H_{ab}^{\;\;\;\;c}[{\cal L}]
  - 2 {\cal H}^{cd} H_{abd}\left[{\cal H}\right]
  \nonumber\\
  && 
  + 2 {\pounds}_{X} H_{ab}^{\;\;\;\;c}[h]
  + \nabla_{a}\nabla_{b}Y^{c}
  + Y^{e} R_{aeb}^{\;\;\;\;\;\;c}
  \nonumber\\
  && 
  - {\pounds}_{X}\left(
    \nabla_{a}\nabla_{b}X^{c}
    + R_{aeb}^{\;\;\;\;\;\;c}X^{e}
  \right).
  \label{eq:kouchan-17.79}
\end{eqnarray}
The components of the tensor 
$H_{ab}^{\;\;\;\;c}\left[{\cal H}\right]$ are summarized in
Appendix of KN2007 and the components of 
$H_{ab}^{\;\;\;\;c}\left[{\cal L}\right]$ can be derived from
the components of $H_{ab}^{\;\;\;\;c}\left[{\cal H}\right]$
by the replacements 
\begin{eqnarray}
  \stackrel{(1)}{\Phi} \rightarrow \stackrel{(2)}{\Phi}, 
  \quad
  \stackrel{(1)}{\nu_{i}} \rightarrow \stackrel{(2)}{\nu_{i}},
  \quad
  \stackrel{(1)}{\Psi} \rightarrow \stackrel{(2)}{\Psi},
  \quad
  \stackrel{(1)}{\chi_{ij}} \rightarrow \stackrel{(2)}{\chi_{ij}}.
\end{eqnarray}


Then, through the last equation in Eqs.~(\ref{eq:kouchan-16.13})
and (\ref{eq:kouchan-17.78}), the first-order perturbation of
the tensor $\bar{A}_{ab}$ in Eq.~(\ref{eq:kouchan-17.98}) is 
decomposed as
\begin{eqnarray}
  \stackrel{(1)}{A_{ab}}
  &=& 
  \stackrel{(1)\;\;\;\;}{{\cal A}_{ab}}
  + {\pounds}_{X}A_{ab}
  ,
  \label{eq:kouchan-17.105}
\end{eqnarray}
where
\begin{eqnarray}
  \stackrel{(1)\;\;\;\;}{{\cal A}_{ab}}
  &:=&
    \nabla_{a}\stackrel{(1)\;\;}{{\cal U}_{b}} 
  - u_{c} H_{ab}^{\;\;\;\;c}\left[{\cal H}\right]
  .
  \label{eq:kouchan-17.106}
\end{eqnarray}
$\stackrel{(1)}{{\cal A}_{ab}}$ in Eq.~(\ref{eq:kouchan-17.106})
is the gauge-invariant part of the first-order perturbation
$\stackrel{(1)}{A_{ab}}$ of the tensor $\bar{A}_{ab}$ and we
have verified that the first-order perturbation
$\stackrel{(1)}{A_{ab}}$ is decomposed as
Eq.~(\ref{eq:matter-gauge-inv-decomp-1.0}). 
Similarly, through the last equation in Eqs.~(\ref{eq:kouchan-16.13}),
(\ref{eq:kouchan-16.18}), (\ref{eq:kouchan-17.78}), and
(\ref{eq:kouchan-17.79}),  the second-order perturbation
$\stackrel{(2)}{A_{ab}}$ in Eq.~(\ref{eq:kouchan-17.98}) of
$\bar{A}_{ab}$ is decomposed as 
\begin{eqnarray}
  \stackrel{(2)}{A_{ab}}
  &=&
  \stackrel{(2)}{{\cal A}_{ab}}
  + {\pounds}_{Y}A_{ab}
  - {\pounds}_{X}^{2}A_{ab}
  + 2 {\pounds}_{X}\stackrel{(1)}{A_{ab}}
  ,
  \label{eq:kouchan-17.111}
\end{eqnarray}
where
\begin{eqnarray}
  \stackrel{(2)}{{\cal A}_{ab}}
  &:=& 
    \nabla_{a}\stackrel{(2)\;\;}{{\cal U}_{b}}
  - u_{c} H_{ba}^{\;\;\;\;c}[{\cal L}]
  + 2 u_{c} {\cal H}^{cd} H_{bad}\left[{\cal H}\right]
  - 2 H_{ba}^{\;\;\;\;c}\left[{\cal H}\right] \stackrel{(1)}{{\cal U}_{c}}
  .
  \label{eq:kouchan-17.112}
\end{eqnarray}
$\stackrel{(2)}{{\cal A}_{ab}}$ in Eq.~(\ref{eq:kouchan-17.111})
is the gauge-invariant part of the second-order perturbation
$\stackrel{(2)}{A_{ab}}$ of $\bar{A}_{ab}$ and we have verified
that the second-order perturbation $\stackrel{(2)}{A_{ab}}$ is
decomposed as Eq.~(\ref{eq:matter-gauge-inv-decomp-2.0}).


Now, we consider the components of gauge-invariant parts of the
perturbations of the tensor $\bar{A}_{ab}$.
First, we consider the components of the background value
$A_{ab}$.
Through Eq.~(\ref{eq:kouchan-17.378}), we obtain 
\begin{eqnarray}
  \label{eq:kouchan-17.406}
  A_{ab}
  = a {\cal H} \gamma_{ij} (dx^{i})_{a} (dx^{j})_{b}
  = a {\cal H} \gamma_{ab},
\end{eqnarray}
where ${\cal H} := \partial_{\eta}a/a$. 
The components of the gauge-invariant parts
$\stackrel{(1)}{{\cal A}_{ab}}$ of the first-order perturbation
of the tensor $\bar{A}_{ab}$ are summarized as 
\begin{eqnarray}
  \label{eq:kouchan-17.412}
  \stackrel{(1)}{{\cal A}_{\eta\eta}}
  &=&
  0
  , \\
  \label{eq:kouchan-17.413}
  \stackrel{(1)}{{\cal A}_{\eta i}}
  &=&
  a \left(
    \partial_{\eta}D_{i}\stackrel{(1)}{v} 
    + \partial_{\eta}\stackrel{(1)}{{\cal V}_{i}}
    + D_{i}\stackrel{(1)}{\Phi}
    + {\cal H} \stackrel{(1)}{\nu_{i}}
  \right)
  , \\
  \label{eq:kouchan-17.414}
  \stackrel{(1)}{{\cal A}_{i\eta}}
  &=&
  - a {\cal H} \left(
      D_{i}\stackrel{(1)}{v}
    + \stackrel{(1)}{{\cal V}_{i}}
    - \stackrel{(1)}{\nu_{i}}
  \right)
  , \\
  \label{eq:kouchan-17.415}
  \stackrel{(1)}{{\cal A}_{ij}}
  &=&
  a \left\{
        D_{i}D_{j}\stackrel{(1)}{v} 
    +   D_{i}\stackrel{(1)}{{\cal V}_{j}}
    -   D_{(i}\stackrel{(1)}{\nu_{j)}}
    -   \left( 
          {\cal H} \stackrel{(1)}{\Phi}
      + 2 {\cal H} \stackrel{(1)}{\Psi}
      +   \partial_{\eta}\stackrel{(1)}{\Psi}
    \right) \gamma_{ij}
  \right.
  \nonumber\\
  && \quad\quad
  \left.
    +   {\cal H} \stackrel{(1)}{\chi_{ij}}
    + \frac{1}{2} \partial_{\eta}\stackrel{(1)}{\chi_{ij}}
  \right\}
  .
\end{eqnarray}
Finally, the components of the gauge-invariant part
$\stackrel{(2)}{{\cal A}_{ab}}$ of the second-order perturbation
of the tensor $\bar{A}_{ab}$ are summarized as 
\begin{eqnarray}
  \stackrel{(2)}{{\cal A}_{\eta\eta}}
  &=&
  2 a
  \left\{
    \partial_{\eta}\left(
        D_{i}\stackrel{(1)}{v}
      + \stackrel{(1)}{{\cal V}_{i}}
    \right)
    + {\cal H} \stackrel{(1)}{\nu_{i}}
    + D_{i}\stackrel{(1)}{\Phi}
  \right\}
  \left(
           \stackrel{(1)}{\nu^{i}}
    -       D^{i}\stackrel{(1)}{v}
    - \stackrel{(1)}{{\cal V}^{i}}
  \right)
  \label{eq:kouchan-17.433}
  , \\
  \stackrel{(2)}{{\cal A}_{i\eta}}
  &=&
  a \left\{
    {\cal H} \left(
        \stackrel{(2)}{\nu_{i}}
      - D_{i}\stackrel{(2)}{v}
      - \stackrel{(2)}{{\cal V}_{i}}
    \right)
    - 2 {\cal H} \stackrel{(1)}{\Phi} \stackrel{(1)}{\nu_{i}}
  \right.
  \nonumber\\
  && \quad\quad
  \left.
    + \left(
        \stackrel{(1)}{\nu^{j}}
      - D^{j}\stackrel{(1)}{v} 
      - \stackrel{(1)}{{\cal V}^{j}}
    \right)
    \left(
        2 D_{i}D_{j}\stackrel{(1)}{v} 
      + 2 D_{i}\stackrel{(1)}{{\cal V}_{j}}
      - 2 \partial_{\eta}\stackrel{(1)}{\Psi} \gamma_{ij}
    \right.
  \right.
  \nonumber\\
  && \quad\quad\quad\quad\quad\quad\quad\quad\quad\quad\quad\quad\quad\quad
  \left.
    \left.
      +   \partial_{\eta}\stackrel{(1)}{\chi_{ij}}
      - 2 D_{(i}\stackrel{(1)}{\nu_{j)}}
    \right)
  \right\}
  \label{eq:kouchan-17.434}
  , \\
  \stackrel{(2)}{{\cal A}_{\eta i}}
  &=&
  a \left\{
    \left(
      \partial_{\eta}D_{i}\stackrel{(2)}{v}
      + \partial_{\eta}\stackrel{(2)}{{\cal V}_{i}}
      + D_{i}\stackrel{(2)}{\Phi}
      + {\cal H} \stackrel{(2)}{\nu_{i}}
    \right)
    - 2 \stackrel{(1)}{\Phi}
    \left(
      D_{i}\stackrel{(1)}{\Phi}
      + {\cal H} \stackrel{(1)}{\nu_{i}}
    \right)
  \right.
  \nonumber\\
  && \quad\quad
  \left.
    + \left(
      D^{j}\stackrel{(1)}{v} 
      + \stackrel{(1)}{{\cal V}^{j}}
      - \stackrel{(1)}{\nu^{j}}
    \right)
    \left(
      2 \partial_{\eta}\stackrel{(1)}{\Psi} \gamma_{ij}
      - \partial_{\eta}\stackrel{(1)}{\chi_{ij}}
      - 2 D_{[i}\stackrel{(1)}{\nu_{j]}}
    \right)
  \right\}
  \label{eq:kouchan-17.435}
  , \\
  \stackrel{(2)}{{\cal A}_{ij}}
  &=&
  a \left[
      D_{i}D_{j}\stackrel{(2)}{v}
    + D_{i}\stackrel{(2)}{{\cal V}_{j}}
    - D_{(i}\stackrel{(2)}{\nu_{j)}}
    - \left(
        2 {\cal H} \stackrel{(2)}{\Psi}
      +   {\cal H} \stackrel{(2)}{\Phi}
      +   \partial_{\eta}\stackrel{(2)}{\Psi}
    \right) \gamma_{ij}
  \right.
  \nonumber\\
  && \quad\quad
  \left.
    + \frac{1}{2} \partial_{\eta}\stackrel{(2)}{\chi_{ij}}
    +             {\cal H} \stackrel{(2)}{\chi_{ij}}
  \right.
  \nonumber\\
  && \quad\quad
  \left.
    + \stackrel{(1)}{\Phi}
    \left\{
        2 D_{(i}\stackrel{(1)}{\nu_{j)}}
      +   \left(
          4 {\cal H} \stackrel{(1)}{\Psi}
        + 3 {\cal H} \stackrel{(1)}{\Phi}
        + 2 \partial_{\eta}\stackrel{(1)}{\Psi}
      \right) \gamma_{ij}
      -   \partial_{\eta}\stackrel{(1)}{\chi_{ij}}
      - 2 {\cal H} \stackrel{(1)}{\chi_{ij}}
    \right\}
  \right.
  \nonumber\\
  && \quad\quad
  \left.
    + \left(
        D^{k} \stackrel{(1)}{v}
      + \stackrel{(1)}{{\cal V}^{k}}
      - \stackrel{(1)}{\nu^{k}}
    \right)
    \left(
        4 D_{(i}\stackrel{(1)}{\Psi} \gamma_{j)k}
      - 2 D_{k}\stackrel{(1)}{\Psi} \gamma_{ij}
    \right.
  \right.
  \nonumber\\
  && \quad\quad\quad\quad\quad\quad\quad\quad\quad\quad\quad\quad\quad\quad
  \left.
    \left.
      - 2 D_{(i} \stackrel{(1)}{\chi_{j)k}}
      +   D_{k} \stackrel{(1)}{\chi_{ij}}
    \right)
  \right.
  \nonumber\\
  && \quad\quad
  \left.
    - {\cal H} \gamma_{ij} \stackrel{(1)}{\nu_{k}} \stackrel{(1)}{\nu^{k}}
    + {\cal H} \gamma_{ij} \left(
        D^{k} \stackrel{(1)}{v}
      + \stackrel{(1)}{{\cal V}^{k}}
    \right)
    \left(
        D_{k} \stackrel{(1)}{v}
      + \stackrel{(1)}{{\cal V}_{k}}
    \right)
  \right]
  ,
  \label{eq:kouchan-17.436}
\end{eqnarray}
where we have used Eqs.~(\ref{eq:components-calHab}),
(\ref{eq:components-calLab}),
(\ref{eq:kouchan-17.378})--(\ref{eq:kouchan-17.399}), and the 
components of $H_{ab}^{\;\;\;\;c}\left[{\cal H}\right]$ and
$H_{ab}^{\;\;\;\;c}\left[{\cal L}\right]$ summarized in Appendix
in KN2007. 
The components $\stackrel{(1)}{{\cal A}_{\eta\eta}}$,
$\stackrel{(1)}{{\cal A}_{i\eta}}$, 
$\stackrel{(2)}{{\cal A}_{\eta\eta}}$, and 
$\stackrel{(2)}{{\cal A}_{i\eta}}$ are also derived from the
perturbations of the identity $\bar{u}^{b}\bar{A}_{ab}=0$.


\subsection{Perturbation of the acceleration 
  $\bar{a}_{b}=\bar{u}^{c}\bar{\nabla}_{c}\bar{u}_{b}$} 
\label{sec:acceleration-appendix}


Here, we consider the perturbations of the acceleration which is
defined by
\begin{eqnarray}
  \label{eq:acceleration-def-full}
  \bar{a}_{b}
  :=
  \bar{u}^{c}\bar{\nabla}_{c}\bar{u}_{b}
  =
  \bar{u}^{c}\bar{A}_{cb}.
\end{eqnarray}


To obtain the perturbations of the acceleration, it is
convenient to introduce perturbations of the contravariant
four-velocity $\bar{u}^{a}$ which is expanded as
\begin{eqnarray}
  \bar{u}^{a}
  &=&
  u^{a}
  + \lambda \stackrel{(1)}{(u^{a})} 
  + \frac{1}{2} \lambda^{2} \stackrel{(2)}{(u^{a})}
  + O(\lambda^{3}),
  \label{eq:contravariant-four-velocity-expansion}
\end{eqnarray}
To obtain the relation between the perturbations $\bar{u}^{a}$
and $\bar{u}_{a}$ of the fluid four-velocity, we have to
consider the perturbations of the inverse metric $\bar{g}^{ab}$ 
which is given by 
\begin{eqnarray}
  \bar{g}^{ab}
  &=&
  g^{ab}
  - \lambda h^{ab}
  + \frac{1}{2} \lambda^{2} \left( 2 h^{ac}h_{c}^{\;\;b} - l^{ab}\right)
  + O(\lambda^{3})
  \label{eq:inverse-metric-expansion}
  .
\end{eqnarray}
Through Eqs.~(\ref{eq:four-velocity-expansion}) and
(\ref{eq:inverse-metric-expansion}), we obtain the relation
between perturbations of $\bar{u}^{a}$ and $\bar{u}_{a}$ of each
order. 
Further, through Eqs.~(\ref{eq:linear-metric-decomp}),
(\ref{eq:second-metric-decomp}), (\ref{eq:kouchan-16.13}), and
(\ref{eq:kouchan-16.18}), the first- and the second-order
perturbations of $\bar{u}^{a}$ are also decomposed into the
gauge-invariant and the gauge-variant parts as
\begin{eqnarray}
  \label{eq:kouchan-16.40}
  \stackrel{(1)}{(u^{a})}
  &=&
  \stackrel{(1)}{{\cal U}^{a}} 
  - {\cal H}^{ab} u_{b}
  + {\pounds}_{X}u^{a}
  ,
  \\
  \stackrel{(2)}{(u^{a})}
  &=&
  \stackrel{(2)}{{\cal U}^{a}}
  - 2 {\cal H}^{ab} \stackrel{(1)}{{\cal U}_{b}}
  + 2 {\cal H}^{ac}{\cal H}_{cb} u^{b}
  -   {\cal L}^{ab} u_{b}
  \nonumber\\
  &&
  + 2 {\pounds}_{X} \left( 
    g^{ab} \stackrel{(1)}{(u_{b})} - h^{ab} u_{b}
  \right)
  + {\pounds}_{Y}u^{a}
  - {\pounds}_{X}^{2}u^{a},
  \label{eq:kouchan-16.41}
\end{eqnarray}
where we used
Eq.~(\ref{eq:first-second-pert-contravariant-gauge-inv-four-velocity-def}). 
These formulae are also given in KN2007 and we also note that
Eqs.~(\ref{eq:kouchan-16.40}) and (\ref{eq:kouchan-16.41}) have
the same form as Eqs.~(\ref{eq:matter-gauge-inv-decomp-1.0}) and
(\ref{eq:matter-gauge-inv-decomp-2.0}), respectively.


Through the expansion formulae (\ref{eq:kouchan-17.98}) and 
(\ref{eq:contravariant-four-velocity-expansion}), the
acceleration $\bar{a}_{b}$ is also expanded as 
\begin{eqnarray}
  \bar{a}_{b}
  &=&
  a_{b}
  + \lambda \stackrel{(1)}{(a_{b})}
  + \frac{1}{2} \lambda^{2} \stackrel{(2)}{(a_{b})}
  + O(\lambda^{3})
  \label{eq:acceleration-expansion}
\end{eqnarray}
and we easily see that each order perturbation of the
acceleration $\bar{a}_{b}$ is given by
\begin{eqnarray}
  \label{eq:kouchan-17.257}
  a_{b}
  &:=&
  u^{a}A_{ab} = u^{a}\nabla_{a}u_{b}
  , \\
  \label{eq:kouchan-17.258}
  \stackrel{(1)}{(a_{b})}
  &:=&
    u^{a}\stackrel{(1)\;\;\;\;}{A_{ab}}
  + \stackrel{(1)}{(u^{a})} A_{ab}
  , \\
  \label{eq:kouchan-17.259}
  \stackrel{(2)}{(a_{b})}
  &:=&
    u^{a}\stackrel{(2)\;\;\;\;}{A_{ab}}
  + 2 \stackrel{(1)}{(u^{a})} \stackrel{(1)\;\;\;\;}{A_{ab}}
  + \stackrel{(2)}{(u^{a})} A_{ab}
  .
\end{eqnarray}
Substituting Eqs.~(\ref{eq:kouchan-17.105}) and
(\ref{eq:kouchan-16.40}) into Eq.~(\ref{eq:kouchan-17.258}), 
we easily see that the first-order perturbation of the
acceleration is decomposed into gauge-invariant and
gauge-variant parts as 
\begin{eqnarray}
  \stackrel{(1)}{(a_{b})}
  &=&
  \stackrel{(1)}{{\cal A}}_{b} + {\pounds}_{X}a_{b}.
  \label{eq:first-acceleration-gauge-inv}
\end{eqnarray}
Further, substituting Eqs.~(\ref{eq:kouchan-17.105}),
(\ref{eq:kouchan-17.111}), (\ref{eq:kouchan-16.40}), and
(\ref{eq:kouchan-16.41}) into Eq.~(\ref{eq:kouchan-17.259}), 
we easily see that the second-order perturbation of the
acceleration is decomposed into gauge-invariant and
gauge-variant parts as 
\begin{eqnarray}
  \stackrel{(2)}{(a_{b})}
  &=&
  \stackrel{(2)}{{\cal A}}_{b}
  + 2 {\pounds}_{X} \stackrel{(1)}{(a_{b})}
  + \left\{
       {\pounds}_{Y}
    -  {\pounds}_{X}^{2}
  \right\}a_{b}
  \label{eq:second-acceleration-gauge-inv}
  .
\end{eqnarray}
Here, we have defined the gauge-invariant parts
$\stackrel{(1)}{{\cal A}}_{b}$ and 
$\stackrel{(2)}{{\cal A}}_{b}$ of the first-
and the second-order perturbations of the accelerations by 
\begin{eqnarray}
  \label{eq:kouchan-17.262}
  \stackrel{(1)}{{\cal A}_{b}}
  &:=&
    u^{a} \stackrel{(1)\;\;\;\;}{{\cal A}_{ab}}
  + \stackrel{(1)}{{\cal U}^{a}} A_{ab}
  - {\cal H}^{ac} u_{c} A_{ab}
  ,
  \\
  \stackrel{(2)}{{\cal A}_{b}}
  &:=&
      u^{a} \stackrel{(2)}{{\cal A}_{ab}}
  +   \stackrel{(2)}{{\cal U}^{a}} A_{ab}
  -   {\cal L}^{ac} u_{c} A_{ab}
  - 2 {\cal H}^{ac} \stackrel{(1)}{{\cal U}_{c}} A_{ab}
  + 2 {\cal H}^{ac}{\cal H}_{cd} u^{d} A_{ab}
  \nonumber\\
  &&
  + 2 \stackrel{(1)}{{\cal A}_{ab}} \stackrel{(1)}{{\cal U}^{a}}
  - 2 \stackrel{(1)}{{\cal A}_{ab}} {\cal H}^{ac} u_{c}
  .
  \label{eq:kouchan-17.265}
\end{eqnarray}
We note again that Eqs.~(\ref{eq:first-acceleration-gauge-inv})
and (\ref{eq:second-acceleration-gauge-inv}) have the same forms
as Eqs.~(\ref{eq:matter-gauge-inv-decomp-1.0}) and
(\ref{eq:matter-gauge-inv-decomp-2.0}), respectively.


Now, we consider the explicit components of the gauge-invariant
parts of the perturbations of the acceleration $\bar{a}_{b}$
associated with the fluid four-velocity $\bar{u}_{a}$ through
Eqs.(\ref{eq:kouchan-17.262}) and (\ref{eq:kouchan-17.265}). 
First, we consider the components of the background value of the
acceleration associated with the four-velocity
\begin{eqnarray}
  a_{b}
  = u^{a}A_{ab}
  = a {\cal H} u^{a} \gamma_{ab}
  = 0,
  \label{eq:background-accerelation}
\end{eqnarray}
where we have used Eq.~(\ref{eq:kouchan-17.406}).
Next, though Eqs.~(\ref{eq:components-calHab}), 
(\ref{eq:kouchan-17.378}), (\ref{eq:kouchan-17.380}),
(\ref{eq:kouchan-17.406})--(\ref{eq:kouchan-17.415}),
and (\ref{eq:kouchan-17.262}), the components of the
gauge-invariant part $\stackrel{(1)}{{\cal A}_{a}}$ of the
first-order perturbation of the acceleration are summarized as
follows: 
\begin{eqnarray}
  \label{eq:kouchan-17.438}
  \stackrel{(1)}{{\cal A}_{\eta}} &=& 0, \\
  \label{eq:kouchan-17.439}
  \stackrel{(1)}{{\cal A}_{i}}
  &=&
    D_{i}\stackrel{(1)}{\Phi}
  + \left(
    \partial_{\eta} + {\cal H}
  \right) \left(
    D_{i}\stackrel{(1)}{v} 
    + \stackrel{(1)}{{\cal V}_{i}}
  \right)
  .
\end{eqnarray}
Finally, we consider the components of the gauge-invariant part
(\ref{eq:kouchan-17.265}) of the second-order perturbation of
the acceleration. 
By the direct calculations through
Eqs.~(\ref{eq:components-calHab}), (\ref{eq:components-calLab}),
(\ref{eq:first-second-pert-contravariant-gauge-inv-four-velocity-def}),
(\ref{eq:kouchan-17.378})--(\ref{eq:kouchan-17.399}),
(\ref{eq:kouchan-17.406})--(\ref{eq:kouchan-17.415}), and 
(\ref{eq:kouchan-17.433})--(\ref{eq:kouchan-17.436}), the
components of the gauge-invariant part
$\stackrel{(2)}{{\cal A}_{a}}$ of the acceleration are given by 
\begin{eqnarray}
  \stackrel{(2)}{{\cal A}_{\eta}}
  &=&
  2 \stackrel{(1)}{{\cal A}_{i}}
  \left(
           \stackrel{(1)}{\nu^{i}}
    -       D^{i}\stackrel{(1)}{v}
    - \stackrel{(1)}{{\cal V}^{i}}
  \right)
  ,
  \label{eq:kouchan-17.446}
  \\
  \stackrel{(2)}{{\cal A}_{i}}
  &=&
  \left(
    \partial_{\eta} + {\cal H}
  \right)
  \left(
    D_{i}\stackrel{(2)}{v} + \stackrel{(2)}{{\cal V}_{i}}
  \right)
  + D_{i}\stackrel{(2)}{\Phi}
  \nonumber\\
  &&
  - 2 \stackrel{(1)}{\Phi}
  \left\{
      2 D_{i}\stackrel{(1)}{\Phi}
    +   \left(
      \partial_{\eta} + {\cal H}
    \right) \left(
      D_{i}\stackrel{(1)}{v} + \stackrel{(1)}{{\cal V}_{i}}
    \right)
  \right\}
  \nonumber\\
  &&
  + 2 \left(
      \stackrel{(1)}{\nu^{j}}
    - D^{j}\stackrel{(1)}{v} 
    - \stackrel{(1)}{{\cal V}^{j}}
  \right)
  \left(
      D_{i}\stackrel{(1)}{\nu_{j}}
    - D_{j}D_{i}\stackrel{(1)}{v} 
    - D_{j}\stackrel{(1)}{{\cal V}_{i}}
  \right)
  .
  \label{eq:kouchan-17.447}
\end{eqnarray}
We can easily check the components
(\ref{eq:kouchan-17.438})--(\ref{eq:kouchan-17.447}) are
consistent with the perturbation of the identity
$\bar{u}^{b}\bar{a}_{b}=0$.


\subsection{Perturbation of the tensor 
  $\bar{B}_{ab}:=\bar{A}_{ab}+\bar{u}_{a}\bar{a}_{b}$}


Here, we consider the perturbations of the tensor field
$\bar{B}_{ab}$ defined by Eq.~(\ref{eq:barBab-definition}),
which is also written by 
\begin{eqnarray}
  \bar{B}_{ab} = \bar{A}_{ab} + \bar{u}_{a}\bar{a}_{b}
  \label{eq:kouchan-17.80}
\end{eqnarray}
through the tensor $\bar{A}_{ab}$ defined by
Eq.~(\ref{eq:kouchan-17.93}). 
The tensor $\bar{B}_{ab}$ is also decomposed into the trace part
and the traceless part.
Further, the traceless part of the tensor $\bar{B}_{ab}$ is also
classified into the symmetric part and the antisymmetric part:
\begin{eqnarray}
  \label{eq:kouchan-17.86}
  \bar{B}_{ab} &=& \frac{1}{3} \bar{q}_{ab} \bar{\theta} 
  + \bar{\sigma}_{ab} + \bar{\omega}_{ab}, \\
  \label{eq:kouchan-17.89}
  \bar{\theta} &:=& \bar{q}^{ab} \bar{B}_{ab} = \bar{\nabla}_{c}\bar{u}^{c}, \\
  \label{eq:kouchan-17.91}
  \bar{\sigma}_{ab} &:=&
  \bar{B}_{(ab)} - \frac{1}{3} \bar{q}_{ab} \bar{\theta}, \\
  \label{eq:kouchan-17.92}
  \bar{\omega}_{ab} &:=& \bar{B}_{[ab]}.
\end{eqnarray}
$\bar{\theta}$, $\bar{\sigma}_{ab}$, and $\bar{\omega}_{ab}$ are
called the expansion, the shear, and the rotation associated
with the four-velocity $\bar{u}_{a}$, respectively.


The perturbative expansions of the tensor $\bar{B}_{ab}$ is
given by
\begin{eqnarray}
  \bar{B}_{ab}
  &=&
  B_{ab}
  + \lambda \stackrel{(1)}{B_{ab}}
  + \frac{1}{2} \lambda^{2} \stackrel{(2)}{B_{ab}}
  + O(\lambda^{3}).
  \label{eq:kouchan-17.701}
\end{eqnarray}
On the other hand, the perturbative expansion of the tensor
$\bar{B}_{ab}$ is also given by the direct expansion of
Eq.~(\ref{eq:kouchan-17.80}).
Comparing these perturbative expansions, we obtain
\begin{eqnarray}
  B_{ab}
  &=&
  A_{ab} + u_{a} a_{b}
  \label{eq:kouchan-17.296}
  , \\
  \stackrel{(1)}{B_{ab}}
  &=&
    \stackrel{(1)}{A_{ab}}
  + u_{a} \stackrel{(1)}{{a}_{b}}
  + \stackrel{(1)}{(u_{a})} a_{b}
  \label{eq:kouchan-17.297}
  , \\
  \stackrel{(2)}{B_{ab}} 
  &=&
    \stackrel{(2)}{A_{ab}}
  + u_{a} \stackrel{(2)}{(a_{b})}
  + \stackrel{(2)}{(u_{a})} a_{b}
  + 2 \stackrel{(1)}{(u_{a})} \stackrel{(1)}{(a_{b})}
  .
  \label{eq:kouchan-17.298}
\end{eqnarray}
Further, substituting Eqs.~(\ref{eq:kouchan-17.105}),
(\ref{eq:first-acceleration-gauge-inv}), and
(\ref{eq:kouchan-16.18}) into Eq.~(\ref{eq:kouchan-17.297}), we
can decompose the first-order perturbation
$\stackrel{(1)}{B_{ab}}$ of $\bar{B}_{ab}$ into gauge-invariant
and gauge-variant parts as
\begin{eqnarray}
  \stackrel{(1)}{B_{ab}} 
  &=&
  \stackrel{(1)}{{\cal B}_{ab}} 
  +
  {\pounds}_{X}B_{ab},
  \label{eq:first-order-calBab-def}
\end{eqnarray}
where the gauge-invariant part $\stackrel{(1)}{{\cal B}_{ab}}$
is defined by 
\begin{eqnarray}
  \stackrel{(1)}{{\cal B}_{ab}}
  &:=&
    \stackrel{(1)}{{\cal A}_{ab}}
  + u_{a} \stackrel{(1)}{{\cal A}_{b}}
  + \stackrel{(1)}{{\cal U}_{a}} a_{b}
  .
  \label{eq:kouchan-17.301}
\end{eqnarray}
Note that Eq.~(\ref{eq:first-order-calBab-def}) has the same
form as the decomposition formula
(\ref{eq:matter-gauge-inv-decomp-1.0}). 
Further, substituting Eqs.~(\ref{eq:kouchan-16.13}),
(\ref{eq:kouchan-16.18}), (\ref{eq:kouchan-17.111}),
(\ref{eq:first-acceleration-gauge-inv}), 
(\ref{eq:second-acceleration-gauge-inv}), the last equation in
Eqs.~(\ref{eq:kouchan-16.13}) and (\ref{eq:kouchan-16.18}) into
Eq.~(\ref{eq:kouchan-17.298}), the second-order perturbation
$\stackrel{(2)}{B_{ab}}$ of the tensor $\bar{B}_{ab}$ is 
decomposed as 
\begin{eqnarray}
  \stackrel{(2)}{B_{ab}} 
  &=&
  \stackrel{(2)}{{\cal B}_{ab}} 
  +
  2 {\pounds}_{X}\stackrel{(1)}{B_{ab}} 
  + \left(
    {\pounds}_{Y}
    - {\pounds}_{X}^{2}
  \right)B_{ab}
  \label{eq:second-order-calBab-def}
  ,
\end{eqnarray}
where the gauge-invariant part $\stackrel{(2)}{{\cal B}_{ab}}$
is defined as
\begin{eqnarray}
  \label{eq:kouchan-17.306}
  \stackrel{(2)\;\;\;\;}{{\cal B}_{ab}}
  &:=&
      \stackrel{(2)\;\;\;\;}{{\cal A}_{ab}}
  +   u_{a} \stackrel{(2)}{{\cal A}_{b}}
  +   \stackrel{(2)\;\;}{{\cal U}_{a}} a_{b}
  + 2 \stackrel{(1)}{{\cal U}_{a}} \stackrel{(1)}{{\cal A}_{b}}
  .
\end{eqnarray}
We also note that Eq.~(\ref{eq:second-order-calBab-def}) has the
same form as Eq.~(\ref{eq:matter-gauge-inv-decomp-2.0}).


Now, we consider the components of the gauge-invariant parts
$\stackrel{(1)}{{\cal B}_{ab}}$ and 
$\stackrel{(2)}{{\cal B}_{ab}}$ of the first- and the
second-order perturbations of the tensor field $\bar{B}_{ab}$.
First, we note that the background value of the tensor field
$\bar{B}_{ab}$ is given by Eq.~(\ref{eq:kouchan-17.406}),
(\ref{eq:background-accerelation}), and
(\ref{eq:kouchan-17.296}):
\begin{eqnarray}
  B_{ab} = a {\cal H} \gamma_{ab}.
  \label{eq:kouchan-17.451}
\end{eqnarray}
Through Eqs.~(\ref{eq:kouchan-17.378}),
(\ref{eq:kouchan-17.380}),
(\ref{eq:background-accerelation})--(\ref{eq:kouchan-17.439}), 
and (\ref{eq:kouchan-17.301}), the components of the
gauge-invariant part $\stackrel{(1)}{{\cal B}_{ab}}$ of the
first-order perturbation of the tensor $\bar{B}_{ab}$ are
summarized as  
\begin{eqnarray}
  \stackrel{(1)}{{\cal B}_{\eta\eta}}
  &=&
  0
  \label{eq:kouchan-17.459}
  , \\
  \stackrel{(1)}{{\cal B}_{\eta i}}
  &=&
  a {\cal H} \left(
      \stackrel{(1)}{\nu_{i}}
    - D_{i}\stackrel{(1)}{v}
    - \stackrel{(1)}{{\cal V}_{i}}
  \right)
  =
  \stackrel{(1)}{{\cal B}_{i\eta}}
  \label{eq:kouchan-17.460}
  , \\
  \stackrel{(1)}{{\cal B}_{ij}}
  &=&
  a \left\{
    D_{i}\left(
      D_{j}\stackrel{(1)}{v} + \stackrel{(1)}{{\cal V}_{j}}
    \right)
    -   D_{(i}\stackrel{(1)}{\nu_{j)}}
    -   \left(
          {\cal H} \stackrel{(1)}{\Phi}
      + 2 {\cal H} \stackrel{(1)}{\Psi}
      +   \partial_{\eta}\stackrel{(1)}{\Psi}
    \right) \gamma_{ij}
  \right.
  \nonumber\\
  && \quad\quad
  \left.
    + \frac{1}{2} \left(
      \partial_{\eta} + 2 {\cal H}
    \right) \stackrel{(1)}{\chi_{ij}}
  \right\}
  \label{eq:kouchan-17.462}
  .
\end{eqnarray}
Finally, through
Eqs.~(\ref{eq:kouchan-17.378})--(\ref{eq:kouchan-17.398}),
(\ref{eq:kouchan-17.433})--(\ref{eq:kouchan-17.436}),
(\ref{eq:kouchan-17.438})--(\ref{eq:kouchan-17.447}),
(\ref{eq:kouchan-17.306}), the components of the gauge-invariant
part $\stackrel{(2)}{{\cal B}_{ab}}$ of the second-order
perturbation of the tensor $\bar{B}_{ab}$ are summarized as
follows: 
\begin{eqnarray}
  \stackrel{(2)}{{\cal B}_{\eta\eta}}
  &=&
  2 a {\cal H}
  \left(
      D_{i}\stackrel{(1)}{v}
    + \stackrel{(1)}{{\cal V}_{i}}
    - \stackrel{(1)}{\nu_{i}}
  \right)
  \left(
      D^{i}\stackrel{(1)}{v}
    + \stackrel{(1)}{{\cal V}^{i}}
    - \stackrel{(1)}{\nu^{i}}
  \right)
  \label{eq:kouchan-17.468}
  , \\
  \stackrel{(2)}{{\cal B}_{\eta i}}
  &=&
  a \left[
    2 \left(
        \stackrel{(1)}{\nu^{j}}
      - D^{j}\stackrel{(1)}{v} 
      - \stackrel{(1)}{{\cal V}^{j}}
    \right)
    \left\{
        D_{j}\left(
          D_{i}\stackrel{(1)}{v} 
        + \stackrel{(1)}{{\cal V}_{i}}
      \right)
      - D_{(i}\stackrel{(1)}{\nu_{j)}}
      - \partial_{\eta}\stackrel{(1)}{\Psi} \gamma_{ij}
      + \frac{1}{2} \partial_{\eta}\stackrel{(1)}{\chi_{ij}}
    \right\}
  \right.
  \nonumber\\
  && \quad\quad
  \left.
    +   {\cal H} \left(
      \stackrel{(2)}{\nu_{i}}
      - D_{i}\stackrel{(2)}{v}
      - \stackrel{(2)}{{\cal V}_{i}}
    \right)
    - 2 {\cal H} \stackrel{(1)}{\Phi} \stackrel{(1)}{\nu_{i}}
  \right]
  \label{eq:kouchan-17.469}
  , \\
  \stackrel{(2)}{{\cal B}_{i\eta}}
  &=&
  a \left[
    2 \left(
        \stackrel{(1)}{\nu^{j}}
      - D^{j}\stackrel{(1)}{v} 
      - \stackrel{(1)}{{\cal V}^{j}}
    \right)
    \left\{
        D_{i}\left(
          D_{j}\stackrel{(1)}{v} 
        + \stackrel{(1)}{{\cal V}_{j}}
      \right)
      - D_{(i}\stackrel{(1)}{\nu_{j)}}
      - \partial_{\eta}\stackrel{(1)}{\Psi} \gamma_{ij}
      + \frac{1}{2} \partial_{\eta}\stackrel{(1)}{\chi_{ij}}
    \right\}
  \right.
  \nonumber\\
  && \quad\quad
  \left.
    + {\cal H} \left(
      \stackrel{(2)}{\nu_{i}}
      - D_{i}\stackrel{(2)}{v}
      - \stackrel{(2)}{{\cal V}_{i}}
    \right)
    - 2 {\cal H} \stackrel{(1)}{\Phi} \stackrel{(1)}{\nu_{i}}
  \right]
  \label{eq:kouchan-17.470}
  , \\
  \stackrel{(2)}{{\cal B}_{ij}}
  &=&
  a \left[
      D_{i}\left(
      D_{j}\stackrel{(2)}{v} + \stackrel{(2)}{{\cal V}_{j}}
    \right)
    -   D_{(i}\stackrel{(2)}{\nu_{j)}}
    -   \left\{
        {\cal H} \left(
          2 \stackrel{(2)}{\Psi}
        +   \stackrel{(2)}{\Phi}
      \right)
      +   \partial_{\eta}\stackrel{(2)}{\Psi}
    \right\} \gamma_{ij}
  \right.
  \nonumber\\
  && \quad\quad
  \left.
    + \frac{1}{2} \left(
      \partial_{\eta} + 2 {\cal H}
    \right) \stackrel{(2)}{\chi_{ij}}
  \right.
  \nonumber\\
  && \quad\quad
  \left.
    + \stackrel{(1)}{\Phi}
    \left\{
        2 D_{(i}\stackrel{(1)}{\nu_{j)}}
      + \left(
          2 \partial_{\eta}\stackrel{(1)}{\Psi}
        + {\cal H} \left(
            4 \stackrel{(1)}{\Psi}
          + 3 \stackrel{(1)}{\Phi}
        \right)
      \right) \gamma_{ij}
      -   \left(
        \partial_{\eta} + 2 {\cal H}
      \right) \stackrel{(1)}{\chi_{ij}}
    \right\}
  \right.
  \nonumber\\
  && \quad\quad
  \left.
    + 2 \left(
        D_{i}\stackrel{(1)}{v}
      + \stackrel{(1)}{{\cal V}_{i}}
    \right)
    \left\{
        D_{j}\stackrel{(1)}{\Phi}
      + \left(
        \partial_{\eta} + {\cal H}
      \right) \left(
          D_{j}\stackrel{(1)}{v}
        + \stackrel{(1)}{{\cal V}_{j}}
      \right)
    \right\}
  \right.
  \nonumber\\
  && \quad\quad
  \left.
    + \left(
        \stackrel{(1)}{\nu^{k}}
      - D^{k} \stackrel{(1)}{v}
      - \stackrel{(1)}{{\cal V}^{k}}
    \right)
    \left(
        2 D_{k}\stackrel{(1)}{\Psi} \gamma_{ij}
      - 4 D_{(i}\stackrel{(1)}{\Psi} \gamma_{j)k}
      + 2 D_{(i} \stackrel{(1)}{\chi_{j)k}}
      -   D_{k} \stackrel{(1)}{\chi_{ij}}
    \right)
  \right.
  \nonumber\\
  && \quad\quad
  \left.
    - {\cal H} \gamma_{ij} \stackrel{(1)}{\nu_{k}} \stackrel{(1)}{\nu^{k}}
    + {\cal H} \gamma_{ij} \left(
        D^{k} \stackrel{(1)}{v}
      + \stackrel{(1)}{{\cal V}^{k}}
    \right)
    \left(
        D_{k} \stackrel{(1)}{v}
      + \stackrel{(1)}{{\cal V}_{k}}
    \right)
  \right]
  \label{eq:kouchan-17.471}
  .
\end{eqnarray}


We also note that the components 
$\stackrel{(1)}{{\cal B}_{\eta\eta}}$, 
$\stackrel{(1)}{{\cal B}_{i\eta}}$,
$\stackrel{(1)}{{\cal B}_{\eta i}}$ for the first-order
perturbation $\stackrel{(1)}{{\cal B}_{ab}}$ and the components     
$\stackrel{(2)}{{\cal B}_{\eta\eta}}$, 
$\stackrel{(2)}{{\cal B}_{i\eta}}$,
$\stackrel{(2)}{{\cal B}_{\eta i}}$ for the second-order
perturbation $\stackrel{(2)}{{\cal B}_{ab}}$ are also derived
from the perturbations of the properties 
\begin{eqnarray}
  \label{eq:kouchan-17.82}
  \bar{u}^{a} \bar{B}_{ab} = \bar{u}^{a} \bar{B}_{ba} = 0.
\end{eqnarray}


\subsubsection{Expansion $\bar{\theta}$}


Now, we consider the perturbation of the expansion
$\bar{\theta}$ defined by Eq.~(\ref{eq:kouchan-17.89}).
Through Eqs.~(\ref{eq:inverse-metric-expansion}) and
(\ref{eq:kouchan-17.701}), the expansion $\bar{\theta}$ is
expanded as 
\begin{eqnarray}
  \bar{\theta}
  &=&
  \theta
  + \lambda \stackrel{(1)}{\theta}
  + \frac{1}{2} \lambda^{2} \stackrel{(2)}{\theta}
  + O(\lambda^{3}),
  \label{eq:theta-expansion}
\end{eqnarray}
where each order perturbations of $\bar{\theta}$ is given by 
\begin{eqnarray}
  \label{eq:kouchan-17.333}
  \theta &=& g^{ab} B_{ab} = g^{ab}A_{ab} = \nabla_{a}u^{a}
  , \\
  \label{eq:kouchan-17.334}
  \stackrel{(1)}{\theta}
  &=&
  g^{ab} \stackrel{(1)}{B_{ab}} - B_{ab} h^{ab}
  , \\
  \label{eq:kouchan-17.335}
  \stackrel{(2)}{\theta}
  &=&
  g^{ab} \stackrel{(2)}{B_{ab}}
  - 2 \stackrel{(1)}{B_{ab}} h^{ab}
  + B_{ab} ( 2 h^{ae}h_{e}^{\;\;b} - l^{ab})
  .
\end{eqnarray}


Through Eqs.~(\ref{eq:linear-metric-decomp}) and
(\ref{eq:first-order-calBab-def}), the first-order perturbation
$\stackrel{(1)}{\theta}$ is decomposed as
\begin{eqnarray}
  \stackrel{(1)}{\theta}
  &=:&
  \stackrel{(1)}{\Theta} + {\pounds}_{X}\theta
  \label{eq:first-theta-gauge-inv}
  ,
\end{eqnarray}
where the gauge-invariant part $\stackrel{(1)}{\Theta}$ of
$\stackrel{(1)}{\theta}$ is given by 
\begin{eqnarray}
  \stackrel{(1)}{\Theta}
  &=:&
    g^{ab} \stackrel{(1)}{{\cal B}_{ab}}
  - B_{ab} {\cal H}^{ab}
  .
  \label{eq:kouchan-17.338}
\end{eqnarray}
Similarly, through Eqs.~(\ref{eq:linear-metric-decomp}),
(\ref{eq:second-metric-decomp}),
(\ref{eq:first-order-calBab-def}), and
(\ref{eq:second-order-calBab-def}), the second-order
perturbation $\stackrel{(2)}{\theta}$ is also decomposed as
\begin{eqnarray}
  \stackrel{(2)}{\theta}
  &:=&
  \stackrel{(2)}{\Theta} 
  + 2 {\pounds}_{X} \stackrel{(1)}{\theta}
  + \left\{
       {\pounds}_{Y}
    -  {\pounds}_{X}^{2}
  \right\} \theta
  \label{eq:second-theta-gauge-inv}
  ,
\end{eqnarray}
where the gauge-invariant part $\stackrel{(2)}{\Theta}$ of
$\stackrel{(2)}{\theta}$ is given by
\begin{eqnarray}
  \stackrel{(2)}{\Theta}
  &:=&
      g^{ab} \stackrel{(2)}{{\cal B}_{ab}}
  -   B_{ab} {\cal L}^{ab}
  - 2 \stackrel{(1)}{{\cal B}_{ab}} {\cal H}^{ab}
  + 2 B_{ab} {\cal H}^{ae} {\cal H}_{e}^{\;\;b}
  .
  \label{eq:kouchan-17.342}
\end{eqnarray}
Here again, we note that the decompositions
(\ref{eq:first-theta-gauge-inv}) and
(\ref{eq:second-theta-gauge-inv}) of the perturbations of the
expansion $\bar{\theta}$ into the gauge-invariant and the
gauge-variant parts have the same forms as
Eqs.~(\ref{eq:matter-gauge-inv-decomp-1.0}) and
(\ref{eq:matter-gauge-inv-decomp-2.0}), respectively.


Now, we derive the explicit expression of the perturbations of
the expansion $\bar{\theta}$.
First, we consider the background value of the expansion
(\ref{eq:kouchan-17.333}).
From Eq.~(\ref{eq:kouchan-17.451}), the background value
$\theta$ of $\bar{\theta}$ is given by
\begin{eqnarray}
  \theta
  = 3 \frac{1}{a} {\cal H}
  = 3 \frac{\partial_{\eta}a}{a^{2}}
  = 3 \frac{1}{a}\frac{da}{d\tau}
  = 3 H
  ,
  \label{eq:kouchan-17.498}
\end{eqnarray}
where $H$ is the Hubble parameter of the background universe and
$d\tau = a d\eta$.
Second, through Eqs.~(\ref{eq:components-calHab}), and
(\ref{eq:kouchan-17.451})--(\ref{eq:kouchan-17.462}), the
gauge-invariant part (\ref{eq:kouchan-17.338}) of the
first-order perturbation $\stackrel{(1)}{\theta}$ is given by
\begin{eqnarray}
  \stackrel{(1)}{\Theta}
  &=&
  \frac{1}{a} \left(
                  \Delta\stackrel{(1)}{v} 
    -          3  {\cal H} \stackrel{(1)}{\Phi}
    -          3  \partial_{\eta}\stackrel{(1)}{\Psi}
  \right)
  .
  \label{eq:kouchan-17.501}
\end{eqnarray}
Finally, through
Eqs.~(\ref{eq:components-calHab}), (\ref{eq:components-calLab}),
and (\ref{eq:kouchan-17.451})--(\ref{eq:kouchan-17.471}), the
gauge-invariant part (\ref{eq:kouchan-17.342}) of the
second-order perturbation of the expansion is given by 
\begin{eqnarray}
  \stackrel{(2)}{\Theta}
  &=&
  \frac{1}{a} \left[
         \Delta \stackrel{(2)}{v}
    -  3 \partial_{\eta}\stackrel{(2)}{\Psi}
    -  3 {\cal H} \stackrel{(2)}{\Phi}
    +  6 \partial_{\eta}\stackrel{(1)}{\Psi} \left(
      \stackrel{(1)}{\Phi} - 2 \stackrel{(1)}{\Psi}
    \right)
  \right.
  \nonumber\\
  && \quad
  \left.
    +  4 \stackrel{(1)}{\Psi} \Delta \stackrel{(1)}{v}
    +  9 {\cal H} \left(\stackrel{(1)}{\Phi}\right)^{2}
    +  2 \stackrel{(1)}{\nu^{k}} D_{k}\stackrel{(1)}{\Psi}
    -  3 {\cal H} \stackrel{(1)}{\nu^{k}} \stackrel{(1)}{\nu_{k}}
  \right.
  \nonumber\\
  && \quad
  \left.
    + \left(
        D^{k}\stackrel{(1)}{v}
      + \stackrel{(1)}{{\cal V}^{k}}
    \right)
    \left\{
        2 D_{k}\left(\stackrel{(1)}{\Phi} - \stackrel{(1)}{\Psi}\right)
      + \left(
        2 \partial_{\eta} + 3 {\cal H}
      \right)
      \left(
        D_{k}\stackrel{(1)}{v} + \stackrel{(1)}{{\cal V}_{k}}
      \right)
    \right\}
  \right.
  \nonumber\\
  && \quad
  \left.
    + \stackrel{(1)}{\chi^{ik}}
    \left(
        2 D_{i}\stackrel{(1)}{\nu_{k}}
      - 2 D_{i}D_{k}\stackrel{(1)}{v} 
      - 2 D_{i}\stackrel{(1)}{{\cal V}_{k}}
      -   \partial_{\eta}\stackrel{(1)}{\chi_{ik}}
    \right)
  \right]
  .
  \label{eq:kouchan-17.505}
\end{eqnarray}


\subsubsection{Shear $\bar{\sigma}_{ab}$}


The perturbations of the shear tensor $\bar{\sigma}_{ab}$
defined by Eq.~(\ref{eq:kouchan-17.91}) are given as follows.
Through Eqs.~(\ref{eq:three-metric-expansion}),
(\ref{eq:kouchan-17.701}), and (\ref{eq:theta-expansion}), the
perturbative expansion of the shear tensor $\bar{\sigma}_{ab}$
is given by
\begin{eqnarray}
  \bar{\sigma}_{ab}
  =:
  \sigma_{ab}
  + \lambda \stackrel{(1)}{(\sigma_{ab})}
  + \frac{1}{2} \lambda^{2} \stackrel{(2)}{(\sigma_{ab})}
  + O(\lambda^{3})
  ,
  \label{eq:kouchan-17.355}
\end{eqnarray}
where
\begin{eqnarray}
  \label{eq:kouchan-17.356}
  \sigma_{ab}
  &:=&
  B_{(ab)} - \frac{1}{3} q_{ab} \theta
  , \\
  \label{eq:kouchan-17.357}
  \stackrel{(1)}{(\sigma_{ab})}
  &:=&
  \stackrel{(1)}{B_{(ab)}}
  - \frac{1}{3} \stackrel{(1)}{(q_{ab})} \theta
  - \frac{1}{3} q_{ab} \stackrel{(1)}{\theta}
  , \\
  \label{eq:kouchan-17.358}
  \stackrel{(2)}{(\sigma_{ab})}
  &:=&
  \stackrel{(2)}{B_{(ab)}}
  - \frac{1}{3} q_{ab} \stackrel{(2)}{\theta}
  - \frac{2}{3} \stackrel{(1)}{(q_{ab})} \stackrel{(1)}{\theta}
  - \frac{1}{3} \stackrel{(2)}{(q_{ab})} \theta
  .
\end{eqnarray}


Through Eqs.~(\ref{eq:first-three-metric-gauge-inv}),
(\ref{eq:first-order-calBab-def}), and 
(\ref{eq:first-theta-gauge-inv}), the first-order perturbation
(\ref{eq:kouchan-17.357}) of the shear tensor is decomposed as 
\begin{eqnarray}
  \label{eq:kouchan-17.361}
  \stackrel{(1)}{(\sigma_{ab})}
  =
  \stackrel{(1)}{\Sigma_{ab}}
  + {\pounds}_{X}\sigma_{ab}
  ,
\end{eqnarray}
where
\begin{eqnarray}
  \label{eq:kouchan-17.362}
  \stackrel{(1)}{\Sigma_{ab}}
  :=
  \stackrel{(1)}{{\cal B}_{(ab)}}
  - \frac{1}{3} \stackrel{(1)}{{\cal Q}_{ab}} \theta
  - \frac{1}{3} q_{ab} \stackrel{(1)}{\Theta}
  .
\end{eqnarray}
Similarly, through Eqs.~(\ref{eq:first-three-metric-gauge-inv}),
(\ref{eq:second-three-metric-gauge-inv}),
(\ref{eq:first-order-calBab-def}),
(\ref{eq:second-order-calBab-def}),
(\ref{eq:first-theta-gauge-inv}), and
(\ref{eq:second-theta-gauge-inv}), the second-order
perturbation (\ref{eq:kouchan-17.358}) of the shear tensor is
decomposed as
\begin{eqnarray}
  \stackrel{(2)}{\sigma_{ab}}
  =
  \stackrel{(2)}{\Sigma_{ab}}
  + 2 {\pounds}_{X}\stackrel{(1)}{(\sigma_{ab})}
  +   {\pounds}_{Y}\sigma_{ab}
  - {\pounds}_{X}^{2}\sigma_{ab}
  ,
  \label{eq:kouchan-17.365}
\end{eqnarray}
where
\begin{eqnarray}
  \stackrel{(2)}{\Sigma_{ab}}
  &:=&
      \stackrel{(2)}{{\cal B}_{(ab)}}
  - \frac{1}{3} q_{ab} \stackrel{(2)}{\Theta}
  - \frac{1}{3} \stackrel{(2)}{{\cal Q}_{ab}} \theta
  - \frac{2}{3} \stackrel{(1)}{Q_{ab}} \stackrel{(1)}{\Theta}
  .
  \label{eq:kouchan-17.364}
\end{eqnarray}
Here again, Eqs.~(\ref{eq:kouchan-17.361}) and
(\ref{eq:kouchan-17.365}) show that the perturbations of the
shear tensor are decomposed into the gauge-invariant and the
gauge-variant parts in the same form as
Eqs.~(\ref{eq:matter-gauge-inv-decomp-1.0}) and
(\ref{eq:matter-gauge-inv-decomp-2.0}), respectively.


Now, we consider the explicit components of the shear tensor of
each order perturbations.
First, from Eqs.~(\ref{eq:background-three-metric-components}),
(\ref{eq:kouchan-17.451}), (\ref{eq:kouchan-17.498}), and
(\ref{eq:kouchan-17.356}), the background shear tensor is given
by
\begin{eqnarray}
  \sigma_{ab} = 0.
\end{eqnarray}
Second, through
Eqs.~(\ref{eq:background-three-metric-components})--(\ref{eq:kouchan-17.517-5}),
(\ref{eq:kouchan-17.459})--(\ref{eq:kouchan-17.462}), 
and (\ref{eq:kouchan-17.501}), the components of the
gauge-invariant part $\stackrel{(1)}{\Sigma_{ab}}$ of the
first-order perturbation of the shear tensor are given by 
\begin{eqnarray}
  \stackrel{(1)}{\Sigma_{\eta\eta}} &=& \stackrel{(1)}{\Sigma_{\eta i}}
  = \stackrel{(1)}{\Sigma_{i\eta}} = 0
  ,
  \label{eq:shear-explicit-first-order-1}
  \\
  \stackrel{(1)}{\Sigma_{ij}}
  &=&
  a \left\{
    \left(D_{i}D_{j}-\frac{1}{3}\gamma_{ij}\Delta\right)\stackrel{(1)}{v} 
    +             D_{(i}\stackrel{(1)}{{\cal V}_{j)}}
    -             D_{(i}\stackrel{(1)}{\nu_{j)}}
    + \frac{1}{2} \partial_{\eta}\stackrel{(1)}{\chi_{ij}}
  \right\}
  \label{eq:shear-explicit-first-order-2}
  .
\end{eqnarray}
Finally, through
Eqs.~(\ref{eq:background-three-metric-components})--(\ref{eq:kouchan-17.517-8}),
(\ref{eq:kouchan-17.468})--(\ref{eq:kouchan-17.471}),
and (\ref{eq:kouchan-17.498})--(\ref{eq:kouchan-17.505}), the
components of the gauge-invariant part
$\stackrel{(2)}{\Sigma_{ab}}$ of the second-order perturbation 
of the shear tensor are summarized as
\begin{eqnarray}
  \stackrel{(2)}{\Sigma_{\eta\eta}}
  &=&
  0
  \label{eq:shear-explicit-second-order-1}
  , \\
  \stackrel{(2)}{\Sigma_{i\eta}}
  &=&
  \stackrel{(2)}{\Sigma_{\eta i}}
  = 
  2 \left(
      \stackrel{(1)}{\nu^{j}}
    - D^{j}\stackrel{(1)}{v} 
    - \stackrel{(1)}{{\cal V}^{j}}
  \right)
  \stackrel{(1)}{\Sigma_{ij}}
  \label{eq:shear-explicit-second-order-2}
  , \\
  \stackrel{(2)}{\Sigma_{ij}}
  &=&
  a \left[
    \left(D_{i}D_{j}-\frac{1}{3}\gamma_{ij}\Delta\right)\stackrel{(2)}{v}
    +             D_{(i}\stackrel{(2)}{{\cal V}_{j)}}
    -             D_{(i}\stackrel{(2)}{\nu_{j)}}
    + \frac{1}{2} \partial_{\eta}\stackrel{(2)}{\chi_{ij}}
  \right.
  \nonumber\\
  && \quad
  \left.
    + \stackrel{(1)}{\Phi}
    \left(
        2 D_{(i}\stackrel{(1)}{\nu_{j)}}
      -   \partial_{\eta}\stackrel{(1)}{\chi_{ij}}
    \right)
  \right.
  \nonumber\\
  && \quad
  \left.
    + 2 \left(
        D^{k}\stackrel{(1)}{v}
      + \stackrel{(1)}{{\cal V}^{k}}
    \right)
    \left\{
      \left(
        \gamma_{k(i} D_{j)} - \frac{1}{3} \gamma_{ij} D_{k}
      \right)
      \left(
        \stackrel{(1)}{\Phi} + 2 \stackrel{(1)}{\Psi}
      \right)
    \right.
  \right.
  \nonumber\\
  && \quad\quad\quad\quad\quad\quad\quad\quad\quad\quad
  \left.
    \left.
      + \partial_{\eta} \left(
          \gamma_{k(i} D_{j)}\stackrel{(1)}{v}
        + \gamma_{k(i} \stackrel{(1)}{{\cal V}_{j)}}
      \right)
    \right.
  \right.
  \nonumber\\
  && \quad\quad\quad\quad\quad\quad\quad\quad\quad\quad
  \left.
    \left.
      - \frac{1}{3} \gamma_{ij} \partial_{\eta}\left(
        D_{k}\stackrel{(1)}{v}
        + \stackrel{(1)}{{\cal V}_{k}}
      \right)
    \right.
  \right.
  \nonumber\\
  && \quad\quad\quad\quad\quad\quad\quad\quad\quad\quad
  \left.
    \left.
      -             D_{(i}\stackrel{(1)}{\chi_{j)k}}
      + \frac{1}{2} D_{k}\stackrel{(1)}{\chi_{ij}}
    \right\}
  \right.
  \nonumber\\
  && \quad
  \left.
    + 2 \stackrel{(1)}{\nu^{k}}
    \left\{
      - 2 \left(
                      \gamma_{k(i} D_{j)}
        - \frac{1}{3} \gamma_{ij} D_{k}
      \right)\stackrel{(1)}{\Psi}
      +             D_{(i}\stackrel{(1)}{\chi_{j)k}}
      - \frac{1}{2} D_{k}\stackrel{(1)}{\chi_{ij}}
    \right\}
  \right.
  \nonumber\\
  && \quad
  \left.
    + \frac{2}{3} \stackrel{(1)}{\chi^{lk}}
    \left\{
      - \gamma_{ij} D_{l}\stackrel{(1)}{\nu_{k}}
      + \gamma_{ij} D_{l}\left(
        D_{k}\stackrel{(1)}{v}
        + \stackrel{(1)}{{\cal V}_{k}}
      \right)
      + \frac{1}{2} \gamma_{ij} \partial_{\eta}\stackrel{(1)}{\chi_{lk}}
    \right.
  \right.
  \nonumber\\
  && \quad\quad\quad\quad\quad\quad
  \left.
    \left.
      - \gamma_{ik} \gamma_{jl} \left(
        \Delta\stackrel{(1)}{v} 
        - \frac{1}{3} \partial_{\eta}\stackrel{(1)}{\Psi}
      \right)
    \right\}
  \right]
  \label{eq:shear-explicit-second-order-3}
  .
\end{eqnarray}


\subsubsection{Rotation $\bar{\omega}_{ab}$}


The perturbations of the rotation defined by
Eq.~(\ref{eq:kouchan-17.92}) are given by as follows.
The perturbative expansion of the rotation is directly derived
from the perturbative expansion (\ref{eq:kouchan-17.701}) of the
tensor field $\bar{B}_{ab}$ and $\bar{\omega}_{ab}$ is expanded
as 
\begin{eqnarray}
  \bar{\omega}_{ab}
  &=:&
    \omega_{ab}
  + \lambda \stackrel{(1)}{\omega_{ab}}
  + \frac{1}{2} \lambda^{2} \stackrel{(2)}{\omega_{ab}}
  + O(\lambda^{3})
  ,
  \label{eq:kouchan-17.367}
\end{eqnarray}
and we have
\begin{eqnarray}
  \label{eq:kouchan-17.368}
  \omega_{ab} := B_{[ab]}
  , \quad
  \stackrel{(1)}{\omega_{ab}} := \stackrel{(1)}{(B_{[ab]})}
  , \quad
  \label{eq:kouchan-17.370}
  \stackrel{(2)}{\omega_{ab}} := \stackrel{(2)}{(B_{[ab]})}
  .
\end{eqnarray}
We can also define the gauge-invariant expression of the first-
and the second-order perturbation of the rotation:
\begin{eqnarray}
  \label{eq:kouchan-17.371}
  \stackrel{(1)}{\omega_{ab}}
  =
  \stackrel{(1)}{\Omega_{ab}} + {\pounds}_{X}\omega_{ab}
  , \quad
  \stackrel{(2)}{\omega_{ab}}
  =
  \stackrel{(2)}{\Omega_{ab}}
  + 2 {\pounds}_{X}\stackrel{(1)}{\omega_{ab}}
  + \left({\pounds}_{Y} - {\pounds}_{X}^{2}\right)\omega_{ab}
  ,
\end{eqnarray}
where the gauge-invariant variables for the first- and the
second-order perturbations of the rotation are given by 
\begin{eqnarray}
  \label{eq:kouchan-17.375}
  \stackrel{(1)}{\Omega_{ab}}
  =
  \stackrel{(1)}{{\cal B}_{[ab]}}
  , \quad
  \stackrel{(2)}{\Omega_{ab}}
  =
  \stackrel{(2)}{{\cal B}_{[ab]}}
  .
\end{eqnarray}


From Eq.~(\ref{eq:kouchan-17.451}), the background value of the
rotation is given by 
\begin{eqnarray}
  \omega_{ab}
  =
  a {\cal H} \gamma_{[ab]}
  =
  0
  \label{eq:kouchan-17.592}
  .
\end{eqnarray}
Through
Eqs.~(\ref{eq:kouchan-17.459})--(\ref{eq:kouchan-17.462}), the
components of the gauge-invariant part
$\stackrel{(1)}{\Omega_{ab}}$ of the first-order perturbation of
the rotation are given by
\begin{eqnarray}
  \label{eq:kouchan-17.598}
  \stackrel{(1)}{\Omega_{\eta\eta}}
  =
  \stackrel{(1)}{\Omega_{\eta i}}
  =
  \stackrel{(1)}{\Omega_{i\eta}}
  = 0
  , \quad
  \stackrel{(1)}{\Omega_{ij}}
  =
  a D_{[i}\stackrel{(1)}{{\cal V}_{j]}}
  .
\end{eqnarray}
Further, through
Eqs.~(\ref{eq:kouchan-17.468})--(\ref{eq:kouchan-17.471}), the
components of the gauge-invariant part of the second-order
perturbation of the rotation are given by
\begin{eqnarray}
  \label{eq:kouchan-17.606}
  \stackrel{(2)}{\Omega_{\eta\eta}}
  &=&
  0
  ,
  \\
  \label{eq:kouchan-17.607}
  \stackrel{(2)}{\Omega_{i\eta}}
  &=&
  - \stackrel{(2)}{\Omega_{\eta i}}
  =
  2 a \left(
      \stackrel{(1)}{\nu^{j}}
    - D^{j}\stackrel{(1)}{v} 
    - \stackrel{(1)}{{\cal V}^{j}}
  \right)
  \left(
    D_{[i}\stackrel{(1)}{{\cal V}_{j]}}
  \right)
  ,
  \\
  \stackrel{(2)}{\Omega_{ij}}
  &=&
  a \left\{
    D_{[i}\stackrel{(2)}{{\cal V}_{j]}}
    - 2 \left(
        D^{k}\stackrel{(1)}{v}
      + \stackrel{(1)}{{\cal V}^{k}}
    \right)
    \left(
        D_{[i}\stackrel{(1)}{\Phi} \gamma_{j]k}
      + \partial_{\eta}D_{[i}\stackrel{(1)}{v} \gamma_{j]k}
    \right.
  \right.
  \nonumber\\
  && \quad\quad\quad\quad\quad\quad\quad\quad\quad\quad\quad\quad\quad\quad
  \left.
    \left.
      + \partial_{\eta}\stackrel{(1)}{{\cal V}_{[i}} \gamma_{j]k}
    \right)
  \right\}
  \label{eq:kouchan-17.609}
  .
\end{eqnarray}



\end{document}